\documentclass[preprint2]{emulateapj}

\input{ulem.sty}
\shorttitle{Spatial Clustering of RASS-AGN: IV. -- More massive SMBHs reside in more massive DMHs}
\shortauthors{Krumpe et al.}

\begin{document}
\def\mpch {$h^{-1}$ Mpc} 
\def\kpch {$h^{-1}$ kpc} 
\def\kms {km s$^{-1}$} 
\def\lcdm {$\Lambda$CDM} 
\def\xir {$\xi(r)$}
\def\wprp {$w_p(r_p)$}
\def\xisp {$\xi(r_p,\pi)$}
\def\xis {$\xi(s)$}
\def\rr {$r_0$}
\def\etal {et al.}

\title{THE SPATIAL CLUSTERING OF \textit{ROSAT} ALL-SKY SURVEY ACTIVE GALACTIC NUCLEI\\ IV. 
More massive black holes reside in more massive dark matter halos}

\author{Mirko Krumpe\altaffilmark{1,2,3,4}, Takamitsu Miyaji\altaffilmark{5,3}, 
        Bernd Husemann\altaffilmark{2,1}, Nikos Fanidakis\altaffilmark{6}, Alison L. Coil\altaffilmark{3}, 
                                    and Hector Aceves\altaffilmark{5}}

\altaffiltext{1}{Leibniz-Institut f\"ur Astrophysik Potsdam (AIP), An der
  Sternwarte 16, D-14482 Potsdam, Germany}
\altaffiltext{2}{European Southern Observatory, ESO Headquarters, 
                 Karl-Schwarzschild-Stra\ss e 2, D-85748 Garching bei 
                 M\"unchen, Germany}
\altaffiltext{3}{University of California, San Diego, Center for Astrophysics and
                 Space Sciences, 9500 Gilman Drive, La Jolla, CA 92093-0424, USA}
\altaffiltext{4}{Max-Planck-Institut f\"ur extraterrestrische Physik, Gie\ss enbachstra\ss e 1, 
                 D-85748 Garching, Germany}
\altaffiltext{5}{IAUNAM-E (Instituto de Astronom\'ia de la Universidad Nacional
                 Aut\'onoma de M\'exico, Ensenada), P.O. Box 439027, San Diego, 
                 CA 92143-9027, USA}
\altaffiltext{6}{Max-Planck-Institut f\"ur Astronomie , K\"onigstuhl 17, 69117 Heidelberg, Germany}
\email{mkrumpe@aip.de}

\begin{abstract}

This is the fourth paper in a series that reports on our investigation of the clustering properties 
of active galactic nuclei (AGNs) identified in the \textit{ROSAT} All-Sky Survey (RASS) and Sloan Digital 
Sky Survey (SDSS). In this paper we investigate the cause of the X-ray luminosity dependence 
of the clustering of broad-line, luminous AGNs at $0.16<z<0.36$.  
We fit the H$\alpha$ line profile in the SDSS spectra for all X-ray and
optically selected broad-line 
AGNs, determine the mass of the supermassive black hole (SMBH), $M_{\rm BH}$, and infer the accretion 
rate relative to Eddington ($L/L_{\rm EDD}$). Since $M_{\rm BH}$ and $L/L_{\rm EDD}$ are correlated, 
we create AGN subsamples in one parameter while maintaining the same
distribution in the other parameter. In both the X-ray and optically selected AGN samples, we detect a 
weak clustering dependence with $M_{\rm BH}$ and no statistically significant dependence on $L/L_{\rm EDD}$. 
We find a difference of up to 2.7$\sigma$ when comparing the objects that belong to the 30\% 
least and 30\% most massive $M_{\rm BH}$ subsamples, in that luminous broad-line AGNs with more 
massive black holes reside in more massive parent dark matter halos at these
redshifts.  
These results provide evidence that higher accretion rates in AGNs do not 
necessarily require dense galaxy environments, in which more galaxy mergers 
and interactions are expected to channel large amounts of gas onto the SMBH.   
We also present semianalytic models that predict a positive $M_{\rm DMH}$ dependence on $M_{\rm BH}$, which is most prominent at
$M_{\rm BH} \sim 10^{8-9}\,\rm{M}_\odot$.


\end{abstract}



\section{Introduction}
\label{introduction}
 
There has been increasing interest in 
large-scale clustering measurements of active galactic nucleus (AGNs) in 
recent years. Such measurements (see review by \citealt{krumpe_miyaji_2014}) 
not only allow one to study the distribution of matter in the Universe out 
to redshifts where it 
becomes very challenging and observationally expensive to detect galaxies in 
large numbers, but they also can be used to constrain theoretical models of AGN/galaxy coevolution, 
feedback mechanisms, AGN host-galaxy properties, the distribution of AGNs with 
dark matter halo (DMH) mass, and the fueling process(es) of supermassive black 
holes (SMBH) (e.g., \citealt{porciani_magliocchetti_2004}; \citealt{gilli_daddi_2005,gilli_zamorani_2009}; 
\citealt{yang_mushotzky_2006}; \citealt{coil_georgakakis_2009}; \citealt{ross_shen_2009}; 
\citealt{cappelluti_ajello_2010}; 
\citealt{krumpe_miyaji_2010,krumpe_miyaji_2012}; \citealt{allevato_2011}; 
\citealt{miyaji_krumpe_2011}; \citealt{mountrichas_georgakakis_2012}; 
\citealt{koutoulidis_plionis_2013}).

Spatial correlation measurements with several tens of thousands of galaxies yield significant 
clustering dependences on galaxy properties such as luminosity, morphological type, spectral 
type, and redshift (e.g., \citealt{norberg_baugh_2002}; \citealt{madgwick_hawkins_2003}; 
\citealt{zehavi_zheng_2005,zehavi_zheng_2011}; \citealt{meneux_fevre_2006,meneux_guzzo_2009}; \citealt{coil_newman_2008}).
These findings confirm the hierarchical model of structure 
formation in which more massive, and hence more luminous, galaxies 
reside in more massive DMHs and are therefore clustered more strongly
(e.g., \citealt{zehavi_zheng_2005, zehavi_zheng_2011}; \citealt{coil_newman_2006}). 
Whether this relation should also apply to AGN luminosity is not trivial. Since the AGN luminosity 
depends primarily on the mass of the SMBH ($M_{\rm BH}$), the accretion rate relative to 
Eddington ($L/L_{\rm EDD}$), and the radiative accretion efficiency, a
relation between clustering and AGN luminosity should ultimately be due to a
connection between DMH mass and one or more of these physical
parameters. 

Based on smoothed-particle hydrodynamic simulations using {\tt GADGET} (\citealt{springel_2005}), 
\cite{booth_schaye_2010} explore the correlation between SMBH mass and the mass of the hosting DMH. 
In the simulations, the black holes grow either by the accretion of ambient gas or mergers. The 
black holes also inject a fixed fraction of the rest mass energy of the gas into the surrounding 
medium. A self-regulating black hole injects enough energy to displace gas from 
the host galaxy on longer timescales. The binding energy of the gas is determined by the DMH potential. 
Thus, \cite{booth_schaye_2010} conclude that the mass of the SMBH is regulated primarily by 
the DMH mass and not the stellar mass of the galaxy.
\cite{fanidakis_baugh_2012} use the semianalytical galaxy-formation model {\tt GALFORM} 
(\citealt{cole_lacey_2000}; more details on their simulations are given below in Sect.~\ref{simulations}) and 
find a correlation between SMBH mass and DMH mass at almost all cosmic times ($z=0.0-6.2$).
\cite{volonteri_natarajan_2011} use the measured black hole mass, velocity dispersion $\sigma$, and
asymptotic circular velocity of 25 local galaxies from \cite{kormendy_bender_2011} to show that, 
although with some scatter, the black hole masses correlate well with the parent DMH masses.

AGN clustering studies can be used to observationally test such predictions. 
Several studies have measured the large-scale clustering dependence on AGN on properties such as X-ray luminosity, with 
varying results. \cite{coil_georgakakis_2009} find no correlation between the 
AGN X-ray luminosity and the clustering strength at $z=0.7-1.4$. 
At $z \sim 0.1$, \cite{mountrichas_georgakakis_2012} also detect no significant 
correlation in their sample of \textit{XMM-Newton}-selected AGNs. 
\cite{yang_mushotzky_2006} reports a tentative ($\sim$1$\sigma$) dependence of the clustering 
signal, such that the brighter sample is more clustered than the 
fainter sample. 
\cite{cappelluti_ajello_2010} and \cite{koutoulidis_plionis_2013} verify this finding using 
different AGN samples at different redshifts. \cite{koutoulidis_plionis_2013} use a large sample 
of $\sim$1500 AGNs from the Chandra Deep Field (CDF) North and South, the extended CDF South, 
COSMOS, and AEGIS. The median redshifts of their low and high $L_{\rm X}$ AGN samples 
are $\langle z \rangle \sim 0.8-1.1$. Except for the CDF South survey, they find weak 
X-ray luminosity dependences at a level of up to $\sim$2$\sigma$. 

One way to reduce the uncertainties involved in clustering measurements --- beyond using larger AGN samples --- is to 
compute the cross-correlation function (CCF) with a dense galaxy sample 
instead of computing the autocorrelation function (ACF) of the AGNs.
Several studies (e.g., \citealt{li_kauffmann_2006}; \citealt{coil_hennawi_2007, coil_georgakakis_2009}; 
\citealt{wake_croom_2008}; \citealt{hickox_jones_2009};
\citealt{mountrichas_georgakakis_2013}; \citealt{georgakakis_mountrichas_2014})
demonstrate the potential of this approach and 
compute the CCF between AGNs and a large sample of galaxies to infer the 
ACF of the AGNs. The significant increase in the number of pairs 
at a given separation, used to measure the clustering strength, 
reduces the uncertainties in the spatial correlation function 
compared to the direct measurements of the AGN ACF. 

In \citet[hereafter Paper I]{krumpe_miyaji_2010}, we use the same technique 
as \cite{coil_georgakakis_2009} to 
measure the CCF between \textit{ROSAT} All-Sky Survey (RASS) AGNs
identified in the Sloan Digital Sky Survey (SDSS) and a large set of SDSS 
luminous red galaxies (LRGs) at $0.16<z<0.36$. The study is based 
on SDSS data release 4+ (DR4+). The unprecedented low uncertainties of the inferred 
broad-line AGN ACF allow us to split the sample into low and high X-ray luminosity 
subsamples and to report a $\sim$2.5$\sigma$ X-ray luminosity dependence of 
broad-line AGN clustering. We find that higher luminosity AGNs cluster more strongly 
than their lower luminosity counterparts. Consequently, higher luminosity X-ray 
AGNs reside, on average, 
in more massive DMHs than do lower luminosity X-ray AGN.

In the second paper of this series \citep[hereafter Paper II]{miyaji_krumpe_2011}, 
we describe a novel method of applying the halo occupation distribution (HOD) modeling 
technique directly to the precise measured CCF between RASS/SDSS AGNs 
and SDSS LRGs to constrain the distribution of AGNs as a function of 
DMH mass. This 
method also allows us to derive the large-scale bias parameter of the 
AGN sample with much lower systematic uncertainties than using a 
phenomenological power-law fit, as is often done.  
As shown in Paper II, the X-ray luminosity dependence is more prominent in the one-halo 
term ($r_p <1$ $h^{-1}$ Mpc). The HOD-based typical DMH masses derived from the two-halo term 
for the high- and low-luminosity RASS/SDSS AGN subsamples differ by $\sim$1.8$\sigma$.
In addition, we find that models where the AGN fraction among satellites decreases with 
DMH mass beyond $M_{\rm DMH} \sim 10^{12}$ $h^{-1}$ $M_{\odot}$ are preferred. This is in 
contrast to what is found for satellite galaxies without AGNs 
(\citealt{zheng_zehavi_2009}; \citealt{zehavi_zheng_2011}).  

In the third paper \citep[hereafter Paper III]{krumpe_miyaji_2012}, we extend the 
cross-correlation measurements to lower and higher redshifts, covering a redshift 
range of $z=0.07-0.50$. We show that the weak X-ray luminosity dependence of 
broad-line AGN clustering is also found if radio-detected AGNs are excluded.
Furthermore, we compute the large-scale clustering for optically selected broad-line SDSS AGNs 
using the final SDSS DR7, but we detect no optical luminosity dependence on the clustering strength, 
although the optical broad-line SDSS AGN sample in $0.16<z<0.36$ contains more than twice 
as many objects as the RASS/SDSS AGN sample. 
We conclude that the most likely explanation for this result 
is the smaller dynamic range probed in optical luminosities compared to 
X-ray luminosities. 

In this paper we focus on the redshift range $0.16<z<0.36$, where 
the CCF of RASS/SDSS AGNs and LRGs has the highest signal-to-noise ratio (S/N). We 
fit the H$\alpha$ line profile in the SDSS spectra of broad-line AGN 
to infer the $M_{\rm BH}$ and the 
$L/L_{\rm EDD}$. Dividing the AGN sample into 
low and high $M_{\rm BH}$ mass, as well as low and high $L/L_{\rm EDD}$, provides insights 
into the main physical driver of the weak detected 
X-ray luminosity dependence of broad-line AGN clustering. 

This paper is organized as follows. In Section~2, we describe the properties of the 
LRG tracer set and the AGN samples. Section~3 provides details on how we fit the 
H$\alpha$ line profile in the optical SDSS AGN spectra, derive the $M_{\rm BH}$, estimate 
$L/L_{\rm EDD}$, and define our AGN subsamples. In Section~4, we briefly summarize 
the cross-correlation technique, 
how the AGN ACF is inferred from this, and how we derive the large-scale bias 
parameters using HOD modeling. Section~5 provides the results of our clustering measurements. 
The detailed results of the HOD modeling of 
the CCFs presented in this paper and in paper III will be included in a future paper 
(T. Miyaji et al. in preparation). Our results are discussed in Section~6, and 
we present our conclusions in Section~7. Throughout the paper, all distances are measured in comoving 
coordinates and given in units of $h^{-1}$\,Mpc, where $h= H_{\rm 0}/100$\,km\,s$^{-1}$\,Mpc$^{-1}$, unless 
otherwise stated. We use a cosmology of $\Omega_{\rm m} = 0.3$, $\Omega_{\rm \Lambda} = 0.7$, and 
$\sigma_8(z=0)=0.8$, which is consistent with the {\it WMAP} data release 7 
(Table~3 of \citealt{larson_dunkley_2011}). The same cosmology is used in Papers I--III.
Luminosities and absolute magnitudes are calculated for $h=0.7$.
We use AB magnitudes throughout the paper. All uncertainties represent 1$\sigma$ 
(68.3\%) confidence intervals unless otherwise stated.


\section{Data}

The data sets used in this study are drawn from the SDSS, which consists 
of an imaging survey in five bands and an extensive spectroscopic
follow-up survey with a fiber spectrograph. 
The selection 
of the optically selected AGN candidates is described in \cite{richards_fan_2002}. 
LRGs are chosen by following \cite{eisenstein_annis_2001}.

\subsection{SDSS Luminous Red Galaxy Sample} 
\label{desc_LRG}

The selection of SDSS LRGs follows the procedure described in Section~2.1 
of Paper I and Section~2.2 of Paper III. Here we briefly summarize the 
sample selection. We extract LRGs from the web-based SDSS Catalog Archive Server 
Jobs System\footnote{http://casjobs.sdss.org/CasJobs/} using the flag 
``galaxy\_red,'' which is based on the selection criteria defined in 
\cite{eisenstein_annis_2001}. We verify that the extracted objects meet all LRG 
selection criteria and create a volume-limited sample with $0.16<z<0.36$ and 
$-23.2<M_g^{0.3}<-21.2$, where $M_g^{0.3}$ is based on the extinction-corrected 
$r^{*}_{\rm petro}$ magnitude, $k$-corrected and passively 
evolved to rest-frame $g^{*}_{\rm petro}$ magnitudes at $z=0.3$. 
We consider only  
LRGs that fall into the SDSS area with a DR7 spectroscopic completeness ratio of 
greater than 0.8 and that have a redshift confidence level of 
greater than 0.95. 
The SDSS geometry and completeness ratio are expressed in terms of spherical 
polygons (\citealt{hamilton_tegmark_2004}). The file is publicly 
available\footnote{http://sdss.physics.nyu.edu/lss/dr72}$^,$\footnote{http://sdss.physics.nyu.edu/lss/dr4plus}
 through the 
New York University Value-Added Galaxy Catalog (NYU-VAGC) website 
(\citealt{blanton_schlegel_2005}).

We correct for the SDSS fiber collision as described in detail in 
\cite{krumpe_miyaji_2010, krumpe_miyaji_2012}. We have to assign to
approximately 2\% of all LRGs in our sample a redshift due to the 
fiber collision problem.

The construction of the random LRG sample is identical to 
the procedure described in Section~3.1 of Paper I.
Our LRG random sample contains 200 times as many objects as the real 
LRG sample. We generate a set of random R.A. and decl. values within DR7 areas
with spectroscopic completeness ratios of greater than 0.8, populate areas 
with higher completeness 
ratios more than ones with lower completeness ratios, and randomly assign redshifts to 
the objects in the sample by using the smoothed redshift profile of the observed 
redshift distribution.


\subsection{RASS/SDSS AGN Samples}
\label{RASS_sample}
The \textit{ROSAT} All-Sky Survey (RASS, \citealt{voges_aschenbach_1999}) is 
currently still the most sensitive all-sky survey in the soft (0.1--2.4 keV) X-ray regime. 
\cite{anderson_voges_2003, anderson_margon_2007} positionally cross-correlate 
RASS sources with SDSS spectroscopic objects and classify RASS- and SDSS-detected 
AGNs based on SDSS DR5. They find 6224 AGN with broad permitted emission lines in excess
of 1000 km\,s$^{-1}$ FWHM and 515 narrow permitted emission line AGNs 
matching RASS sources within 1 arcmin. More details on the sample selection 
are given in Section~2.2 of Paper I and \cite{anderson_voges_2003, anderson_margon_2007}.
Since \textit{ROSAT} observed the sky in the soft energy band (0.1-2.4 keV), the 
RASS/SDSS AGN sample is biased toward AGNs with little to no X-ray absorption. 
The vast majority of the optical counterparts are therefore AGNs with broad emission 
lines and UV excess. There is no overlap between the RASS/SDSS AGNs and the LRG sample.

To study the X-ray luminosity dependence of the clustering of these AGNs, 
we have to limit the 
SDSS footprint to the publicly available DR4+ geometry. We split our sample 
in the redshift range $0.16<z<0.36$ into subsamples according to X-ray 
luminosity. We use a 0.1--2.4 keV observed luminosity cut (assuming a 
photon index $\Gamma =2.5$, corrected for Galactic absorption) of 
log $(L_{\rm X}/[\rm{erg}\,\rm{s}^{-1}])=44.29$. The observed flux in the 0.1--2.4 keV 
band has a large soft-excess contribution that is not representative of the 
underlying intrinsic, hard power-law X-ray spectrum. Thus, we use the template 
\textit{XMM-Newton} spectrum of powerful radio-quiet \textit{ROSAT} QSOs from 
\cite{krumpe_lamer_2010} to estimate the corresponding flux in the 0.5--10 keV 
and 2--10 keV energy ranges. Using the median redshift of the sample ($z=0.27$), 
the cut at 0.1--2.4 keV corresponds to a cut at 
log $(L_{\rm 0.5-10\,keV}/[\rm{erg}\,\rm{s}^{-1}])=43.7$ and 
log $(L_{\rm 2-10\,keV}/[\rm{erg}\,\rm{s}^{-1}])=43.4$. 

As an alternative estimate, we match our RASS AGN sample with the 
3XMM-DR5 catalog (\citealt{rosen_webb_2015}). We fit a regression line between 
the RASS 0.1--2.4 keV fluxes and the {\it XMM-Newton} 2--12 keV fluxes. Using 
{\it Xspec} (\citealt{arnaud_1996}), we model the same ratio 
with a broken
power law and determine the luminosity ratios in the ranges 0.1--2.4 keV, 
0.5--10 keV, and 2--10 keV. We find somewhat smaller corrections: log 
$(L_{\rm X}/[\rm{erg}\,\rm{s}^{-1}])=44.29$ corresponds to 
log $(L_{\rm 0.5-10\,keV}/[\rm{erg}\,\rm{s}^{-1}])=43.9$ and 
log $(L_{\rm 2-10\,keV}/[\rm{erg}\,\rm{s}^{-1}])=43.7$. 
This approach has the disadvantage that the observations 
were taken over a decade apart, and temporal variation in the X-ray
luminosity/flux of the objects will affect the estimate. However, one 
could argue that such variations are effectively averaged over 
in a large sample of X-ray objects as used here. Fits to the 
0.5--10 keV {\it XMM-Newton} spectra verified that 
the vast majority ($\sim$95\%) of the cross-matched RASS/XMM AGNs are unabsorbed in the 
X-rays ($N_{\rm H} < 10^{21}$ cm$^{-2}$), while the remaining sources show absorption
at a level of only a few 10$^{21}$ cm$^{-2}$.

We calculate the comoving number densities as described in detail in Paper I. 
For a given R.A. and decl., we compute the limiting observable RASS count 
rate and infer the absorption-corrected flux limit versus survey area for RASS/SDSS AGNs.
We then compute the comoving volume available to each object ($V_{\rm a}$) to be included in the sample (\citealt{avni_bahcall_1980}). The comoving number density follows by 
computing the sum of the available volume over each object, $n_{\rm AGN}= \sum_i 1/V_{{\rm a},i}$. 
The comoving number densities for the total, high $L_{\rm X}$, and low $L_{\rm X}$ subsamples are 6.0, 0.12, 
and 5.8 $\times 10^{-5}$ $h^{3}$ Mpc$^{-3}$, respectively.


\section{Deriving $M_{\rm BH}$ and $L/L_{\rm EDD}$ through H$\alpha$ line profile fits}
\label{M_BH__LLedd}
The so-called virial method is routinely employed to estimate $M_\mathrm{BH}$ 
from single-epoch spectra based solely on the width of broad emission 
lines together 
with an appropriate continuum luminosity (\citealt{kaspi_smith_2000}; \citealt{peterson_wandel 2000};  
\citealt{vestergaard_2002}). The FWHM of H$\beta$ 
($\mathrm{FWHM}_{\mathrm{H}\beta}$) and the rest-frame continuum luminosity at 5100\,\AA\ ($L_{5100}$) 
are most commonly used to compute $M_\mathrm{BH}$ for low-redshift AGNs, because the corresponding 
calibrations are the best-studied ones to date. On the other hand, the H$\alpha$ line offers a 
higher S/N than H$\beta$ and tight relations between $\mathrm{FWHM}_{\mathrm{H}\beta}$ and 
$\mathrm{FWHM}_{\mathrm{H}\alpha}$, as well as between $L_{5100}$ and $L_{\mathrm{H}\alpha}$, have been established 
(\citealt{greene_ho_2005}, \citealt{schulze_wisotzki_2010}). Consequently, accurate $M_\mathrm{BH}$ 
estimates can also be obtained from the broad H$\alpha$ line. 

The simultaneous use of the broad H$\alpha$ line as a proxy for the 
velocity dispersion in the broad-line region (BLR) 
and for the AGN luminosity has two major advantages. First, 
we avoid any potential 
contamination of $L_{5100}$ by host galaxy light and the influence of the broad \ion{Fe}{2} emission in type I AGNs 
on the broad H$\beta$ FWHM measurement (\citealt{greene_ho_2005}). Second, we 
obtain reliable BH mass estimates even for those objects 
where the S/N of $\mathrm{H}\beta$ would be too low.

\subsection{Fitting the H$\alpha$ Line Profile}
\label{halpha_fit}
\begin{figure}
  \centering
  \includegraphics[width=8.5cm]{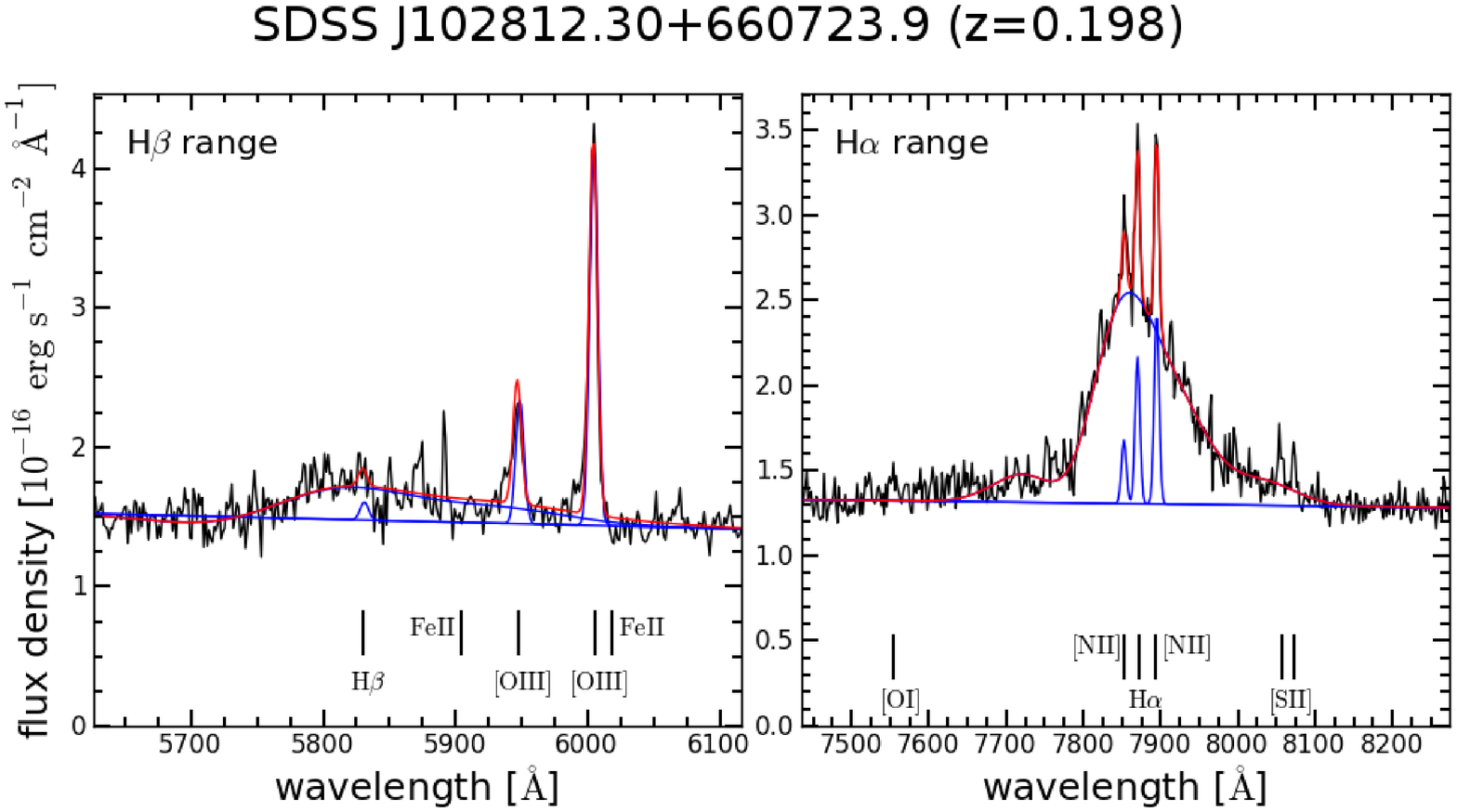}\\
  \includegraphics[width=8.5cm]{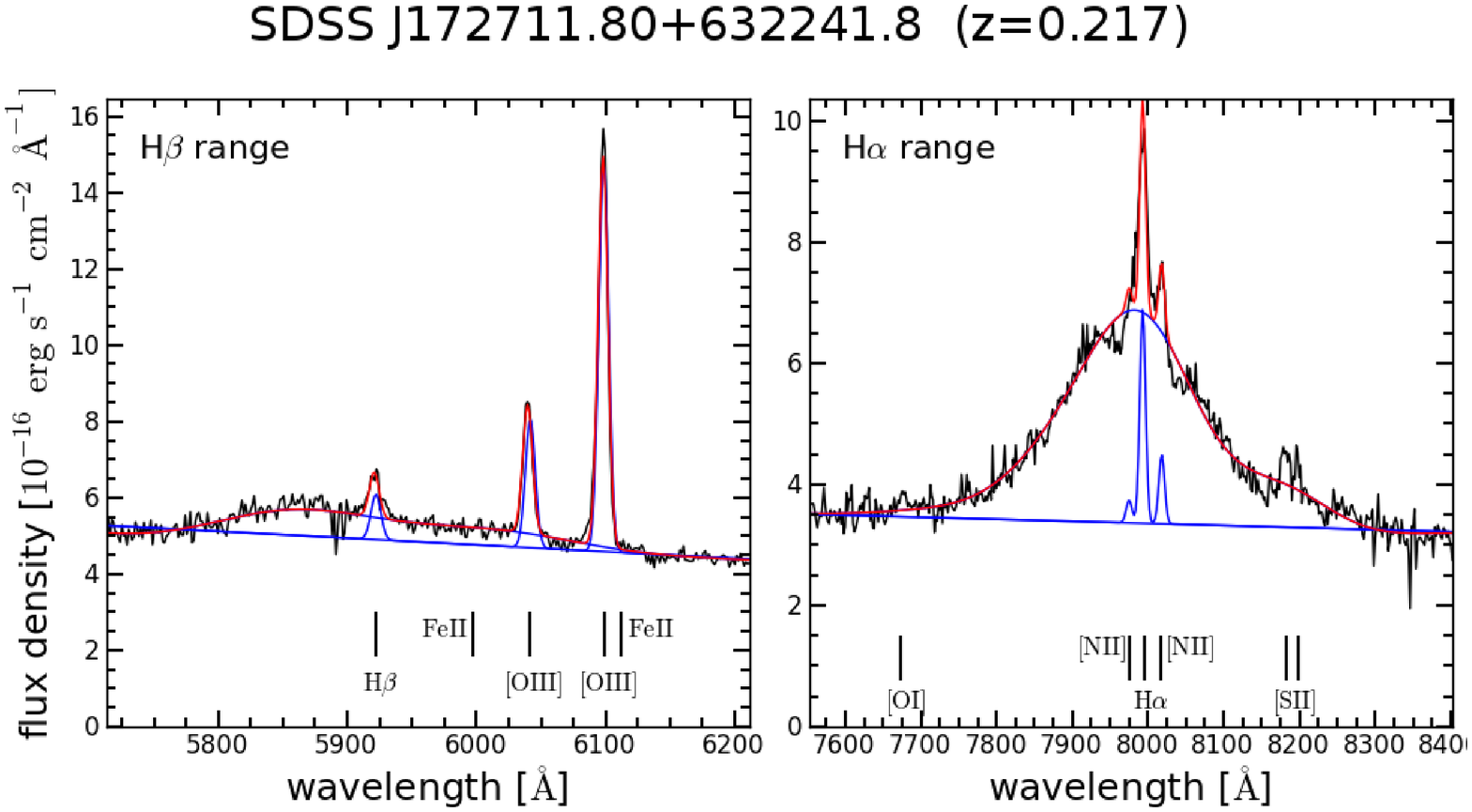}\\
  \includegraphics[width=8.5cm]{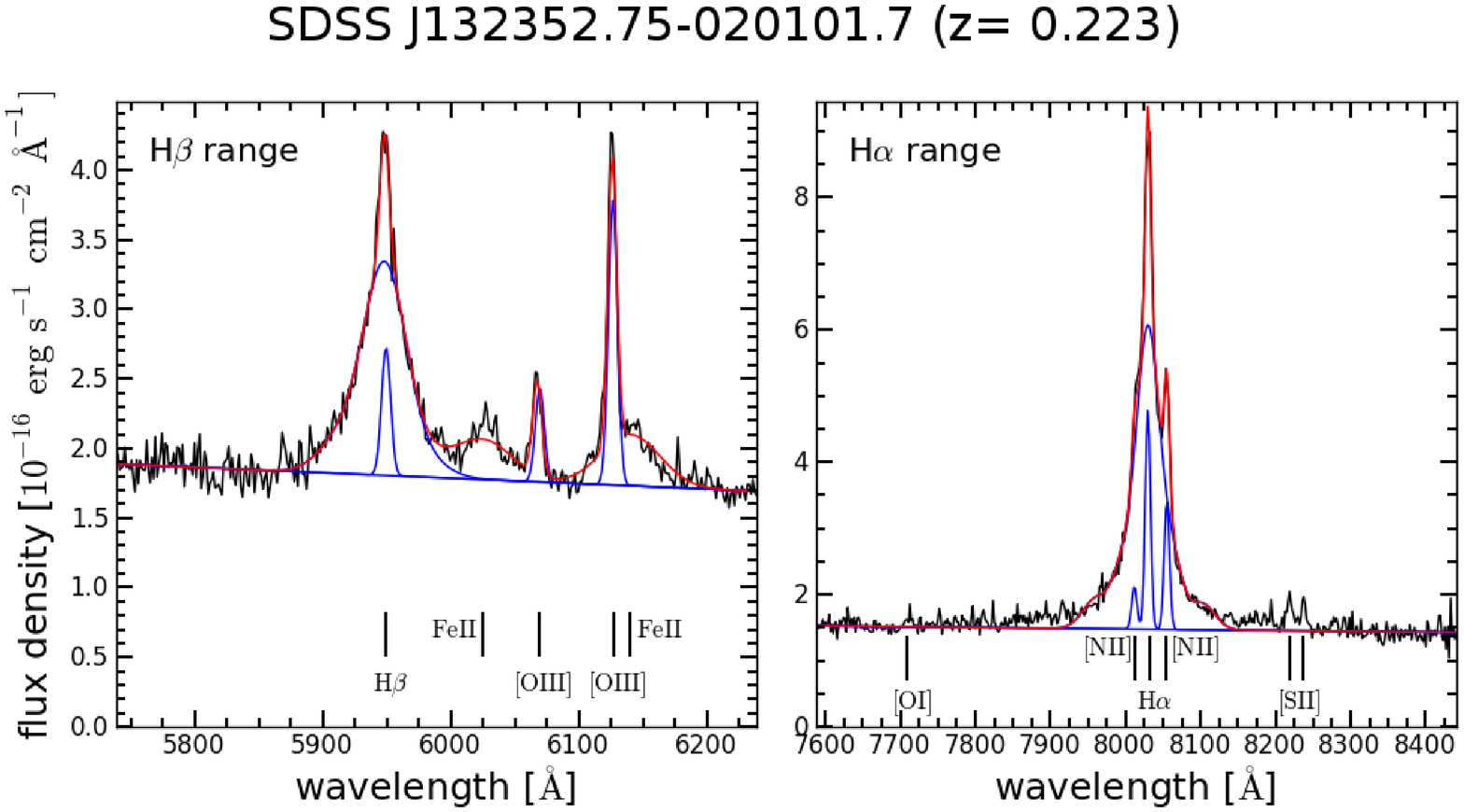}\\
 \includegraphics[width=8.5cm]{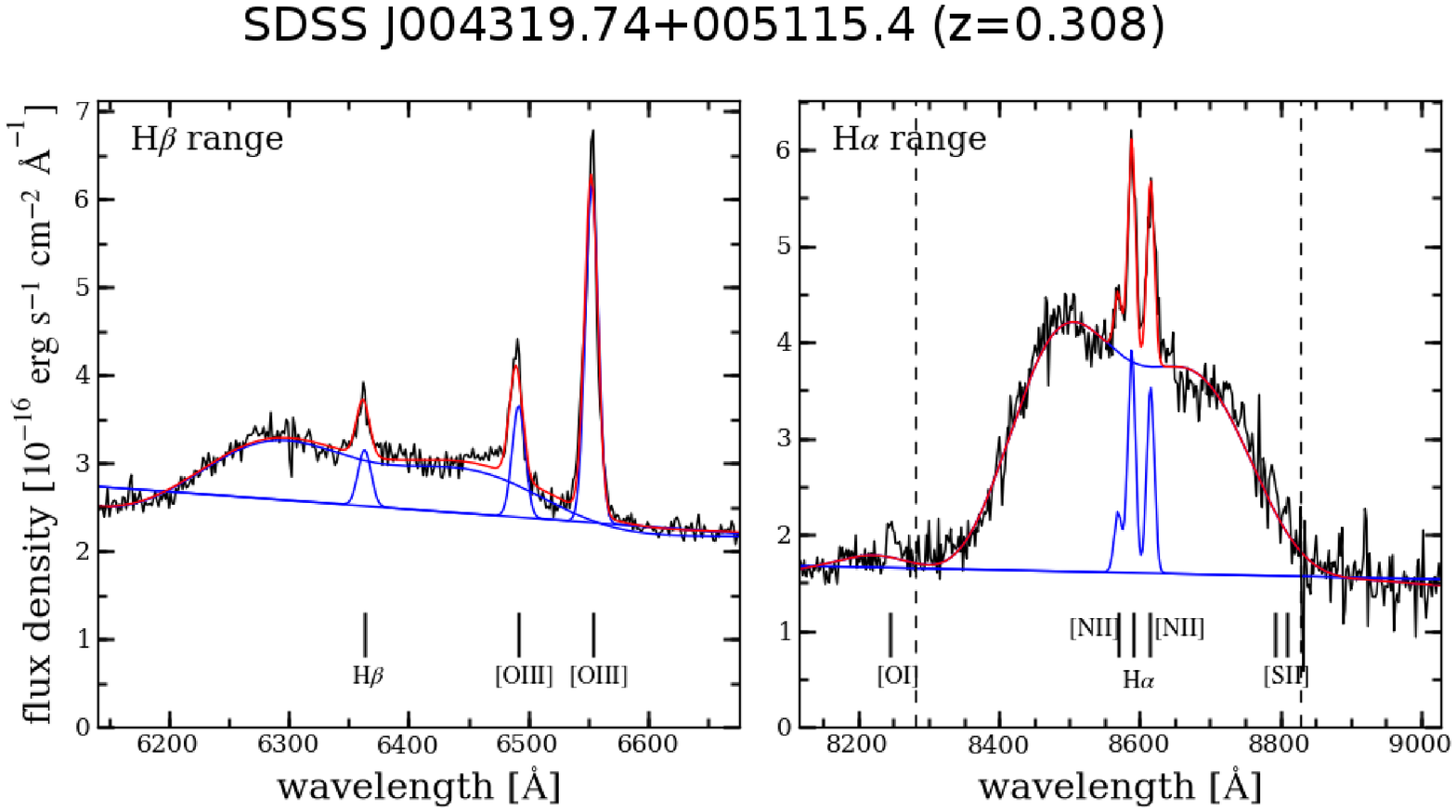}
      \caption{Observed SDSS spectra (black solid lines) and the corresponding models (red solid lines) in the 
               observed H$\beta$ (left panels) and H$\alpha$ (right panels) wavelength ranges. 
                The fitted individual line components are indicated by blue solid lines 
               (underlying continuum, narrow and broad lines).
                The most prominent line features 
                are labeled.  
               }
         \label{AGN_SDSS_SPEC_FIT}
\end{figure}

We retrieve the individual SDSS DR7 optical spectra for our entire RASS/SDSS AGN sample in order to measure 
the luminosity and FWHM of H$\alpha$ for each AGN. Although the FWHM and integrated flux of H$\alpha$ 
could in principle be measured directly from the spectra, the narrow H$\alpha$ and [\ion{N}{2}] lines on 
top of the broad H$\alpha$ line have a significant impact on the estimated $M_{\rm BH}$ and $L/L_{\rm EDD}$ 
values (Fig.~\ref{AGN_SDSS_SPEC_FIT}, left panels), as highlighted by
\cite{denney_peterson_2009}. 
The narrow H$\alpha$ originates from star formation in the host galaxy and the narrow-line region (NLR) 
photoionized by the AGN outside of the central. Deblending the broad and narrow emission lines
is therefore essential to minimize systematic effects by separating the
contribution from the BLR and the NLR/host in the H$\alpha$ line.

One major complication in modeling the AGN spectra is the commonly complex, non-Gaussian shape of 
the broad H$\alpha$ line. Previous studies have used multiple Gaussian components  
(e.g., \citealt{brotherton_wills_1994}; \citealt{sulentic_marziani_2002}; \citealt{shen_greene_2008}; 
\citealt{schulze_wisotzki_2009}) or high-order Gauss-Hermite polynomials (e.g., 
\citealt{salviander_shields_2007}; \citealt{mcgill_woo_2008}; \citealt{stern_laor_2012}) to describe the 
asymmetries of the line. We test both approaches to fit the SDSS spectra and
conclude that the 
latter is more robust against the choice of initial parameter estimates.

We simultaneously model the broad H$\beta$ and H$\alpha$ lines together with the narrow 
H$\beta$, [\ion{O}{3}]\,$\lambda4960, \lambda5007$, H$\alpha$, and [\ion{N}{2}]\,$\lambda6548, \lambda6583$ 
emission lines. We briefly describe our algorithm of modeling the broad H$\alpha$ line 
below and follow an approach similar to \cite{stern_laor_2012}. 
The model of the H$\beta$ region (left panels in Fig.~\ref{AGN_SDSS_SPEC_FIT}) is only used to better constrain the parameters 
of narrow lines. 
The well-isolated narrow [\ion{O}{3}] line in the H$\beta$ region gives
parameter restrictions that are superior to the narrow-line profiles in the H$\alpha$ region, where the narrow and broad lines can be strongly blended.

As a first step, the approximate underlying continua are 
independently subtracted from the H$\beta$ and H$\alpha$ lines by linear interpolation between the adjacent 
rest-frame spectral regions 4720--4760 \AA\ and 5070--5130 \AA, as well as 6150--6230 \AA\ and 6750--6950 \AA , respectively. 
The fit requires initial guesses for the broad and narrow line components. The initial guesses for the FWHM and 
line flux of the broad lines are determined by a fourth-order Gauss-Hermite polynomial of the entire H$\alpha$ line
and assuming standard H$\alpha$ and H$\beta$ ratios. 
The initial guesses of the narrow line components are 
based on the total flux of the observed [\ion{O}{3}] line and typical narrow-line AGN ratios.    
An eighth-order Gauss-Hermite polynomial provides a good fit for the full model of complex broad H$\beta$ and H$\alpha$ line shapes 
in almost all cases. All narrow 
emission lines are assumed to have ordinary Gaussian profiles that are coupled in redshift, 
velocity dispersion, and line ratios for doublets. 
While [\ion{O}{3}] can exhibit a blueshifted wing in cases with
an outflow (e.g., \citealt{mullaney_alexander_2013}), the narrow component 
usually strongly dominates the line flux, and therefore a single-component 
line fit is generally robust when coupled to
lower ionization lines like [\ion{N}{2}]. In cases where 
the outflow component dominates, we 
decouple [\ion{O}{3}] from the fit to other narrow lines, as described below.
We also add the two dominant \ion{Fe}{2} lines in the H$\beta$ 
wavelength range as Gaussians to our model. The best-fit model is determined by using a Levenberg--Marquardt minimization 
algorithm, where spectral regions of unconsidered weak narrow emission lines, e.g., [\ion{O}{1}]$\lambda 6300$ and 
[\ion{S}{2}]$\lambda6718, \lambda6732$, have been masked out along with pixels assigned as bad by the SDSS spectroscopic 
pipeline. 

We deviate slightly from the scheme above in the following three cases: (1) the broad H$\beta$ line is too weak to 
be modeled, (2) the [\ion{O}{3}] line exceeds a width of 600 km sec$^{-1}$, or 
(3) the narrow H$\alpha$ line flux is negative in the 
initial best-fit model. 
In the first case, we repeat the model without 
any broad H$\beta$ and \ion{Fe}{2} components. In the second and third case, the [\ion{O}{3}] line is
likely to be affected by outflows and may not match the line profile of the other narrow lines. Therefore, 
we allow the [\ion{O}{3}] line to have an independent redshift and velocity dispersion with respect to the other 
narrow lines.

One percent of the RASS/SDSS spectra do not allow such an analysis because significant 
parts of the spectral range around H$\alpha$ or [\ion{O}{3}] are masked out.  We do 
not expect any significant impact on our study, due to the low number of
objects for which this is the case (see also Sect.~\ref{robustness} below).

Examples of the best-fit models for typical S/N AGN spectra are shown in 
Figure~\ref{AGN_SDSS_SPEC_FIT}. From those best-fit models we measure the FWHM and 
integrated flux of the broad-line H$\alpha$ component, as these are the parameters 
required to estimate $M_{\rm BH}$ and $L/L_{\rm EDD}$.  Our model is not designed to provide 
a perfect fit to H$\alpha$ line profiles for the most complex cases. However, the model deviations 
concerning the H$\alpha$ width and flux are much smaller than the systematics of 
virial $M_{\rm BH}$ estimates (see discussion in \citealt{denney_peterson_2009}). 
To assess the uncertainties of the derived parameters and to identify 
unreliable models, we generate 100 realizations of the same spectrum by replacing each pixel 
within its 1$\sigma$ variance given by the SDSS error spectrum and analyze the artificial spectra in the 
same manner. The uncertainty for each quantity is then taken as the
dispersion (1$\sigma$ error corresponding to 68.3\%) in the best-fit value of all 100 measurements.

\begin{figure}
  \centering
 \resizebox{\hsize}{!}{ 
  \includegraphics{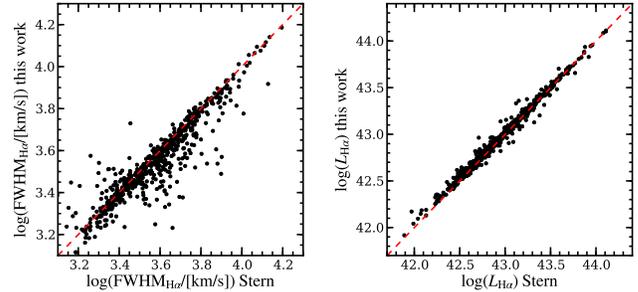}} 
      \caption{Comparison of the fitted FWHM and luminosity of the broad H$\alpha$ line component between 
               our work and \cite{stern_laor_2012},
for sources in both studies.
               The dashed red lines show the unity correlations. The 1 $\sigma$ scatter about these lines is 0.08 dex (left) and 0.04 dex (right), respectively.}
         \label{Ha_comparison}
\end{figure}

We compared our measurements with those of \cite{stern_laor_2012} to cross-check our results with 
a completely independent algorithm (Fig.~\ref{Ha_comparison}) and to provide an estimate of the systematic 
uncertainties. In contrast to our study, their sample contains lower luminosity AGNs at a slightly 
different redshift range. Thus, the samples only overlap partially. For the broad H$\alpha$ FWHM and luminosity, 
we find relatively tight correlations around the unity relation. 
The measurements of the broad H$\alpha$ FWHM exhibit a larger scatter and 
are slightly skewed toward lower values using our method. We inspected those spectra
and identify a common characteristic in these cases: the narrow
lines are broader and weaker than for the other AGNs in the sample. 
Thus, deblending the
narrow [NII]+H$\alpha$ from the broad H$\alpha$ line can be challenging. This 
is particularly true for broad H$\alpha$ lines with a FWHM less than 
$4000$\,km s$^{-1}$.
In summary, our line-profile fitting method provides robust estimates for the 
FWHM and line flux within their systematic uncertainties and without the
need for human interaction.

\begin{figure}
  \centering
 \resizebox{\hsize}{!}{ 
  \includegraphics{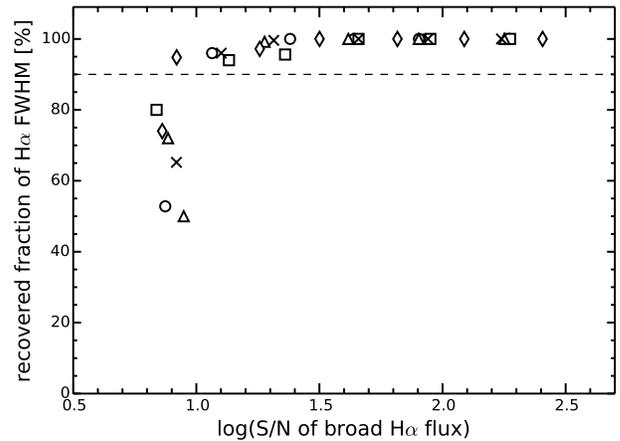}} 
      \caption{Accuracy in recovering the FWHM value of the broad H$\alpha$ line as a function 
               of the signal-to-noise ratio (S/N) of the integrated flux of the broad H$\alpha$ 
               line.
               We use high S/N spectra to simulate 250 realizations at various degraded S/N levels. 
               Within each S/N bin, we determine the fraction of spectra for which the FWHM 
               is within 40\% of the original high S/N spectrum (see text for details).
               We show the results for five different objects indicated by different symbols.}  
         \label{Ha_SN_sim_results}
\end{figure}

\subsection{Estimating $M_{\rm BH}$}
\label{estimating_MBH}
The virial black hole mass is estimated by combining the empirically calibrated
BLR size-luminosity relation, derived from reverberation mapping monitoring 
of nearby AGNs, and assuming virialized BLR motions, 
$M_{\rm BH}=f_\mathrm{vir}v^2R_\mathrm{BLR}G^{-1}$, 
where $v$ is the FWHM of broad lines in the AGN spectrum and $f_\mathrm{vir}$ is the virial factor 
representing the BLR kinematics. Several different estimators for $M_{\rm BH}$ 
have independently been reported in the literature using different BLR size-luminosity relations 
and virial factors based on different assumptions. Here we employ two different calibrations to 
check their potential impact on our results. The first is from \cite{mclure_dunlop_2004}:
\begin{equation}
 \frac{M_\mathrm{BH}}{M_\sun}=4.7\left(\frac{L_{5100}}{10^{44}\mathrm{erg}\,\mathrm{s}^{-1}}\right)^{0.61}\left(\frac{\mathrm{FWHM}_{\mathrm{H}\beta}}{\mathrm{km}\,\mathrm{s}^{-1}}\right)^2\label{eq:BH_Shen}
\end{equation}
which is also used by \cite{shen_strauss_2009} to estimate $M_{\rm BH}$ for all 
unobscured, optically selected SDSS AGN. 
The second calibration is based on the most recent BLR size-luminosity relation 
of \cite{bentz_peterson_2009}, assuming a virial factor of $f_\mathrm{vir}=5.5$, which is 
empirically determined by \cite{onken_ferrarese_2004}:
\begin{equation}
 \frac{M_\mathrm{BH}}{M_\sun}=8.13\left(\frac{L_{5100}}{10^{44}\mathrm{erg}\,\mathrm{s}^{-1}}\right)^{0.52}\left(\frac{\mathrm{FWHM}_{\mathrm{H}\beta}}{\mathrm{km}\,\mathrm{s}^{-1}}\right)^2.\label{eq:BH_Bentz}
\end{equation}

Throughout the paper, we use the second calibration (\citealt{bentz_peterson_2009}) for the 
estimates of the black hole masses. In Sect.~\ref{robustness} below we test the robustness of our clustering 
results if the first calibration method is used instead. As we demonstrate below, 
we find very similar results using either calibration.

In order to use our measurements from the H$\alpha$ line to estimate $M_\mathrm{BH}$, we have to replace $L_{5100}$ 
by $L_{\rm H\alpha}$ and $\mathrm{FWHM}_{\mathrm{H}\beta}$ by $\mathrm{FWHM}_{\mathrm{H}\alpha}$ in Equations~(\ref{eq:BH_Shen}) 
and (\ref{eq:BH_Bentz}), following \cite{greene_ho_2005} (their Equations~1 and 3).

\cite{denney_peterson_2009} point out that the inclusion of low S/N 
spectra adds a systematic offset in $M_{\rm BH}$ estimates because the line 
width is systematically underestimated. Therefore, we enforce a lower cutoff 
in the measured S/N level of the broad H$\alpha$ line flux. We determine the cutoff 
S/N by selecting the five AGN spectra 
with the highest S/N. Then, we artificially degrade the S/N gradually down to a 
continuum S/N of 1. For each S/N level, we generate 250 spectra and model these 
individual spectra with our line-profile fitting method. 

At an H$\alpha$ FWHM uncertainty level of greater than 40\%, the measurement uncertainties 
exceed the commonly assumed systematic uncertainties for virial $M_\mathrm{BH}$ 
estimates 
of $\sim$0.3 dex. More than 90\% of all of our generated spectra  
have FWHM uncertainties of less than 40\% if we restrict the H$\alpha$ S/N to values 
greater than 10 (Figure~\ref{Ha_SN_sim_results}). 
Consequently, we adapt this 
S/N threshold as a limit for selecting objects that yield reliable $M_\mathrm{BH}$ estimates.
This cutoff leads to a removal of 189 out of 1538 objects (12\% of the sample).
We apply this S/N cut and compare our measured FWHM values to the ones from
\cite{stern_laor_2012}.
The observed scatter in FWHM between both methods is 0.08 dex, which
corresponds to an $M_{\rm BH}$ error of 0.16. This can be interpreted as a 
simple estimate of the minimum systematic uncertainty in the $M_\mathrm{BH}$
determination, which is reflected in the commonly assumed total $\sim$0.3 dex 
systematic uncertainty of this method.

The main purpose of this paper is to study the clustering of AGN samples with 
low and high $M_\mathrm{BH}$ and $L/L_{\rm EDD}$, respectively. Any uncertainty in 
the calibration of $M_\mathrm{BH}$ will affect our entire sample similarly. Thus, 
our results will not depend on the absolute accuracy of the $M_\mathrm{BH}$ 
calibration because we are interested only in relative
$M_\mathrm{BH}$ values such that we can divide the full sample 
into low and high $M_\mathrm{BH}$ subsamples.

\subsection{Estimating $L/L_{\rm EDD}$}

The Eddington ratio $L/L_{\rm EDD}$ is the ratio between the bolometric and the Eddington luminosity 
($L_{\rm edd}= 1.26 \times 10^{38}\, M_{\rm BH}/\mathrm{M}_{\odot}$ erg s$^{-1}$).
The quantity is commonly derived by using the inferred $M_{\rm BH}$ and  
the optical continuum luminosity at 5100\,\AA, adopting
a certain bolometric correction factor. Different bolometric correction factors used 
to estimate the bolometric luminosity $L_{\rm bol}$ from $L_{5100}$ are quoted in the literature
(\citealt{kaspi_smith_2000}; \citealt{mclure_dunlop_2004}; \citealt{richards_lacy_2006}). 
While these correction factors are subject to uncertainties, they are required to 
estimate the bolometric luminosity if only a single-epoch optical spectrum 
is available. We use a correction factor of $f=10.3$ (\citealt{richards_lacy_2006}), 
which is consistent with values used by other studies. Again, we use the 
relation from \cite{greene_ho_2005} to estimate
$L_{5100}$ from $L_{\rm H\alpha}$ for the computation of $L_{\rm bol}$. 

\subsection{Defining X-ray Selected AGN Subsamples}
\label{matchingSamples}
\begin{figure}
  \centering
 \resizebox{\hsize}{!}{ 
  \includegraphics[bbllx=57,bblly=363,bburx=554,bbury=718]{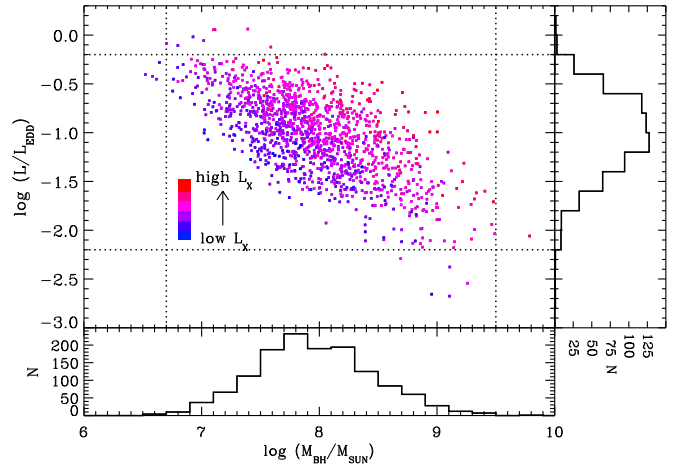}} 
      \caption{$M_{\rm BH}$ vs. $L/L_{\rm EDD}$ for luminous, unabsorbed, broad-line, 
               X-ray selected RASS/SDSS AGNs with a broad H$\alpha$ luminosity 
               of $S/N>10$.  The color of the symbols 
               encodes the X-ray luminosity of the sources. The lower and right
               panels show histograms of $M_{\rm BH}$ and $L/L_{\rm EDD}$, respectively,
               for the full sample. The dotted lines mark the restriction of
               the parameter space in our analysis to remove extreme objects.}
         \label{RASS_AGN_MBH_LLEDD}
\end{figure}

The RASS/SDSS AGNs in our sample do not uniformly populate the $M_{\rm BH}$ -- $L/L_{\rm EDD}$ plane 
(Figure~\ref{RASS_AGN_MBH_LLEDD}). Although there is substantial scatter, on average, 
higher $L/L_{\rm EDD}$ are found in AGNs with lower $M_{\rm BH}$. 
While the absence of objects in the upper right corner of this plane is not an observational 
bias, but reflects the nonexistence of (unabsorbed) RASS/SDSS AGNs with both 
high $M_{\rm BH}$ and high $L/L_{\rm EDD}$ in the redshift range studied here, the lack 
of objects in the lower left corner reflects an observational bias. AGNs with low 
$L/L_{\rm EDD}$ and low $M_{\rm BH}$ are too weak to produce a significant broad 
H$\alpha$ line relative to the host galaxy continuum. 
The H$\alpha$ flux S/N cutoff used here primarily removes 
objects with log ($M_{\rm BH}$/$M_{\odot}$) $\sim$ 7--8. 

Since $M_{\rm BH}$ and $L/L_{\rm EDD}$ are correlated in our full sample, and we aim to reveal 
which of these quantities drives the observed X-ray luminosity dependence of AGN clustering,  
doing a simple cut of the full sample into low and high $M_{\rm BH}$ subsamples, 
as well as high and low $L/L_{\rm EDD}$ subsamples, will 
not be useful. When creating subsamples that depend on one parameter, 
the distribution of the second parameter in both subsamples must be the same. 
This ``matching'' of the subsamples is a commonly used method in clustering 
measurements (e.g., \citealt{coil_georgakakis_2009}). However, one has to test 
that the procedure of creating matched subsamples is not introducing a bias to the 
clustering results. We will do so below by testing different methods to produce 
matched subsamples (see Sect.~\ref{robustness}). 

\begin{figure}
  \centering
 \resizebox{\hsize}{!}{ 
  \includegraphics[bbllx=64,bblly=360,bburx=554,bbury=718]{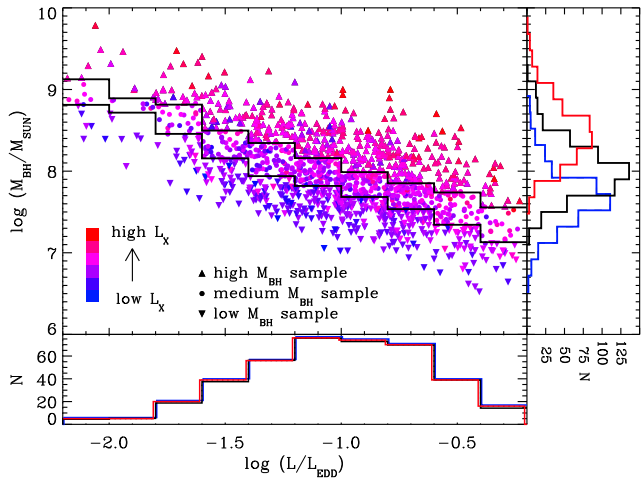}} 
      \caption{Low (filled down-pointing triangles), medium (filled circles), and 
               high (filled up-pointing triangles) $M_{\rm BH}$ RASS/SDSS 
               AGN subsamples with matched $L/L_{\rm EDD}$ distributions.
               The top panel shows the $L/L_{\rm EDD}$--$M_{\rm BH}$ plane. 
               The color of the points encodes the X-ray luminosity from lowest $L_{\rm X}$ (blue)
               to highest $L_{\rm X}$ (red). The solid lines are for visual guidance and 
               mark the division of the full AGN sample into low, medium, and high $M_{\rm BH}$
               subsamples. The bottom and right panels show the $L/L_{\rm EDD}$ and $M_{\rm BH}$ histograms of
               the low (blue), medium (black), and high (red) $M_{\rm BH}$
               subsamples, respectively. Since the medium $M_{\rm BH}$
               subsample contains more objects than the low and high
               subsamples, we normalized the total number of this sample in the $L/L_{\rm EDD}$
               histogram to match the other samples. For clarity, 
               we slightly shift the $x$ values of the low and
               high $M_{\rm BH}$ subsamples in the histograms.}
         \label{high_low_MBH}
\end{figure}

Before creating low and high $M_{\rm BH}$ subsamples, objects that lie outside
the range $-2.2 < {\rm log} (L/L_{\rm EDD}) < -0.2$ are removed.
We then determine the number of
objects in each $L/L_{\rm EDD}$ bin, using a bin width of 0.2 (logarithmic scale).
In each  $L/L_{\rm EDD}$ bin, we create three subsamples: 
30\% of objects with the lowest $M_{\rm BH}$,
40\% with medium $M_{\rm BH}$, and 30\% of objects with the highest $M_{\rm BH}$
(Figure~\ref{high_low_MBH}).
The procedure creates low, medium, and high $M_{\rm BH}$ AGN subsamples
with extremely similar (``matched'') $L/L_{\rm EDD}$ distributions 
but different median $M_{\rm BH}$ (see Table~\ref{xagn_acf}). 
All $M_{\rm BH}$ AGN subsamples have 
median $\langle {\rm log}\, (L/L_{\rm EDD}) \rangle =-1.00$. 

The low (30\%), medium (40\%), and high (30\%) $L/L_{\rm EDD}$ RASS/SDSS AGN subsamples with 
matched $M_{\rm BH}$ distributions are created by first applying a 
cut of $6.7 < {\rm log} (M_{\rm BH}/M_{\odot}) < 9.5$ to remove extreme sources and 
then following the same approach as described above, using 
bins of $M_{\rm BH}$ with a bin width of 0.2 (logarithmic scale). All $L/L_{\rm EDD}$ AGN subsamples have 
median $\langle$log ($M_{\rm BM}$/$M_\odot$)$\rangle =7.92-7.93$.

Figures~\ref{RASS_AGN_MBH_LLEDD} and \ref{high_low_MBH} show that divisions into  
low, medium, and high $M_{\rm BH}$ as well as $L/L_{\rm EDD}$ are very similar to divisions 
into low, medium, and high $L_{\rm X}$.
Thus, one cannot anticipate if the X-ray luminosity clustering dependence is related 
to $L/L_{\rm EDD}$ or $M_{\rm BH}$, as we aim to test here.

\begin{figure*}
 \begin{minipage}[b]{0.48\textwidth}
\centering
 \resizebox{\hsize}{!}{
  \includegraphics[bbllx=74,bblly=364,bburx=546,bbury=700]{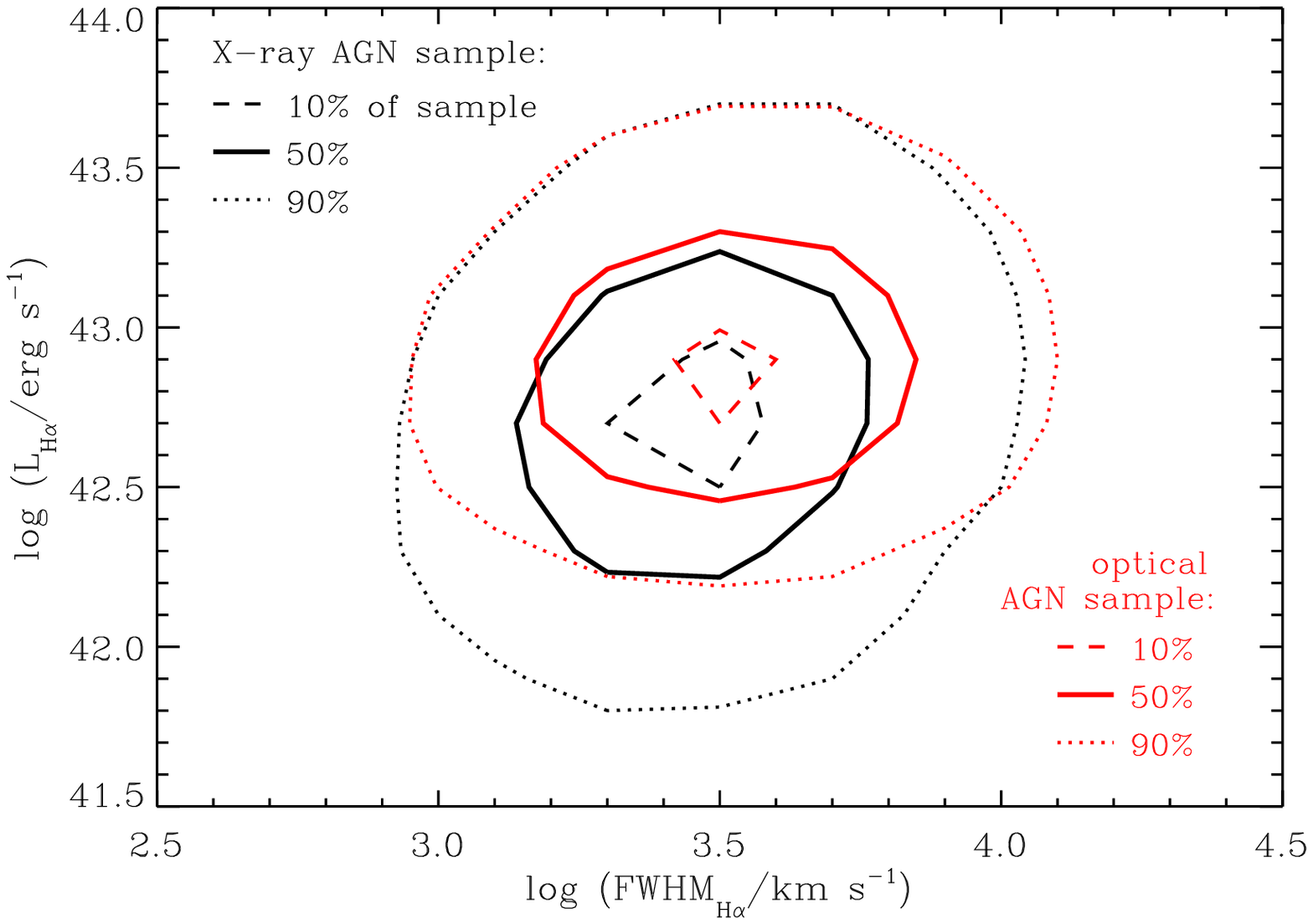}}
\end{minipage}
\hfill
\begin{minipage}{0.48\textwidth}
\vspace*{-5.6cm}
\centering
\resizebox{\hsize}{!}{
  \includegraphics[bbllx=74,bblly=364,bburx=546,bbury=700]{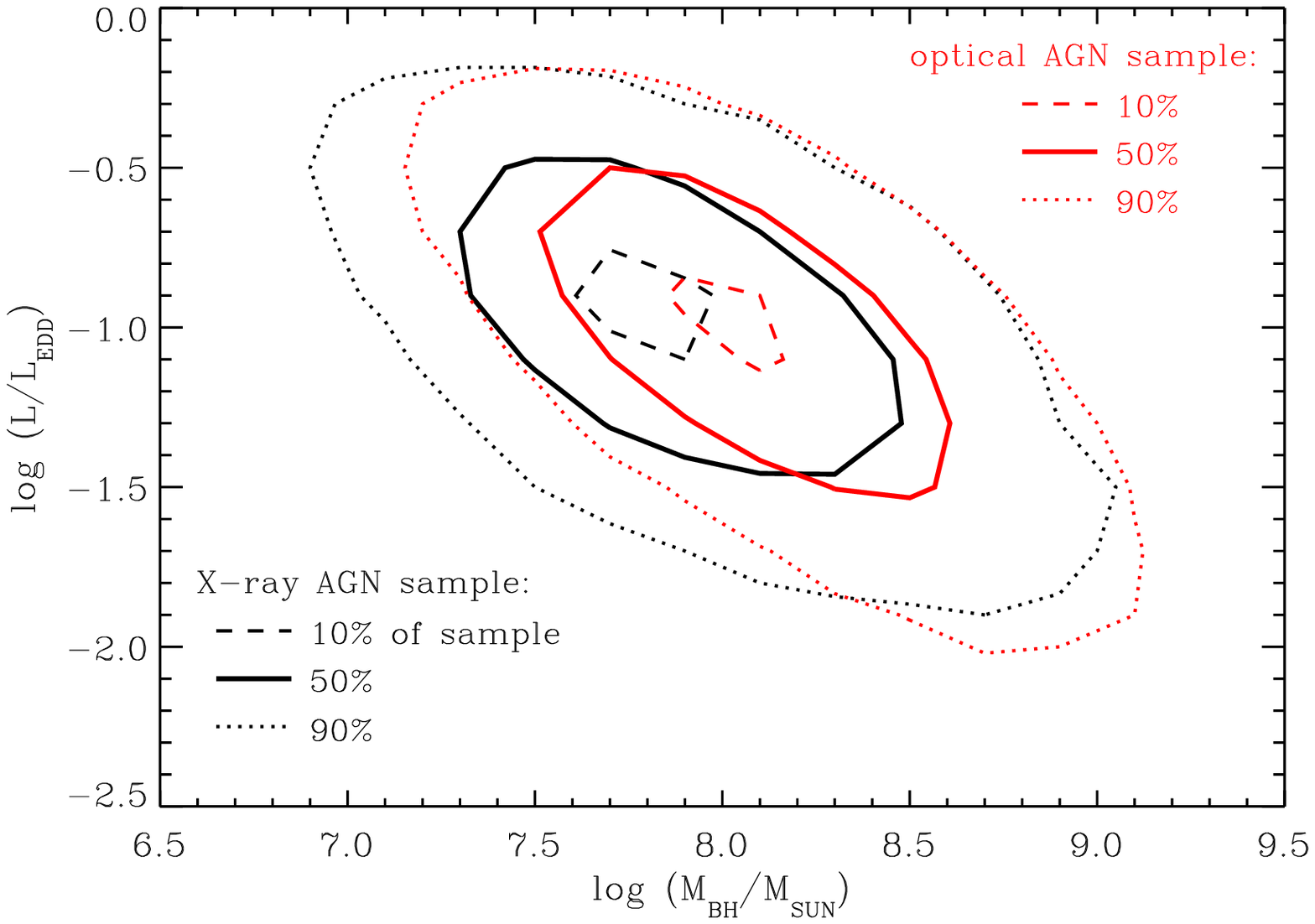}}
\vspace*{-0.0cm}
\end{minipage}
\caption{Comparison between the X-ray-selected RASS/SDSS AGNs (black lines) and the 
optically selected SDSS AGN samples (red lines). The contours show the location of 10\% (dashed lines), 
50\% (solid lines), and 90\% (dotted line) of 
the full sample, respectively. The left figure shows the observed parameters, 
luminosity and FWHM of the H$\alpha$ line, while the right figure 
shows the derived parameters $L/L_{\rm EDD}$ and $M_{\rm BH}$. 
In both figures, a boxcar smoothing with a width of two has been applied.}
\label{contour_compare}
\end{figure*}

The directly observed parameters FWHM and $L_{\rm H\alpha}$ do not correlate
with each other as strongly as 
$M_{\rm BH}$ and $L/L_{\rm EDD}$ (see Fig. 6). 
However, we decide to use the same method to create subsamples defined by 
FWHM and $L_{\rm H\alpha}$ as well. Thus, we again split the full sample into three subsamples defined using one parameter, 
while maintaining the same distribution in the other parameter of interest.
To create subsamples in FWHM, we first limit to objects with 
$41.7 < {\rm log} (L_{\rm H\alpha}/[$erg\,s$^{-1}]) < 44.1$
and then use a bin width of 0.2. All FWHM AGN subsamples have 
$\langle {\rm log} (L_{\rm H\alpha}/[$erg\,s$^{-1}]) \rangle =42.72-42.73$.
To create subsamples in $L_{\rm H\alpha}$, we first limit to objects with 
$2.9 < {\rm log} (\rm FWHM/[$km\,s$^{-1}]) < 4.1$, and then use bins of width 0.1, 
 using the logarithmic values of FWHM.  This choice of bin width ensures a 
similar numbers of bins as compared to the subsamples defined in $L_{\rm H\alpha}$.
The median FWHM of all $L_{\rm H\alpha}$ subsamples is $\langle {\rm FWHM} \rangle =2850-2960$ km\,s$^{-1}$.
 
As we have to exclude $\sim$13\% of the RASS/SDSS AGN sample studied in 
Papers I--III to fulfill our S/N requirement of the broad H$\alpha$ line flux
for reliable $M_{\rm BH}$ estimates, we create low and high 
X-ray luminosity subsamples of the reduced AGN sample used here. 
We use a simple 0.1-2.4 keV luminosity cut 
(with no matching of the distribution of another parameter) of 
log $(L_{\rm X}/[$erg\,s$^{-1}])=44.29$, identical to that used in Papers I--III.

We also apply simple $M_i$ cuts at $M_i=-22.10$ mag and $M_i=-22.70$ mag to divide 
the X-ray-selected AGN sample into a faint (33\% of all objects), medium (34\%), and 
luminous (33\%) $M_i$ subsamples.  
Finally, we compute from the SDSS optical spectra the rest-frame absolute magnitude in the 
band between 5500 and 6800 \AA, which includes the H$\alpha$ line for all objects. We do this as 
the X-ray and optically selected sample covers a redshift range of $0.16<z<0.36$, while 
the H$\alpha$ line is redshifted outside the SDSS $i$-band filter above $z\sim 0.28$. 
The division into 
three subsamples of similar size is realized by applying cuts at 
$M_{5500-6800}^{\rm rest}=-21.72$ mag and $M_{5500-6800}^{\rm rest}=-22.33$ mag. We do not match 
the distribution of any other parameter for these subsamples. 
All samples used in this paper are presented in Table~\ref{xagn_acf}.

\subsection{Defining the Optically Selected AGN Subsamples}

The optically selected SDSS AGNs (called ``quasars'' in the SDSS literature) are drawn 
from the catalog provided by \cite{schneider_richards_2010}. Instead of using 
the classic selection in the $B$-band, \cite{schneider_richards_2010} use the SDSS $i$-band, 
as it is 
less affected by Galactic absorption. However, this comes with the significant 
disadvantage that host-galaxy light might represent a significant fraction of the 
total flux in the $i$-filter. \cite{schneider_richards_2010} apply 
an apparent magnitude cut of $M_{\rm i} \le -22$ and require that objects have 
at least one emission line exceeding an FWHM of 1000 km\,s$^{-1}$. 
Unlike our RASS/SDSS AGN sample, this sample of 3500 objects 
($0.16 \le z \le 0.36$) has the footprint 
of SDSS Data Release 7. The DR4+ and DR7 areas are 5468 deg$^2$ and 
7670 deg$^2$, respectively, when we consider only the area that has a DR7 
completeness ratio of $f_{\rm compl} >0.8$. Thus the area occupied by the 
optically selected AGN sample is 1.4 times larger than the area covered by the 
X-ray-selected AGN sample. Additionally, the number density (per square degree) 
of optically selected AGNs is 1.6 times higher than for the X-ray-selected AGNs. 

We retrieve the individual SDSS spectra of
the optically selected SDSS AGNs and derive the FWHM of the broad H$\alpha$
line and the $L_{\rm H\alpha}$ through spectral fits (see
Sect.~\ref{M_BH__LLedd}). This procedure is identical to the one used for 
the RASS/SDSS AGN sample (see Sect.~\ref{halpha_fit}).
We select only objects with $S/N>10$ in the flux of the broad H$\alpha$ line. We derive 
the individual black hole masses and $L/L_{\rm EDD}$ for the remaining 2831 AGNs. 
The X-ray and optically selected AGN samples have 807 objects in common, which is 29\% of the total 
optically selected AGN sample. Thus, there is substantial difference between the X-ray 
and optically selected AGN samples.

Figure~\ref{contour_compare} compares how the X-ray-selected RASS/SDSS AGN sample 
and the optically selected SDSS AGN sample span the observed parameter space 
of $L_{\rm H\alpha}$ and FWHM (left panel) and the derived parameter space of 
$L/L_{\rm EDD}$ and $M_{\rm BH}$ (right panel). 
In the observed parameter space, there are two obvious differences between
these samples.  First, the 
RASS/SDSS AGN sample extends to lower $L_{\rm H\alpha}$. This is likely 
 a consequence of rejecting AGNs with 
$M_{\rm i} > -22$ in order to exclude objects with a strong starlight component from the 
host galaxy because an AGN has to have a certain luminosity to outshine its host galaxy. 
Clearly, X-ray selection of AGNs (RASS/SDSS) allows us 
to extend to fainter AGN luminosities, below the optical cut. Thus, at X-ray 
wavelengths 
we are able to detect AGNs that might not outshine their host galaxy in the
optical (see, e.g., \citealt{hopkins_hickox_2009}).
At log ($L_{\rm H\alpha}/[{\rm erg}\,{\rm s}^{-1}])\,> 42.5$, both samples span a similar 
parameter space, and the dynamic range in FWHM is almost identical in both samples. 

Differences in the observed parameter space naturally translate into differences 
in the derived $M_{\rm BH}$--$L/L_{\rm EDD}$ space 
(Fig.~\ref{contour_compare}, right panel). Compared with the X-ray-selected AGN sample, 
the optical AGN sample has a deficiency of AGNs with low $M_{\rm BH}$. 
The optically selected sample also extends marginally to higher $M_{\rm BH}$ and lower 
$L/L_{\rm EDD}$.

We create optically selected AGN subsamples with respect to
$M_{\rm BH}$, $L/L_{\rm EDD}$, FWHM, and $L_{\rm H\alpha}$ (with matched distribution 
in the other parameter of interest; see Sect.~\ref{matchingSamples}). 
We have to use slightly different limits for some of the parameters because
the X-ray and optically selected AGN samples cover slightly different parameter spaces. 
Thus, when we create the low, medium, and high $M_{\rm BH}$ subsamples with matched 
$L/L_{\rm EDD}$ distributions, we use an $L/L_{\rm EDD}$ limit of 
$-2.3 < {\rm log} (L/L_{\rm EDD}) < 0.1$. The optical $M_{\rm BH}$ subsamples have 
median $\langle {\rm log}\, (L/L_{\rm EDD}) \rangle =-1.04$ to $-1.00$,  
while the $L/L_{\rm EDD}$ 
subsamples have $\langle$log ($M_{\rm BM}$/$M_\odot$)$\rangle = 8.12-8.13$. 

When we create three subsamples in FWHM, we limit 
$L_{\rm H\alpha}$ to $42.1 < {\rm log} (L_{\rm H\alpha}/[$erg\,s$^{-1}]) < 44.1$. 
The FWHM subsamples have $\langle {\rm log} (L_{\rm H\alpha}/[$erg\,s$^{-1}]) \rangle =42.88$.
For the $L_{\rm H\alpha}$ subsamples, we apply a limit of 
$2.8 < {\rm log} (\rm FWHM/[$km\,s$^{-1}]) < 4.2$. 
We find $\langle {\rm FWHM} \rangle =3240-3280$ km\,s$^{-1}$ for the low, medium, and high 
$L_{\rm H\alpha}$ subsamples. All bin widths are identical 
to the ones used for defining the X-ray-selected AGN subsamples. 

We produce optical AGN subsamples divided by $M_{\rm i}$. 
We use simple $M_{\rm i}$ cuts of $M_{\rm i} =-22.27$ and $M_{\rm i} =-22.67$ to
construct faint, medium, and luminous $M_{\rm i}$ subsamples with similar numbers of sources.
Finally, we create $M_{5500-6800}^{\rm rest}$ subsamples of the optically selected AGN. 
We use cuts at $M_{5500-6800}^{\rm rest}=-22.13$ mag and $M_{5500-6800}^{\rm rest}=-22.53$ mag. 
As with the X-ray-selected AGNs, we do not match the distribution of any other 
parameter when creating the subsamples defined by $M_{\rm i}$ and $M_{5500-6800}^{\rm rest}$. 
All samples are presented in Table~\ref{xagn_acf}. 


\section{Methodology}

\subsection{Clustering Measurements}
We measure the two-point correlation function $\xi(r)$ (\citealt{peebles_1980}), which measures 
the excess probability $dP$ above a Poisson distribution of finding an object in a volume 
element $dV$ at a distance $r$ from another randomly chosen object. The  
ACF measures the spatial clustering of objects in the same sample, 
while the CCF measures the clustering of objects in 
two different samples. We use the same approach as 
described in detail in Section~3 of Paper I and Section~4 in Paper III. 
Here we explain the essential elements of our method.

We use the correlation estimator of \cite{davis_peebles_1983} in the form
\begin{equation}
\label{DD_DR}
 \xi(r)= \frac{DD(r)}{DR(r)} -1\ ,
\end{equation}
where $DD(r)$ is the number of data--data pairs with a separation $r$, and $DR(r)$ 
is the number of data--random pairs. Both pair counts have been normalized by the number density of
data and random points. 
To separate the effect of redshift distortions, the correlation 
function is measured as a function of two components of the separation vector 
between two objects, that is, one perpendicular to ($r_p$) and the other along 
($\pi$) the line of sight. The parameter $\xi(r_p,\pi)$ is thus extracted by counting 
pairs on a two-dimensional grid of separations $r_p$ and $\pi$. 
We obtain the  projected correlation function $w_p(r_p)$ by integrating $\xi(r_p,\pi )$ along
the $\pi$ direction.

As in Paper I, we infer the AGN ACF from the CCF between AGN and the corresponding galaxy tracer set 
and the ACF of the tracer set following 
\cite{coil_georgakakis_2009}: 
\begin{eqnarray}
\label{acf_rassagn}
 w_p({\rm AGN}|{\rm  AGN}) = \frac{\left[w_p({\rm  AGN}|{\rm  trace})\right]^2}{w_p({\rm  trace}|{\rm  trace})}\,,
\end{eqnarray}
where $w_p({\rm  AGN}|{\rm  AGN})$ and $w_p({\rm trace}|{\rm  trace})$ are the ACFs of the 
AGN and the corresponding tracer set, respectively, and $w_p({\rm  AGN}|{\rm trace})$ is the 
CCF between the AGN and the tracer set (i.e., a galaxy sample). 

The CCF is computed using 
\begin{eqnarray}
\label{DD_DR2}
 \xi_{\rm AGN-trace} = \frac{D_{\rm AGN}\,D_{\rm trace}}{D_{\rm AGN}\,R_{\rm trace}}-1.
\end{eqnarray}
For our purposes, the use of this simple estimator has several major advantages and results in only 
a marginal loss in the S/N when compared to more advanced estimators 
(e.g., \citealt{landy_szalay_1993}). The estimator in Equation~(\ref{DD_DR2}) 
requires the generation of a random catalog only for the tracer set. 
Since the random catalog should exactly 
match all observational biases to 
minimize the systematic uncertainties, well-understood selection effects are key to 
creating proper random samples. The tracer sets have well-defined selection 
functions, while the AGN samples suffer from selection functions that are very
complex and difficult to model. 

\subsection{Error Analysis}

The measurements in the adjacent bins of the correlation function are not independent.
Poisson errors will significantly underestimate the uncertainties 
and should not be used for error calculations. Instead, we use
the jackknife resampling technique to estimate the measurement errors
based on the covariance matrix $M_{ij}$, which reflects the degree to which bin $i$ is 
correlated with bin $j$. 

In our jackknife resampling, we divide the survey area into 
$N_{\rm T}=100$ subsections for the DR4+ geometry (X-ray-selected AGN sample), 
each of which is $\sim$50--60 deg$^2$. Since the optically selected AGN sample 
uses the DR7 geometry, we divide this area into $N_{\rm T}=131$ subsections of roughly
equal area. 
These $N_{\rm T}$ jackknife-resampled correlation functions define the 
covariance matrix: 
\begin{eqnarray}
\label{jackknife}
 M_{ij} = \frac{N_{\rm T} -1}{N_{\rm T}} \left[\sum_{k=1}^{N_{\rm T}} \bigg(w_k(r_{p,i})-<w(r_{p,i})>\bigg)\right.\nonumber\\
          \times \bigg(w_k(r_{p,j})-<w(r_{p,j})>\bigg)\bigg] \,  
\end{eqnarray}
We calculate $w_p(r_p)$ $N_{\rm T}$ times, where each jackknife sample excludes one 
section and $w_k(r_{p,i})$ and $w_k(r_{p,j})$ are from the $k$th jackknife samples of the 
AGN ACF and $<w(r_{p,i})>$, $<w(r_{p,j})>$ are the averages over all of the 
jackknife samples. The uncertainties represent 1$\sigma$ (68.3\%) confidence intervals. 

The generation of the covariance matrix for the inferred AGN ACF considers 
the $N_{\rm T}$ jackknife-resampled correlation functions of the CCF (AGN and 
corresponding tracer set) and the tracer set ACF. For each jackknife sample, 
we calculate the inferred AGN ACF by using Equation~\ref{acf_rassagn}. The resulting 
$N_{\rm T}$ $w_p(r_p)$ jackknife-resampled projected correlation functions of the inferred ACFs 
are then used to compute the covariance matrix of the inferred AGN ACF.

\begin{figure*}
 \begin{minipage}[b]{0.99\textwidth}
\centering
 \resizebox{\hsize}{!}{
  \includegraphics[bbllx=55,bblly=360,bburx=617,bbury=784,angle=0]{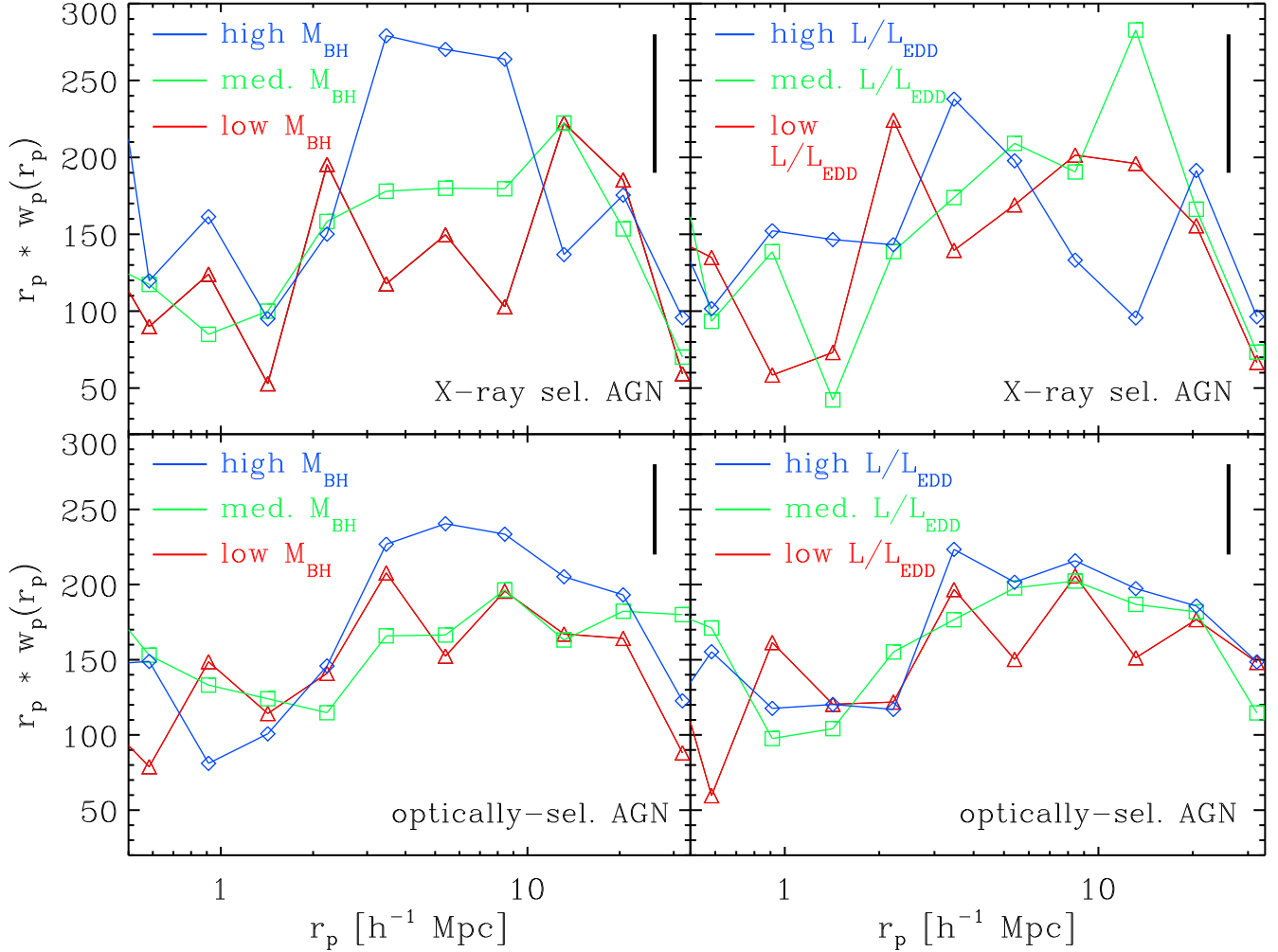}}
\end{minipage}
\caption{Directly measured cross-correlation functions between AGN subsamples for different parameters 
with matched distributions in the other parameter of interest. In the top row we show the results 
for the X-ray-selected AGNs, while the bottom row shows results for the
optically selected AGNs. In each panel, the color encodes the low (red), medium (green), and high (blue) subsample in the 
corresponding parameter. For the sake of clarity, we show only a typical error bar in each  
panel (black solid vertical line in right upper corner of each panel). Note
that we show on the $y$ axis $r_{\rm p} \times w(r_{\rm p})$ instead of
$w(r_{\rm p})$ to allow for an easier comparison between the different subsamples
and to make the values easier to read in the figure.
}
\label{CCFs}
\end{figure*}

\section{Results}

\label{inferringACF}
We measure high-accuracy CCFs of the different AGN samples with the 
LRG tracer set (see Fig.~\ref{CCFs}). 
In addition, we compute the high-precision ACF of the LRG sample. 
In both cases, we measure $r_p$ in the range 0.05--40 $h^{-1}$ Mpc in 15 logarithmic bins, 
identical to those used in Paper III.
We compute $\pi$ in steps of 5 $h^{-1}$ Mpc in the range $\pi=0-200$ $h^{-1}$ Mpc. 

To derive $\pi_{\rm max}$, we compute $w_p(r_p)$ for a set of $\pi_{\rm max}$ ranging 
from 10 to 160 $h^{-1}$ Mpc in steps of 10 $h^{-1}$ Mpc. We then fit $w_p(r_p)$ over 
an $r_p$ range of 0.3--40 $h^{-1}$ Mpc with a fixed $\gamma = 1.9$ and determine the 
correlation length $r_{\rm 0}$ for the individual $\pi_{\rm max}$ measurements. As in 
Papers I--III, we find that the LRG ACF saturates at $\pi_{\rm max}=80$ $h^{-1}$ Mpc. 
All CCFs saturate at $\pi_{\rm max}=40$ $h^{-1}$ Mpc. Power-law fits for the ACFs and CCFs 
are based on 
\begin{eqnarray}
\label{powerlaw}
 w_p(r_p) &=& r_p\left(\frac{r_{\rm 0}}{r_p}\right)^{\gamma}\,\frac{\Gamma(1/2)\Gamma((\gamma-1)/2)}{\Gamma(\gamma/2)},
\end{eqnarray}
where $\Gamma(x)$ is the Gamma function. 

To derive the clustering properties of the AGN samples, we follow two different approaches:\\ 

\textit{(1) Power-law fits to the inferred AGN ACF:}\\ 
We use Eq.~\ref{acf_rassagn} to infer the ACF for the 
individual AGN sample from the CCF of this sample with the LRG tracer set and the ACF of the LRG sample. 
We fit the data points of the inferred AGN ACFs with the expression given in Eq.~\ref{powerlaw} and 
derive best-fit $r_{\rm 0}$ and $\gamma$ values. The data are fitted over the range $r_p=0.3-15$ $h^{-1}$ Mpc 
to be consistent with Papers I--III. 
Since we measure the CCF to infer the ACF, the resulting effective redshift 
distribution for the clustering signal is determined by both the redshift distribution 
of the tracer set and the AGN sample: $N_{\rm CCF}(z) = N_{\rm tracer}(z)*N_{\rm AGN}(z)$.

In Table~\ref{xagn_acf} we list the redshift range, the median effective redshift of 
$N_{\rm CCF}(z)$ for the corresponding AGN samples, the derived best $r_{\rm 0}$ and $\gamma$ values 
based on power-law fits, and $r_{\rm 0}$ for a power-law fit with a fixed
slope of $\gamma=1.9$ (for ease of comparison). \\

\textit{(2) Bias from HOD modeling:}
In Paper II we develop a novel method to infer the HOD 
of RASS/SDSS AGNs directly from the well-constrained CCF 
of RASS/SDSS AGNs with LRGs. In performing the HOD modeling, we consider that 
galaxies and AGNs are associated with DMHs. A DMH may contain one or more galaxies 
or AGNs that are included in our samples. Using the HOD of the LRGs as a template, we constrain 
the HOD of the AGN by fitting the CCF between AGNs and LRGs. A few of the CCFs do not have enough 
pairs on small scales for applying $\chi^2$ statistics. Thus to derive consistent constraints for all 
CCF, we consider only bins with $r_{\rm p}>0.7$ $h^{-1}$ Mpc for all CCF fits (as in Paper III).

\begin{deluxetable*}{lcccccccc}
\tabletypesize{\normalsize}
\tablecaption{Properties of Inferred AGN ACFs and Derived Quantities\label{xagn_acf}}
\tablewidth{0pt}
\tablehead{
\colhead{AGN Sample} &\colhead{Number} &\colhead{Median} &\colhead{10th,50th,90th} & \colhead{$r_{\rm 0}$}   & \colhead{$\gamma$}         &\colhead{$r_{\rm 0,\gamma =1.9}$}     &\colhead{$b(z)$}     &\colhead{log $M_{\rm DMH}^{\rm typ}$}\\
\colhead{Name} & \colhead{of Objects} & \colhead{$z_{\rm eff}$} & \colhead{Percentile} & \colhead{($h^{-1}$ Mpc)} & \colhead{} &\colhead{($h^{-1}$ Mpc)       }              &\colhead{HOD}        &\colhead{($h^{-1}$ $M_{\odot}$)}}
\startdata
\\
\multicolumn{9}{c}{X-ray Selected AGN --- RASS/SDSS AGN --- SDSS Data Release 4+}\\
\\
total RASS AGN         & 1349  & 0.27 & 43.70,44.17,44.68 &4.02$^{+0.45}_{-0.57}$ & 1.62$^{+0.15}_{-0.14}$& 4.09$^{+0.35}_{-0.39}$  &1.32$^{+0.11}_{-0.09}$ &13.20$^{+0.13}_{-0.13}$\\
low $L_{\rm X}$ RASS    &  858  & 0.25 & 43.63,43.99,44.23 &3.12$^{+0.57}_{-0.89}$ & 1.55$^{+0.25}_{-0.21}$& 3.28$^{+0.41}_{-0.46}$  &1.22$^{+0.14}_{-0.12}$ &13.08$^{+0.19}_{-0.20}$\\
high $L_{\rm X}$ RASS   &  491  & 0.29 & 44.34,44.53,44.87 &5.38$^{+0.68}_{-0.87}$ & 1.88$^{+0.18}_{-0.18}$& 5.41$^{+0.65}_{-0.73}$  &1.53$^{+0.15}_{-0.17}$ &13.41$^{+0.13}_{-0.18}$\\
                                 &                                            &                      &         &                    & \\
{\it low $M_{\rm BM}$ RASS}      &  410  & 0.22 & 7.07,7.54,8.01 &2.71$^{+0.53}_{-1.11}$ & 2.07$^{+1.46}_{-0.64}$& 2.65$^{+0.56}_{-0.71}$  &0.96$^{+0.24}_{-0.12}$ &12.60$^{+0.48}_{-0.39}$\\
{\it medium $M_{\rm BM}$ RASS}   &  525  & 0.27 & 7.55,7.92,8.36 &4.34$^{+0.64}_{-0.90}$ & 1.61$^{+0.19}_{-0.18}$& 4.32$^{+0.52}_{-0.58}$  &1.38$^{+0.15}_{-0.17}$ &13.27$^{+0.16}_{-0.23}$\\
{\it high $M_{\rm BM}$ RASS}     &  400  & 0.29 & 7.92,8.38,8.91 &4.60$^{+0.82}_{-1.11}$ & 1.85$^{+0.21}_{-0.22}$& 4.67$^{+0.78}_{-0.92}$  &1.62$^{+0.20}_{-0.16}$ &13.49$^{+0.16}_{-0.15}$\\
                                 &             &                        &                     &               &                     & \\
{\it low $L/L_{\rm EDD}$ RASS}   &  418  & 0.23 & -1.79,-1.36,-0.98 &3.00$^{+0.78}_{-1.20}$ & 1.45$^{+0.26}_{-0.21}$& 3.17$^{+0.54}_{-0.64}$ &1.30$^{+0.18}_{-0.18}$ &13.21$^{+0.21}_{-0.27}$\\
{\it medium $L/L_{\rm EDD}$ RASS}&  522  & 0.27 & -1.36,-1.00,-0.71 &2.87$^{+1.97}_{-1.50}$ & 1.59$^{+0.40}_{-0.33}$& 2.89$^{+0.46}_{-0.54}$ &1.29$^{+0.13}_{-0.15}$ &13.16$^{+0.16}_{-0.23}$\\
{\it high $L/L_{\rm EDD}$ RASS}  &  404  & 0.29 & -1.01,-0.68,-0.35 &4.36$^{+0.67}_{-0.79}$ & 2.14$^{+0.22}_{-0.19}$& 3.92$^{+0.74}_{-0.89}$ &1.18$^{+0.15}_{-0.14}$ &12.97$^{+0.22}_{-0.26}$\\
                                 &             &                        &                     &              &                     & \\
{\it low FWHM RASS}          &  415  & 0.27 & 1280,1910,2470 &2.69$^{+0.61}_{-0.80}$ & 2.30$^{+0.72}_{-0.40}$& $2.32^{+0.71}_{-1.01}$ &1.09$^{+0.12}_{-0.12}$ &12.84$^{+0.20}_{-0.27}$\\
{\it medium FWHM RASS}       &  525  & 0.26 & 2340,2890,3690 &2.90$^{+1.25}_{-2.35}$ & 1.35$^{+0.24}_{-0.28}$& $3.70^{+0.64}_{-0.76}$ &1.47$^{+0.17}_{-0.18}$ &13.38$^{+0.16}_{-0.21}$\\
{\it high FWHM RASS}         &  404  & 0.27 & 3650,4840,7920 &3.52$^{+0.92}_{-1.57}$ & 1.53$^{+0.25}_{-0.26}$& $3.82^{+0.65}_{-0.77}$ &1.53$^{+0.17}_{-0.18}$ &13.43$^{+0.15}_{-0.20}$\\
                                 &             &                        &                     &               &                     & \\
{\it low $L_{\rm H\alpha}$ RASS}   & 414  & 0.22 & 41.99,42.29,42.53 &2.14$^{+0.96}_{-1.72}$ & 1.39$^{+0.56}_{-0.32}$& 2.69$^{+0.54}_{-0.66}$ &1.21$^{+0.18}_{-0.19}$ &13.10$^{+0.23}_{-0.35}$\\
{\it medium $L_{\rm H\alpha}$ RASS}& 526  & 0.27 & 42.52,42.74,42.96 &1.96$^{+1.00}_{-1.97}$ & 1.83$^{+1.32}_{-0.36}$& 2.13$^{+0.64}_{-0.71}$ &1.31$^{+0.16}_{-0.16}$ &13.18$^{+0.19}_{-0.24}$\\
{\it high $L_{\rm H\alpha}$ RASS}  & 403  & 0.29 & 42.93,43.22,43.63 &3.52$^{+0.92}_{-1.57}$ & 1.53$^{+0.25}_{-0.26}$& 3.82$^{+0.65}_{-0.78}$ &1.36$^{+0.18}_{-0.16}$ &13.23$^{+0.19}_{-0.23}$\\
                                 &             &                        &                     &              &                     & \\
faint $M_i$ RASS          &  444  & 0.23 & -21.26,-21.80,-22.04&3.21$^{+0.69}_{-1.58}$ & 1.69$^{+0.69}_{-0.42}$& $3.32^{+0.58}_{-0.70}$ &1.10$^{+0.26}_{-0.14}$ &12.90$^{+0.39}_{-0.31}$\\
medium $M_i$ RASS         &  460  & 0.27 & -22.15,-22.39,-22.63&3.51$^{+0.71}_{-1.16}$ & 1.71$^{+0.36}_{-0.29}$& $3.65^{+0.61}_{-0.72}$ &1.21$^{+0.19}_{-0.16}$ &13.04$^{+0.25}_{-0.29}$\\
luminous $M_i$ RASS       &  445  & 0.29 & -22.76,-23.13,-23.96&3.04$^{+0.76}_{-1.12}$ & 1.47$^{+0.17}_{-0.18}$& 2.94$^{+0.54}_{-0.65}$ &1.48$^{+0.17}_{-0.16}$ &13.36$^{+0.16}_{-0.19}$\\
                                 &             &                        &                     &              &                     & \\
faint $M_{5500-6800}^{\rm rest}$ RASS   &  444  & 0.22 & -20.84,-21.39,-21.66&3.18$^{+0.63}_{-1.12}$ & 1.79$^{+0.56}_{-0.41}$& $3.19^{+0.59}_{-0.71}$ &1.13$^{+0.25}_{-0.19}$ &12.97$^{+0.35}_{-0.42}$\\
med. $M_{5500-6800}^{\rm rest}$ RASS  &  460  & 0.27 & -21.79,-22.02,-22.27&2.44$^{+1.15}_{-1.94}$ & 1.36$^{+0.29}_{-0.28}$& $3.39^{+0.57}_{-0.67}$ &1.25$^{+0.16}_{-0.17}$ &13.10$^{+0.20}_{-0.28}$\\
lum. $M_{5500-6800}^{\rm rest}$ RASS&  445  & 0.30 & -22.42,-22.81,-23.67&4.23$^{+0.76}_{-1.00}$ & 1.70$^{+0.15}_{-0.16}$& 4.20$^{+0.67}_{-0.79}$ &1.44$^{+0.16}_{-0.17}$ &13.31$^{+0.15}_{-0.21}$\\\hline
\\
\multicolumn{9}{c}{Optically Selected AGN --- SDSS AGN --- SDSS Data Release 7}\\
\\
total SDSS AGN           & 2831 & 0.29 & -22.08,-22.45,-23.34 &5.07$^{+0.26}_{-0.28}$ & 1.80$^{+0.09}_{-0.09}$& 5.06$^{+0.25}_{-0.26}$ &1.32$^{+0.07}_{-0.06}$ &13.18$^{+0.08}_{-0.09}$\\
faint $M_i$ SDSS         & 932 & 0.28 & -22.03,-22.13,-22.24 &5.34$^{+0.42}_{-0.51}$ & 1.80$^{+0.20}_{-0.19}$& 5.34$^{+0.40}_{-0.43}$  &1.47$^{+0.09}_{-0.14}$ &13.36$^{+0.09}_{-0.16}$\\
medium $M_i$ SDSS         & 966 & 0.29 & -22.30,-22.45,-22.63 &3.79$^{+0.44}_{-0.59}$ & 1.73$^{+0.16}_{-0.16}$& 3.97$^{+0.36}_{-0.39}$ &1.20$^{+0.09}_{-0.10}$ &13.00$^{+0.14}_{-0.17}$\\      
luminous $M_i$ SDSS        & 933 & 0.30 & -22.73,-23.06,-23.86 &4.48$^{+0.51}_{-0.64}$ & 1.63$^{+0.12}_{-0.12}$& 4.52$^{+0.41}_{-0.45}$&1.40$^{+0.12}_{-0.12}$ &13.26$^{+0.13}_{-0.15}$\\      
                         &      &      &                   &                     &                      &                      &                     & \\
{\it low $M_{\rm BM}$ SDSS}     & 857  & 0.27 & 7.37,7.82,8.35 &4.24$^{+0.46}_{-0.56}$ & 1.74$^{+0.15}_{-0.15}$& 4.26$^{+0.41}_{-0.45}$ &1.28$^{+0.12}_{-0.13}$ &13.14$^{+0.15}_{-0.20}$\\      
{\it medium $M_{\rm BM}$ SDSS}  & 1114 & 0.30 & 7.70,8.08,8.59 &4.97$^{+0.40}_{-0.48}$ & 1.98$^{+0.17}_{-0.18}$& 4.88$^{+0.41}_{-0.44}$ &1.18$^{+0.13}_{-0.09}$ &12.96$^{+0.19}_{-0.16}$\\      
{\it high $M_{\rm BM}$ SDSS}    & 845  & 0.31 & 8.04,8.45,8.97 &4.72$^{+0.63}_{-0.82}$ & 1.50$^{+0.12}_{-0.12}$& 4.76$^{+0.43}_{-0.47}$ &1.60$^{+0.10}_{-0.08}$ &13.45$^{+0.09}_{-0.07}$\\      
                         &      &      &                 &                     &                      &                      &                     & \\
{\it low $L/L_{\rm EDD}$ SDSS}   & 859  & 0.28 & -1.84,-1.31,-0.88 &3.40$^{+0.55}_{-0.77}$ & 1.58$^{+0.14}_{-0.15}$& 3.62$^{+0.39}_{-0.44}$ &1.21$^{+0.11}_{-0.10}$ &13.03$^{+0.16}_{-0.17}$\\      
{\it medium $L/L_{\rm EDD}$ SDSS}& 1113 & 0.30 & -1.53,-1.04,-0.67 &4.68$^{+0.58}_{-0.73}$ & 1.62$^{+0.13}_{-0.13}$& 4.89$^{+0.43}_{-0.46}$ &1.40$^{+0.11}_{-0.11}$ &13.26$^{+0.12}_{-0.14}$\\      
{\it high $L/L_{\rm EDD}$ SDSS}  & 845  & 0.31 & -1.17,-0.74,-0.40 &4.75$^{+0.60}_{-0.78}$ & 1.64$^{+0.14}_{-0.15}$& 4.73$^{+0.50}_{-0.55}$ &1.49$^{+0.15}_{-0.12}$ &13.35$^{+0.14}_{-0.13}$\\      
                                &              &                        &                     &                      &                      &                     & \\
{\it low FWHM SDSS}            & 857  & 0.30 & 1490,2110,2560 &4.18$^{+0.49}_{-0.56}$ & 1.92$^{+0.19}_{-0.19}$& $4.17^{+0.49}_{-0.55}$ &1.11$^{+0.13}_{-0.08}$ &12.84$^{+0.21}_{-0.17}$\\      
{\it medium FWHM SDSS}         & 1120 & 0.29 & 2680,3260,4040 &5.03$^{+0.43}_{-0.53}$ & 1.75$^{+0.13}_{-0.13}$& $5.12^{+0.38}_{-0.40}$ &1.41$^{+0.10}_{-0.09}$ &13.28$^{+0.11}_{-0.11}$\\      
{\it high FWHM SDSS}           & 847  & 0.29 & 4300,5500,8460 &4.62$^{+0.50}_{-0.63}$ & 1.68$^{+0.14}_{-0.14}$& 4.71$^{+0.42}_{-0.46}$ &1.44$^{+0.10}_{-0.12}$ &13.32$^{+0.10}_{-0.15}$\\       
                                &              &                        &                     &                      &                      &                     & \\
{\it low $L_{\rm H\alpha}$ SDSS}   & 859  & 0.27 & 42.40,42.61,42.71 &3.72$^{+0.51}_{-0.72}$ & 1.63$^{+0.16}_{-0.17}$& 3.79$^{+0.41}_{-0.45}$ &1.25$^{+0.13}_{-0.08}$ &13.10$^{+0.17}_{-0.12}$\\      
{\it medium $L_{\rm H\alpha}$ SDSS}& 1119 & 0.30 & 42.76,42.89,43.04 &4.25$^{+0.49}_{-0.68}$ & 1.69$^{+0.15}_{-0.16}$& 4.47$^{+0.39}_{-0.42}$ &1.32$^{+0.12}_{-0.10}$ &13.16$^{+0.15}_{-0.14}$\\        
{\it high $L_{\rm H\alpha}$ SDSS}  & 846  & 0.31 & 43.07,43.26,43.63 &4.52$^{+0.61}_{-0.80}$ & 1.56$^{+0.14}_{-0.14}$& 4.50$^{+0.46}_{-0.51}$ &1.52$^{+0.13}_{-0.13}$ &13.38$^{+0.12}_{-0.14}$\\         
                                 &             &                        &                     &              &                     & \\
faint $M_{5500-6800}^{\rm rest}$ SDSS   &  933  & 0.27 & -21.70,-21.97,-22.10&5.96$^{+0.41}_{-0.44}$ & 1.88$^{+0.13}_{-0.14}$& $5.95^{+0.40}_{-0.43}$ &1.50$^{+0.09}_{-0.14}$ &13.40$^{+0.09}_{-0.15}$\\      
med. $M_{5500-6800}^{\rm rest}$ SDSS  &  964  & 0.30 & -22.17,-22.31,-22.49&3.36$^{+0.71}_{-1.18}$ & 1.55$^{+0.20}_{-0.22}$& $3.96^{+0.40}_{-0.45}$ &1.14$^{+0.10}_{-0.11}$ &12.89$^{+0.16}_{-0.22}$\\
lum. $M_{5500-6800}^{\rm rest}$ SDSS&  934  & 0.31 & -22.59,-22.89,-23.66&4.57$^{+0.55}_{-0.66}$ & 1.69$^{+0.13}_{-0.13}$& 4.56$^{+0.48}_{-0.53}$ &1.46$^{+0.12}_{-0.13}$ &13.32$^{+0.12}_{-0.12}$
\tablecomments{\small Samples displayed in italics are those 
that are created to match 
distributions in the other parameter of interest (e.g, samples split in $M_{\rm BM}$ have matched distributions in $L/L_{\rm EDD}$ and vice versa; samples 
split in FWHM have matched distributions in $L_{\rm H\alpha}$). 
The fourth column lists either the log ($L_{\rm X}$/[erg\,s$^{-1}$]),  log ($M_{\rm BM}$/$M_\odot$), log ($L/L_{\rm EDD}$), FWHM (km\,s$^{-1}$), 
log ($L_{\rm H\alpha}$/[erg\,s$^{-1}$]), absolute magnitude in the $i$-band ($M_i$ in mag), or absolute rest-frame magnitude in the 5500--6800 \AA\ band 
($M_{5500-6800}^{\rm rest}$ in mag) of the studied sample. In this
column, the 50th percentile corresponds to the median value. For the total sample, we state in 
this column the $L_{\rm X}$ or the $M_i$ value. All samples cover a redshift range of $0.16<z<0.36$ and contain only objects with S/N $>10$ in H$\alpha$ flux. 
Values of $r_{\rm 0}$, $\gamma$, and $r_{\rm 0,\gamma =1.9}$ are obtained from a power-law fit to $w_{\rm p}(r_{\rm p})$ over the range $r_p=$ 0.3--15 $h^{-1}$ Mpc, 
using the full error covariance matrix and minimizing the 
correlated $\chi^2$ values.}
\end{deluxetable*}

Using the best-fit model of the AGN HODs, 
we derive the bias parameter of the AGN sample by applying
\begin{equation}
  b_{\rm AGN}= \frac{\int b_{\rm h}(M_{\rm h}) \langle N_{\rm AGN} \rangle (M_{\rm h})
\phi(M_{\rm h}) dM_{\rm h}}{\int \langle N_{\rm AGN}\rangle (M_{\rm h}) \phi(M_{\rm h})dM_{\rm h}},
  \label{eq:b_agn}
\end{equation}
where $b_{\rm h}(M_{\rm h})$ is the bias of DMHs with a mass $M_{\rm h}$, 
and $\phi(M_{\rm h})$ is the DMH mass function. In Paper II, we discuss different realizations of the HOD 
modeling depending on how central and satellite AGNs in DMHs are included or excluded. Here, we apply 
only ``Model A,'' assuming that, at low redshifts, black hole accretion occurs more frequently 
in galaxies with lower stellar mass than in typical LRGs. 
Thus, we assume that (1) the LRG (highest stellar mass) 
occupies the center of the DMH and (2) all AGNs are satellites within the same DMH.
This is motivated by observations of AGN ``downsizing'' (\citealt{ueda_akiyama_2003}; \citealt{hasinger_2008}; 
\citealt{miyaji_2015}).

As we are interested here (and in Paper III) only in the large-scale bias 
values and the estimate of the typical halo mass, our results 
do not depend significantly on the choice of the one-halo term model (model A,
B, or C). For example, in the case of the total RASS AGN sample, the best-fit 
HOD-derived bias values change by less than 0.02 among models A, B, and
C. This is a minor fraction of the statistical uncertainty in the
bias measurement ($\pm$0.09).

The bias parameters derived by power-law fits and HOD modeling can deviate significantly. 
As discussed in Paper III (see Section~5.4. of Paper III for more details), 
HOD-derived bias parameters are strongly 
preferred over those from power-law fits for various reasons. 

Inconsistencies between the HOD-derived $b(z)$ and the results from power-law fits are 
caused by (1) obtaining power-law fits to the inferred AGN ACFs, which have much larger 
uncertainties than the directly measured CCFs between AGNs and LRGs (the HOD uses CCFs instead of 
the AGN ACFs), and (2) fitting the inferred AGN ACF in a range $r_p=$ 0.3--15 $h^{-1}$ Mpc. The 
latter is also applied by other studies because the clustering signal is not 
well constrained above $r_p\sim$ 15 $h^{-1}$ Mpc. However, such a fit includes scales 
in the nonlinear region in which the bias-DMH mass relation, based on linear theory,
should not be applied. We decide to do so nevertheless to be consistent with Papers I 
and III, as well to allow the reader to perform a direct comparison to other studies. 

Based on the HOD-derived bias parameter ($b(z)$ HOD), 
we derive the typical DMH mass occupied by the AGN sample. 
Using Equation~(8) of \cite{sheth_mo_2001} and the improved fit for this equation given by 
\cite{tinker_weinberg_2005}, we compute the expected large-scale Eulerian bias factor for different 
DMH masses at different redshifts. Comparing the observed $b$ value from HOD modeling
with the DMH bias factor from $\Lambda$CDM cosmological simulations provides an estimate 
of the typical DMH mass ($b_{\rm DMH}(M_{\rm DMH}^{\rm typ}) = b_{\rm OBS,HOD}$) in which the 
different AGN samples reside, as listed in the last column of Table~\ref{xagn_acf}. 
Small differences in the median effective redshift between different
subsamples do not significantly change the derived typical DMH mass. The low
and high $M_{\rm BH}$ samples of the X-ray-selected AGN sample have the
largest difference in median effective redshift when we consider only
samples that have matched distributions in the other parameter of
interest. Computing the typical DMH mass at these different 
redshifts leads to a difference of only $\Delta \log (M_{\rm DMH}^{\rm typ}) = 0.08$
(e.g., for a bias value of 1.20, 
log ($M_{\rm DMH,z=0.22}^{\rm typ}/[h^{-1}$ $M_{\odot}])$= 12.83 and  
log ($M_{\rm DMH,z=0.30}^{\rm typ}/[h^{-1}$ $M_{\odot}])$= 12.75).

We emphasize that $M_{\rm DMH}^{\rm typ}$ should \textit{not} be
compared between different studies. Various conversions are used in the literature 
to derive $M_{\rm DMH}^{\rm typ}$ from $b_{\rm DMH}$. In particular, 
\textit{for the same bias parameter}, studies 
based on optically selected AGN samples derive 
$M_{\rm DMH}^{\rm typ}$ values up to $\sim$0.6 dex lower than studies using X-ray-selected 
AGN samples (depending on the actual bias factor and redshift of the sample). 
The use of the improved fit from \cite{tinker_weinberg_2005} compared to 
Equation~(8) of \cite{sheth_mo_2001} alone can account for an $M_{\rm DMH}^{\rm typ}$ difference 
of $\sim$0.3 dex. Thus, instead of blindly using the derived $M_{\rm DMH}^{\rm typ}$ values from different studies, 
one should recompute these values from the bias parameters in the same manner.

For the X-ray and optically selected AGN samples, we find a clustering
dependence on $M_{\rm BH}$, in that subsamples with high $M_{\rm BH}$ cluster 
more strongly than their low $M_{\rm BH}$ counterparts. No 
significant clustering dependence on $L/L_{\rm EDD}$ is observed. 
We also detect a weak FWHM$_{\rm H\alpha}$ clustering dependence, in that 
AGN with low FWHM$_{\rm H\alpha}$ are less clustered than their high 
FWHM$_{\rm H\alpha}$ counterparts. For the X-ray AGN sample, which has 
a larger dynamical range in $L_{\rm H\alpha}$ than the optical sample, 
we do not find a significant dependence on $L_{\rm H\alpha}$.

\subsection{Robustness of the Clustering Measurements}
\label{robustness}

In this section we verify the reliability of our clustering results by altering the 
selection of the different AGN subsamples with respect to different parameters.
Moderate changes should not influence the clustering results significantly.
Therefore this should serve as a test of the robustness of our results. 

In Section~\ref{estimating_MBH}, we clarify why we do not use black hole mass estimates 
from objects with H$\alpha$ luminosity S/N $<$ 10. For the first 
check, we lower this threshold to S/N$=$5, generate the various AGN subsamples, and compute 
their clustering. Comparing the subsamples with respect to the same parameter (e.g., X-ray 
luminosity, $M_{\rm BH}$, $L/L_{\rm EDD}$), the newly derived constraints agree well 
within their 1$\sigma$ uncertainties with the findings of the S/N=$10$ samples. The 
same applies when we apply a very conservative threshold of S/N=$20$.

The intrinsic scatter of approximately 0.3 dex in the SMBH mass estimate
could have a substantial effect on the observed correlations between $M_{\rm DMH}$ and $M_{\rm BH}$ as well as $L/L_{\rm EDD}$.
In Section~\ref{simulations}, we use simulations to explore the impact of SMBH
mass errors on the clustering correlations and show that the expected
modification of the correlations is very mild. 

In this section, we investigate possible effects of the scatter in the SMBH 
mass estimates from the
observational point of view. Two different approaches for estimating 
$M_{\rm BH}$ based on the optical spectra were discussed in Section~\ref{estimating_MBH}. 
Throughout the paper, we use the relation given in 
\cite{bentz_peterson_2009} (see Eq.~\ref{eq:BH_Bentz}). Here 
we recompute $M_{\rm BH}$ based on \cite{mclure_dunlop_2004} (see Eq.~\ref{eq:BH_Shen}).
For the X-ray and optically selected AGN samples, the $M_{\rm BH}$ values estimated from 
\cite{mclure_dunlop_2004} are on average 0.2 dex smaller than the ones from 
\cite{bentz_peterson_2009}. We repeat the clustering measurement of low and high 
$M_{\rm BH}$ and $L/L_{\rm EDD}$ AGN subsamples, respectively. We find trends
identical to those reported 
in Table~\ref{xagn_acf}. The results for all subsamples based on \cite{mclure_dunlop_2004} agree well 
with their corresponding subsamples based on \cite{bentz_peterson_2009}, 
within the 1$\sigma$ uncertainties. Thus, different estimators for the SMBH
mass, which are subject to different systematic errors, give extremely similar
results. We conclude that, in our study, uncertainties related to the clustering
measurement itself (i.e., the moderate number of AGNs) dominate over 
systematic SMBH mass errors.

A key ingredient of our paper is that we split the AGN sample in such a way that we 
match the distribution of another parameter of interest (see Sect.~\ref{matchingSamples}).
To test this, we alter the procedure of creating the different subsamples. 
First, we fit a regression line to the data and split the sample into objects 
above and below the regression line. The resulting subsamples have moderately deviating 
distributions in the other parameter of interest. This approach yields consistent clustering 
results with respect to our final results presented in Table~\ref{xagn_acf}. 

Second, we repeat the generation of low, medium, and high $M_{\rm BH}$ and $L/L_{\rm EDD}$ 
samples with matched distributions in the other parameter of interest. This time, we split 
the sample into 40\% (low), 20\% (medium), and 40\% (high). 
All corresponding samples 
agree with the original samples, within the combined 1$\sigma$ uncertainties. 
Except for the optical $L/L_{\rm EDD}$ subsamples, we also find identical trends for all 
other parameters in the X-ray and optically selected AGN samples. 
For the optical $L/L_{\rm EDD}$ subsamples, the split into 30\% (low), 
40\% (medium), and 30\% (high) suggests a weak and tentative positive $L/L_{\rm EDD}$
clustering dependence ($\sim$1.7$\sigma$ between the low and high $L/L_{\rm EDD}$ 
subsamples; see Table~\ref{xagn_acf}). 
When we split the same sample into 40\% (low), 20\% (medium), and 40\% (high), we find 
$b_{\rm low}=1.35^{+0.11}_{-0.10}$, $b_{\rm medium}=1.33^{+0.17}_{-0.16}$, and 
$b_{\rm high}=1.35^{+0.14}_{-0.11}$. Thus, we find no evidence for an $L/L_{\rm EDD}$
clustering dependence.

Third, we generate subsamples that have perfectly identical distributions in the 
other parameter of interest (instead of very similar distributions). 
To do so, we compute in each bin which of the subsamples  
(low, medium, or high) contains the most objects. We then randomly select and add again 
objects from the subsamples with the lower number of objects until all three subsamples 
contain an equal number of objects in this bin. In other words, we scale up the number of
objects in a subsample to that of the subsample with the highest number of objects. 
This method allows multiple entries of single objects in a given subsample. Ultimately, 
in each bin only one or two objects are added to subsamples. Thus, these subsamples 
are extremely similar to the matched samples that we use throughout the paper. 
Consequently, it is not surprising that both methods result in 
virtually identical clustering signals.

In Fig.~\ref{high_low_MBH} (right panel histogram) the low, medium, and high
$M_{\rm BH}$ subsamples overlap considerably. 
As a fourth test, we now create low and high $M_{\rm BH}$ and $L/L_{\rm EDD}$ subsamples that have 
completely separated distributions, but still match in the $L/L_{\rm EDD}$ and
$M_{\rm BH}$ distribution, respectively. We divide the X-ray and optical AGN samples at $M_{\rm BH} = 10^{8}\,\rm{M}_\odot$ 
and log $(L/L_{\rm EDD}) = -1$. We weight the samples by 
adding sources in a subsample multiple times until a matched distribution is
created. One major disadvantage of this
approach is that we have to limit substantially the parameter space of the
other parameter to match its distribution in both subsamples (see
Figs.~\ref{RASS_AGN_MBH_LLEDD},\,\ref{high_low_MBH}).
For the X-ray-selected RASS/SDSS AGN samples, we find
$b_{\rm low M_{BH}}=1.05^{+0.18}_{-0.16}$, $b_{\rm high M_{BH}}=1.26^{+0.21}_{-0.20}$,
$b_{\rm low L/L_{EDD}}=1.13^{+0.18}_{-0.18}$, and $b_{\rm high L/L_{EDD}}=1.17^{+0.16}_{-0.14}$.
The bias values for the optically selected SDSS AGN samples are 
$b_{\rm low M_{BH}}=1.26^{+0.14}_{-0.16}$, $b_{\rm high M_{BH}}=1.38^{+0.16}_{-0.15}$,
$b_{\rm low L/L_{EDD}}=1.32^{+0.10}_{-0.13}$, and $b_{\rm high L/L_{EDD}}=1.28^{+0.21}_{-0.21}$.
Due to significant restriction of the parameter space and thus a
substantial decrease in the number of AGN, this approach leads to less
significant results.

Fifth, we combine the X-ray and optically selected AGN
samples. We divide the combined sample into the highest 30\%, middle
40\%, and lowest 
30\% with respect to the corresponding parameter. We also combine the middle 
and lowest samples (70\%). Again, the different
subsamples have matched distributions in the other parameter of interest.
We find a 2.6$\sigma$ difference when comparing the AGNs with the 30\% highest 
to the 70\% lowest SMBH masses 
($b_{\rm 70\% low M_{BH}}=1.19^{+0.07}_{-0.08}$, $b_{\rm high M_{BH}}=1.50^{+0.14}_{-0.10}$).

\begin{deluxetable}{lccc }
\tabletypesize{\normalsize}
\tablecaption{Clustering Properties with Respect to $M_{\rm BH}$ and $L/L_{\rm EDD}$ When 
Using Unmatched Distributions in the Other Parameter of Interest\label{xagn_straightcut}}
\tablehead{
\colhead{AGN Sample} &\colhead{Median log} &\colhead{Median log}       &\colhead{$b(z)$}  \\
\colhead{Name} & \colhead{$(M_{\rm BH}/M_\odot)$}  &\colhead{$(L/L_{\rm EDD})$}   &\colhead{HOD}} 
\startdata
\multicolumn{4}{c}{X-ray-selected RASS/SDSS AGN -- Data Release 4+}\\\hline
low $M_{\rm BM}$ RASS      &  7.49 & -0.76&1.00$^{+0.18}_{-0.15}$\\      
medium $M_{\rm BM}$ RASS   &  7.92 & -1.01&1.36$^{+0.20}_{-0.18}$\\      
high $M_{\rm BM}$ RASS     &  8.46 & -1.33&1.49$^{+0.15}_{-0.16}$ \\     
                      &       &      & \\
low $L/L_{\rm EDD}$ RASS   &  8.29 & -1.42&1.41$^{+0.17}_{-0.16}$\\      
medium $L/L_{\rm EDD}$ RASS&  7.92 & -1.00&1.49$^{+0.20}_{-0.15}$\\      
high $L/L_{\rm EDD}$ RASS  &  7.62 & -0.64&1.09$^{+0.10}_{-0.12}$\\\hline 
\multicolumn{4}{c}{Optically Selected AGN -- SDSS AGN -- SDSS Data Release 7}\\\hline
low $M_{\rm BM}$ SDSS      &  7.72 & -0.72&1.28$^{+0.11}_{-0.16}$\\      
medium $M_{\rm BM}$ SDSS   &  8.12 & -1.04&1.15$^{+0.08}_{-0.07}$\\      
high $M_{\rm BM}$ SDSS     &  8.58 & -1.42&1.58$^{+0.07}_{-0.06}$ \\     
                      &       &      & \\
low $L/L_{\rm EDD}$ SDSS   &  8.51 & -1.46&1.43$^{+0.11}_{-0.10}$\\    
medium $L/L_{\rm EDD}$ SDSS&  8.08 & -1.03&1.42$^{+0.11}_{-0.12}$\\    
high $L/L_{\rm EDD}$ SDSS  &  7.76 & -0.65&1.12$^{+0.15}_{-0.06}$      
\tablecomments{An explanation of the columns is given in Table~\ref{xagn_acf}.}
\end{deluxetable}

Last, we drop the requirement of having subsamples with matched distributions in
the other parameter of interest. 
We naively create samples of low, medium, and high $M_{\rm BH}$, $L/L_{\rm EDD}$, FWHM, 
and $L_{\rm H\alpha}$ simply by dividing the samples into the highest 30\%, medium 40\%, and lowest 
30\% with respect to the corresponding parameter. Dividing the sample according to $M_{\rm BH}$, 
for example, in this manner leads to subsamples with very different median $L/L_{\rm EDD}$ values. 
We present the results for subsamples in $M_{\rm BH}$ and $L/L_{\rm EDD}$ with unmatched 
distributions in the other parameter (for X-ray and optical AGN) in Table~\ref{xagn_straightcut}. 
Despite the ``mixing'' of the two parameters, $M_{\rm BH}$ and $L/L_{\rm EDD}$, the trends seen in 
the HOD-based bias parameters are similar, though not identical, to the 
subsamples with matched distributions.
We will discuss these results below in Sect.~\ref{origin} in more detail.   

As described above, we consider only data bins for all CCFs with $r_{\rm p}>0.7$ $h^{-1}$ Mpc for the 
computation of the HOD-based bias parameter.
As a last test, we recompute the HOD bias parameter using bins with 
$r_{\rm p}>0.5$ $h^{-1}$ Mpc. Again, the results agree well within the 1$\sigma$ uncertainties. 
The few exceptions are caused by a very low data point at $r_{\rm p} \sim 0.6$ $h^{-1}$ Mpc. In all
of these cases, there are at most 10 pairs that contribute to this data point. Thus, 
$\chi^2$ statistics, as mentioned before, should not be applied.
The combination of the various tests listed here provides convincing evidence that our 
results are not significantly influenced by systematic effects and demonstrates their robustness
with respect to moderate changes in our methodology and sample selection.

\begin{figure*}
\vspace*{0.2cm}
 \begin{minipage}[b]{0.99\textwidth}
\centering
 \resizebox{\hsize}{!}{
  \includegraphics[bbllx=210,bblly=47,bburx=470,bbury=741,angle=-90]{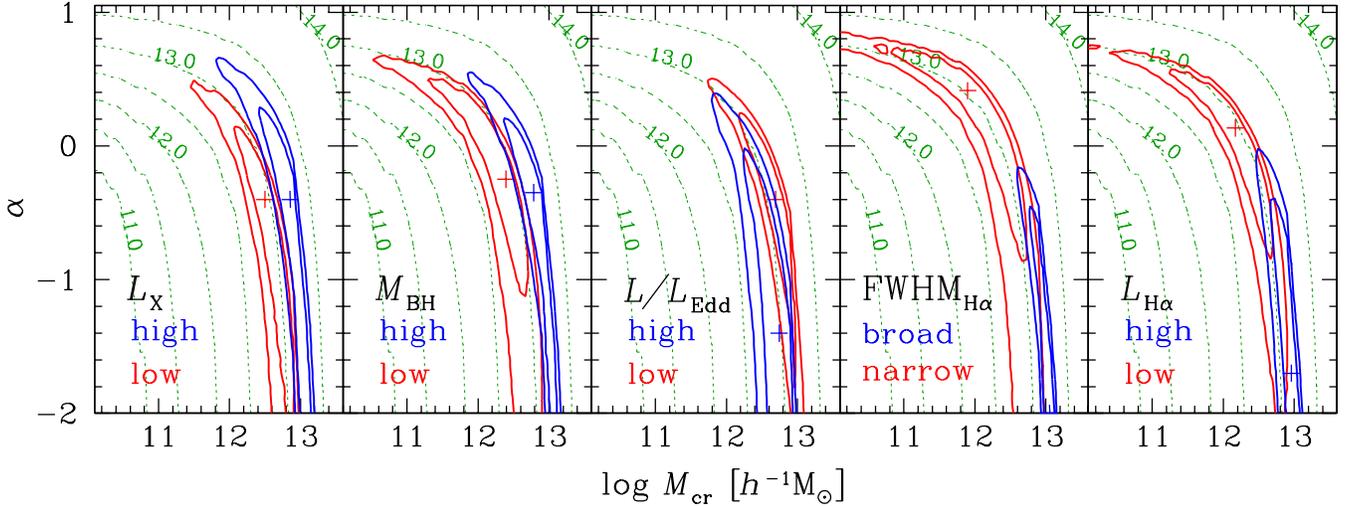}}
\end{minipage}
\caption{Confidence contours of the HOD-derived parameters $\alpha$ and log $M_{\rm cr}$ for 
the low and high $L_{\rm X}$, $M_{\rm BH}$, $L/L_{\rm EDD}$, FWHM, and $L_{\rm H\alpha}$ RASS/SDSS (X-ray-selected) 
luminous, broad-line AGN subsamples (from left to right). 
The blue contours show the high AGN subsamples in one parameter, and the red contours show 
the low subsamples. The confidence intervals correspond to the 
$\Delta \chi^2 = 1.0$ and 2.3 levels. 
The green dashed lines show the mean DMH mass in log ($M_{\rm DMH}/[\rm{h}^{-1}\,M_{\odot}]$) 
derived from the model parameters. Every other contour level is labeled.}
\label{HOD_results}
\vspace{0.4cm}
\end{figure*}

\begin{figure*}
\vspace*{0.2cm}
 \begin{minipage}[b]{0.99\textwidth}
\centering
 \resizebox{\hsize}{!}{
  \includegraphics[bbllx=210,bblly=47,bburx=470,bbury=741,angle=-90]{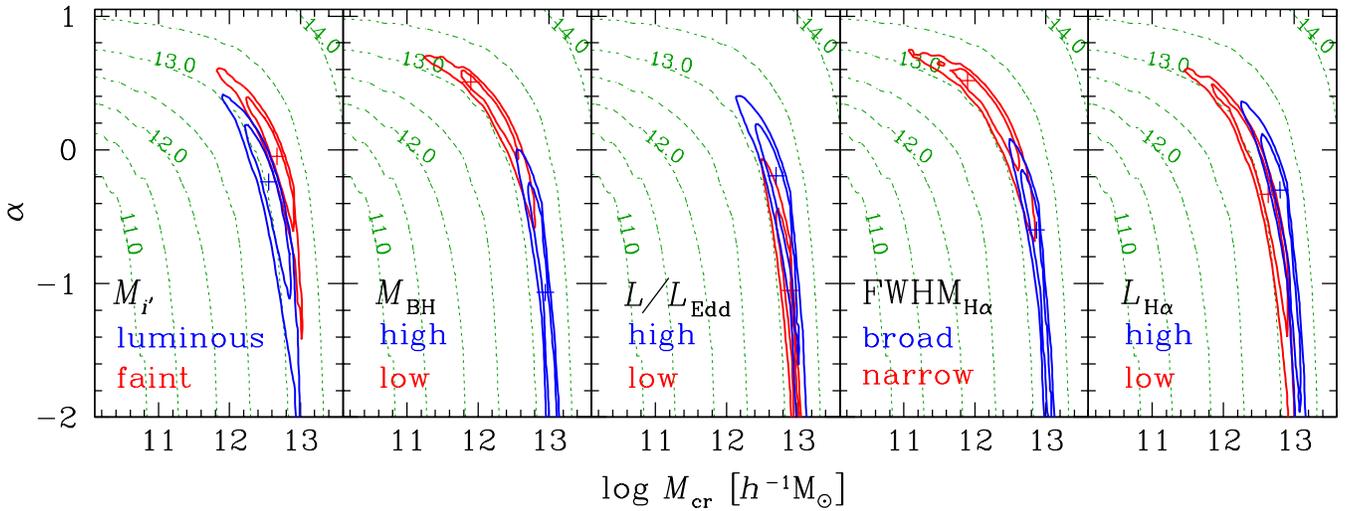}}
\end{minipage}
\caption{Similar to Fig.~\ref{HOD_results}, here showing the confidence contours 
of the HOD parameters $\alpha$ and log $M_{\rm cr}$ for 
the low and high $M_i$, $M_{\rm BH}$, $L/L_{\rm EDD}$, FWHM, and $L_{\rm H\alpha}$ optically selected SDSS AGN samples (from left to right).}
\label{HOD_results_opt}
\vspace{0.4cm}
\end{figure*}

\subsection{Exploring the One-halo Term Dependence of Different AGN Parameters}

The HOD modeling introduced in Paper II of this series allows us to explore 
properties of the one-halo term, that is, at small separations where two objects occupy 
the same DMH. More importantly, we are able to provide the full distribution of the number of 
AGNs as a function of $M_{\rm DMH}$, instead of quoting only a typical value. 
As a parameterization of the AGN HOD, we use a simple 
truncated power law, assuming that all AGNs are in satellites ('Model A' in Paper II, for more 
details see Paper II):
\begin{equation}
\langle N_{\rm AGN,s}\rangle \propto {M_{\rm DMH}}^\alpha\;\Theta(M_{\rm DMH}- M_{\rm cr})
\label{eq:agnhod}
\end{equation}
where $\Theta(x)$ is the step function (=1 at $x\ge 0$; =0 at $x<0$), $M_{\rm cr}$ is a 
critical (minimum) DMH mass below which the HOD is zero (where the DMH does not contain an AGN), 
$\alpha$ is the power-law slope of the HOD above $M_{\rm cr}$, and $N_{\rm AGN,s}$ the 
number of (satellite) AGNs in the same DMH.

\begin{deluxetable}{lcccc}
\tabletypesize{\normalsize}
\tablecaption{Clustering Properties with Respect to $M_{\rm BH}$ and $L/L_{\rm EDD}$ 
for Divisions into Two Subsamples with Matched Distributions\label{sample_2split}}
\tablehead{
\colhead{AGN Sample} &\colhead{Median} &\colhead{Median\,log} &\colhead{Median\,log}       &\colhead{$b(z)$}  \\
\colhead{Name} & \colhead{$z_{\rm eff}$} & \colhead{$(M_{\rm BH}/M_\odot)$}  &\colhead{$(L/L_{\rm EDD})$}   &\colhead{HOD}} 
\\
\multicolumn{5}{c}{X-ray-selected RASS/SDSS AGN -- Data Release 4+}\\\hline
\startdata
\multicolumn{5}{c}{X-ray selected AGN -- RASS/SDSS AGN -- SDSS Data Release 4+}\\\hline
low $M_{\rm BM}$     & 0.24 &  7.66 & -1.00 &1.14$^{+0.13}_{-0.14}$\\ 
high $M_{\rm BM}$    & 0.29 &  8.23 & -1.00 &1.42$^{+0.15}_{-0.15}$\\  
                   &      &       &       &                    \\
low $L/L_{\rm EDD}$  & 0.25 &  7.93 & -1.24 &1.30$^{+0.15}_{-0.13}$\\ 
high $L/L_{\rm EDD}$ & 0.28 &  7.92 & -0.77 &1.28$^{+0.13}_{-0.14}$\\\hline
                   &      &Median &Median\,log&       \\
                   &      & FWHM  &$L_{\rm H\alpha}$&   \\\hline 
low FWHM          &0.26  &  2170 & 42.72 &1.15$^{+0.19}_{-0.12}$\\
high FWHM          &0.27  &  4050 & 42.72 &1.47$^{+0.13}_{-0.13}$\\
                   &      &       &        &                    \\
low $L_{\rm H\alpha}$ &0.24  & 2890 & 42.45 &1.15$^{+0.16}_{-0.14}$\\ 
high $L_{\rm H\alpha}$&0.29  & 2890 & 43.04 &1.44$^{+0.13}_{-0.14}$\\ 
                   &      &       &       &                    \\\hline
\multicolumn{5}{c}{Optically Selected AGN -- SDSS AGN -- SDSS Data Release 7}\\\hline
low $M_{\rm BM}$    &0.28 &  7.89 & -1.02&1.29$^{+0.09}_{-0.10}$\\
high $M_{\rm BM}$    &0.30 &  8.32 & -1.04&1.43$^{+0.10}_{-0.09}$ \\
                   &    &       &      & \\
low $L/L_{\rm EDD}$  &0.28 &  8.12 & -1.22&1.38$^{+0.08}_{-0.09}$\\
high $L/L_{\rm EDD}$ &0.30 &  8.12 & -0.83&1.38$^{+0.09}_{-0.11}$\\\hline 
                   &      &Median &Median\,log&       \\
                   &      & FWHM  &$L_{\rm H\alpha}$&   \\\hline 
low FWHM          &0.26  &  2440 & 42.88 &1.27$^{+0.10}_{-0.08}$\\
high FWHM         &0.27  &  4550 & 42.88 &1.45$^{+0.08}_{-0.11}$\\
                   &      &       &        &                    \\
low $L_{\rm H\alpha}$ &0.24  & 3250 & 42.69 &1.30$^{+0.09}_{-0.10}$\\ 
high $L_{\rm H\alpha}$&0.29  & 3280 & 43.10 &1.42$^{+0.09}_{-0.09}$   
\tablecomments{An explanation of the columns is given in Table~\ref{xagn_acf}.
The units of FWHM are km\,s$^{-1}$ and log $L_{\rm H\alpha}$ are log ($L_{\rm H\alpha}$/[erg\,s$^{-1}$]).
Note that the units and column descriptions of the optically selected SDSS AGNs are identical 
to the ones used for the X-ray-selected RASS/SDSS AGNs.}
\end{deluxetable}

While a model with an adequate mix of central and satellite AGNs is more
realistic (models B and C in Paper II), we here show the results of our
simplest model because the purpose of this section is to highlight the
differences between samples, including the constraints from both the 
one- and two-halo terms, while the constraints on bias values come only from 
the two-halo term. 

To achieve adequate constraints, we divide the full AGN sample into only two subsamples (instead of 
three, as above) for each 
parameter of interest. We follow the description of Sect.~\ref{matchingSamples} and create 
low and high samples for 
$M_{\rm BH}$, $L/L_{\rm EDD}$, FWHM, and $L_{\rm H\alpha}$ with matched distributions in $L/L_{\rm EDD}$,
$M_{\rm BH}$, $L_{\rm H\alpha}$, and FWHM, respectively. We list their clustering properties in 
Table~\ref{sample_2split}. The low and high $L_{\rm X}$ and $M_i$ samples 
use a simple cut (log $(L_{\rm 0.1-2.4\,keV}/[\rm{erg}\,\rm{s}^{-1}])=44.29$, $M_i=-22.4$).
For the faint $M_i$ sample ($\langle M_i \rangle = -22.18$), we find $b_{\rm faint}=1.45^{+0.10}_{-0.12}$, 
while the luminous $M_i$ sample ($\langle M_i \rangle = -22.79$) yields
$b_{\rm lum.}=1.29^{+0.07}_{-0.10}$ (1.2$\sigma$ difference). The $M_i$
clustering dependence will be discussed in detail in Sect.~\ref{why_optical}.

We run the HOD modeling for all low and high AGN subsamples using bins with $r_{\rm p}>0.5$ $h^{-1}$ Mpc. 
To do so, we assume that $\chi^2$ statistics can be applied if a data point contains at 
least 15 pairs. Lowering the $r_{\rm p}$ limit allows us to explore the one-halo term via HOD modeling 
in more detail. We derive the best fit for all of the low and high parameters individually and show their 
results in Figs.~\ref{HOD_results} and \ref{HOD_results_opt}. These two-dimensional plots show a clearer 
difference between the subsamples than collapsing the information into one dimension such as the bias parameter. 

All low and high subsamples (independent of the studied parameter) are consistent with 
 $\alpha < 1$. As pointed out in Paper II, HOD analyses of galaxies over a wide range of absolute 
magnitudes and redshifts find $\alpha \sim 1.0-1.2$. Comparing the AGN and galaxy results implies 
that models are preferred in which the fraction of satellite AGNs to satellite galaxies 
decreases with increasing DMH mass. Fig.~\ref{HOD_results} shows evidence that high 
FWHM and broad-line $L_{\rm H\alpha}$ AGNs show an even steeper decrease with DMH mass (i.e., lower $\alpha$).

\begin{figure*}
 \begin{minipage}[b]{0.48\textwidth}
\centering
 \resizebox{\hsize}{!}{
  \includegraphics[bbllx=84,bblly=364,bburx=545,bbury=700]{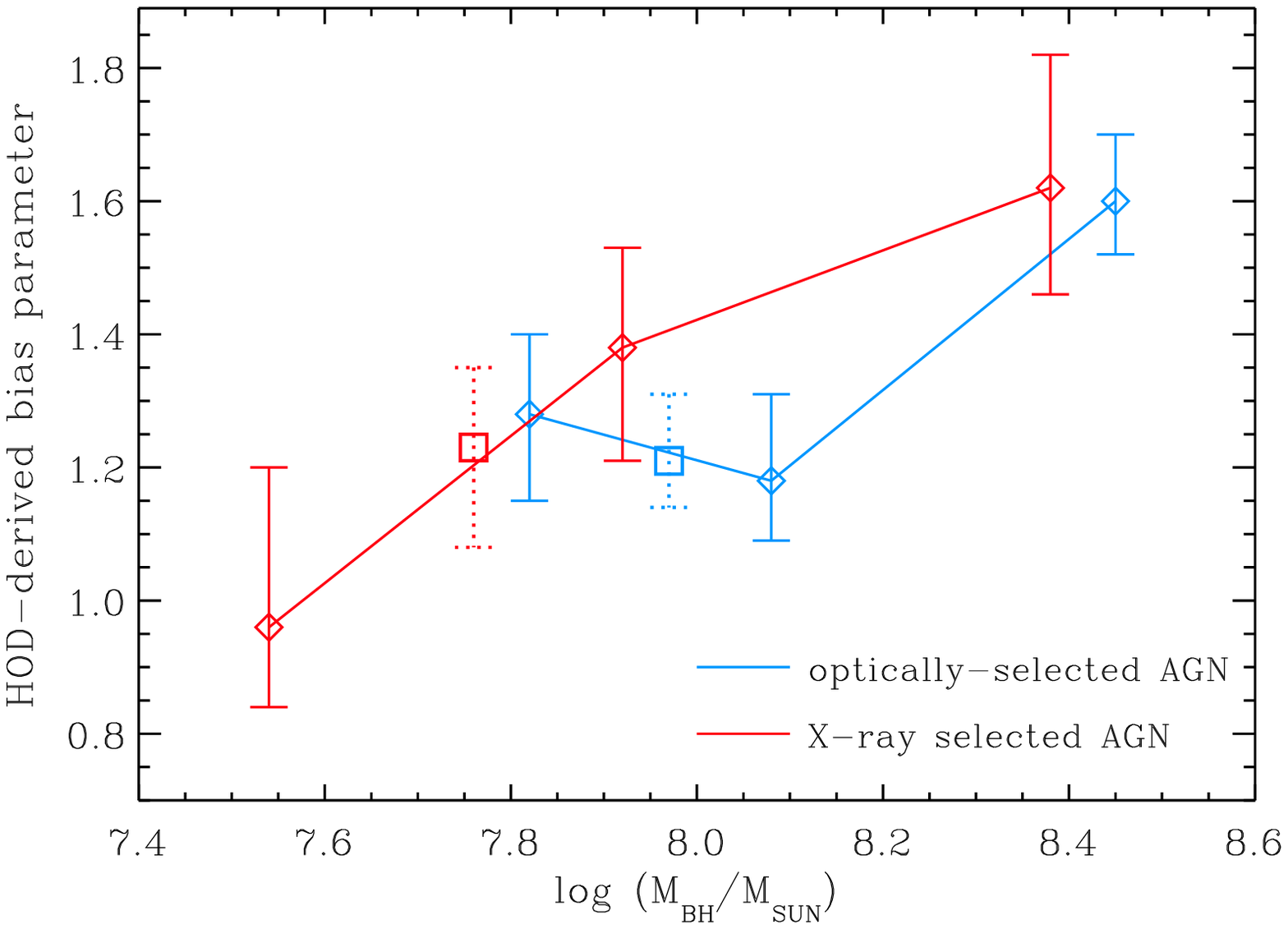}} 
\end{minipage}
\hfill
\begin{minipage}{0.48\textwidth}
\vspace*{-5.6cm}
\centering
\resizebox{\hsize}{!}{
   \includegraphics[bbllx=84,bblly=364,bburx=545,bbury=700]{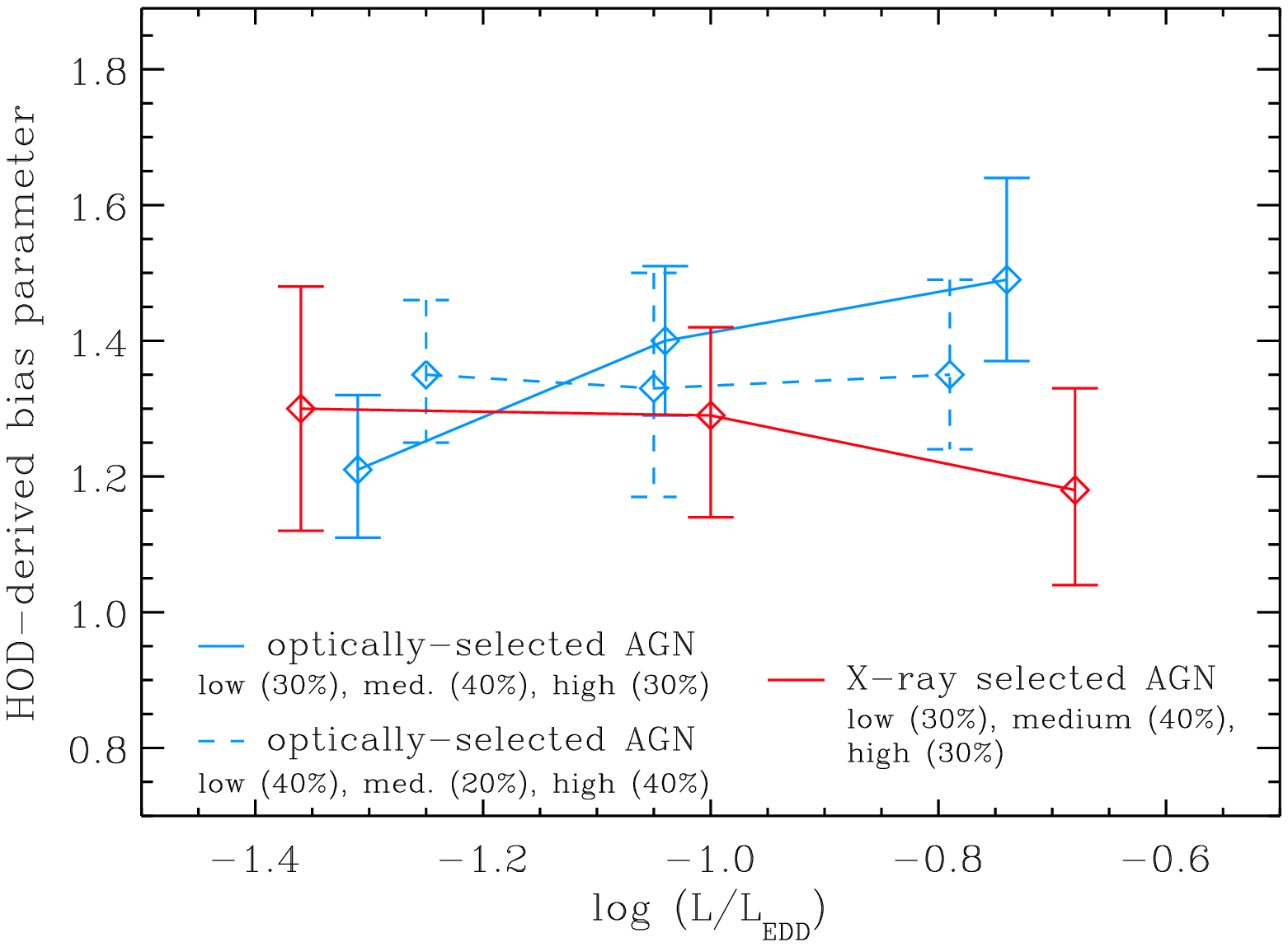}} 
\vspace*{-0.0cm}
\end{minipage}
\caption{\textit{Left:} HOD-derived bias parameter as a function of SMBH mass for 
               the optically selected SDSS (blue) and the X-ray-selected RASS/SDSS AGN 
               sample (red). The diamond points show a 
               division of the full sample in $M_{\rm BH}$ into 30\% high, 40\% medium, and 30\% low
               $M_{\rm BH}$ subsamples. The boxes (with dotted error bars) indicate the combined  
               medium and low $M_{\rm BH}$ subsamples. All data points are plotted 
               at the median $M_{\rm BH}$ of the corresponding subsample.\\
          \textit{Right:} similar to the left panel, showing the HOD-derived 
                          bias parameter as a function of Eddington ratio. 
    For the optically selected AGN sample, we display the division of the full sample into
    30\% high, 40\% medium, and 30\% low (solid blue line) and the division into 40\%
    high, 20\% medium, and 40\% low (dashed blue line). For the sake of clarity,
    for the X-ray-selected AGN sample we show only the division into 30\% high, 40\% medium,
    and 30\% low, which is extremely similar to the other divisions for this sample.}
         \label{MBH_LLedd_bias}
\end{figure*}

The confidence contours of the optically selected AGN subsamples (Fig.~\ref{HOD_results_opt}) are 
narrower than for the X-ray-selected sample. This is due to the larger number of optical AGNs 
than X-ray-selected AGNs. 
However, the general two-dimensional appearance of the confidence contours are very 
similar between the X-ray and optically selected 
AGN samples. We note that, for the $M_{\rm BH}$ subsamples, the low and high
optically selected AGNs 
have different $\alpha$ and $M_{\rm cr}$ values, while the low and high
$M_{\rm BH}$ X-ray-selected 
AGN subsamples have consistent $\alpha$ and different $M_{\rm cr}$ values.\\

\section{Discussion}

In this section, we use our results to discuss the physical origin of the observed
 $L_{\rm X}$ clustering dependence. We also compare our results to other studies and discuss why we 
do not detect an $M_i$ clustering dependence when the clustering results 
between the X-ray and optical AGN samples are so similar. Finally, we compare our results 
to predictions from state-of-the-art semianalytic cosmological simulations.  

\subsection{$M_{\rm BH}$ as the Origin of the $L_{\rm X}$ Clustering Dependence}
\label{origin}
We first verify that the weak X-ray luminosity dependence of the 
clustering strength found in Paper I is still present in the reduced AGN sample when we 
exclude $\sim$13\% of all objects to reliably estimate $M_{\rm BH}$.
The dependence is detected at the 1.4$\sigma$ level when considering ``$b(z)$ HOD'' in Table~\ref{xagn_acf} (and is 2.5$\sigma$ comparing the $r_0$ values at fixed $\gamma$). 
Using the full X-ray sample, we reported a 1.8$\sigma$ detection in Paper II using HOD-derived bias values.  

We split the X-ray-selected AGN sample into three subsamples with respect to 
$M_{\rm BH}$ (with matched  $L/L_{\rm EDD}$ distributions) 
and show the results in Fig.~\ref{MBH_LLedd_bias} (left) as a blue line. The X-ray sample shows 
a steady increase of the bias parameter with $M_{\rm BH}$. The difference between the lowest 
and highest $M_{\rm BH}$ subsamples in the X-ray-selected AGN sample is 2.3$\sigma$. 

For the X-ray and optically selected AGN sample (red line), we detect the highest clustering in the 
subsample that contains the 30\% most massive black holes. For both AGN populations, 
we combine the low and the medium subsamples and compute their combined clustering strength. 
As a result of the increased sample size, the uncertainties decrease. 
The bias difference compared to the high 
 $M_{\rm BH}$ subsample is 2.0$\sigma$ and 2.7$\sigma$ for the X-ray and optical AGN samples, 
respectively. We show these data points at the median $M_{\rm BH}$ values of each subsample 
in Fig.~\ref{MBH_LLedd_bias}. 

The X-ray AGN sample shows no hints of a correlation between clustering strength and 
 $L/L_{\rm EDD}$ (Fig.~\ref{MBH_LLedd_bias}, right panel). The optical 
AGN sample has a slight positive correlation in that 
AGNs with higher $L/L_{\rm EDD}$ cluster are slightly more clustered 
than their lower $L/L_{\rm EDD}$ counterparts. 
However, this $\sim$1.7$\sigma$ difference disappears when we split the 
sample into 40\% (low), 20\% (medium), and 40\% (high) bins in 
$L/L_{\rm EDD}$ (see Fig.~\ref{MBH_LLedd_bias}, right panel). In Sect.~\ref{why_optical} below we discuss the nonnegligible selection 
biases of the optical SDSS AGN sample; as a result, we regard the X-ray-selected AGN sample as 
less contaminated by observational selection effects. Thus, we do not 
find convincing statistical evidence for a clustering dependence with $L/L_{\rm EDD}$. 

For the X-ray-selected AGNs, we still detect a 2.0$\sigma$ difference between the low and high $M_{\rm BH}$ samples 
when we do not require that the subsamples have matched distributions in $L/L_{\rm EDD}$.
Using simple cuts in $L/L_{\rm EDD}$ without required matched distributions in $M_{\rm BH}$ 
yields a 2.2$\sigma$ difference in that low and medium $L/L_{\rm EDD}$ AGNs cluster
 more strongly than high $L/L_{\rm EDD}$ AGNs. If we assume that 
this $L/L_{\rm EDD}$ AGN clustering dependence is real and that there is no clustering dependence 
on $M_{\rm BH}$, this finding would contradict the observed X-ray luminosity clustering dependence 
that high $L_{\rm X}$ AGNs (and thus high $L/L_{\rm EDD}$) are more strongly clustered than low 
$L_{\rm X}$ AGNs. Figures~\ref{RASS_AGN_MBH_LLEDD} and~\ref{high_low_MBH} show that 
the AGNs with the lowest $L/L_{\rm EDD}$ AGN also have high $M_{\rm BH}$. Thus, we conclude 
that differences in $M_{\rm BH}$ (in these naively defined subsamples) are responsible 
for the observed differences in the clustering of the low and high $L/L_{\rm EDD}$ AGN subsamples. 
Similar results are found for the optically selected AGN sample when we do simple divisions of the full 
sample without matching the distributions in the other parameter of interest. 
 
Since $M_{\rm BH}$ is estimated from the observed parameters FWHM and $L_{\rm H\alpha}$, we also investigate 
the dependence of the clustering strength when we divide the full samples according to these parameters 
(see Table~\ref{xagn_acf}). As before, we divide the sample in one parameter in such a 
way that we conserve the distribution in the other parameter in all subsamples.
In our sample of luminous broad-line (X-ray-selected) RASS/SDSS AGNs, we find a 2.0$\sigma$ clustering 
difference between the lowest and highest FWHM samples and only a 0.6$\sigma$ difference between 
the lowest and highest H$\alpha$ luminosity samples. For the optical AGN sample, we find a 
1.9$\sigma$ difference for the low and high FWHM samples and a 1.5$\sigma$
difference for the low and high H$\alpha$ luminosity 
samples. Both parameters are used to estimate $M_{\rm BH}$, though the scaling
is much stronger with FWHM than with $L_{\rm H\alpha}$ (converted to
$L_{5100}$, see Equations~\ref{eq:BH_Shen} and~\ref{eq:BH_Bentz}).  
The similar FWHM clustering dependence in the X-ray and optical AGN samples 
provides further evidence that $M_{\rm BH}$ is the key parameter driving 
the observed $L_{\rm X}$ dependence of the clustering strength.

Analysis of the two-dimensional parameter space in Fig.~\ref{HOD_results} further supports our claim 
that the $L_{\rm X}$ dependence of the clustering strength originates from an $M_{\rm BH}$ dependence. 
For the X-ray-selected sample (Fig.~\ref{HOD_results}), the contours of the low and high 
$M_{\rm BH}$ samples are extremely similar to the $L_{\rm X}$ contours, while the $L/L_{\rm EDD}$ 
contours differ significantly from the $L_{\rm X}$ contours. 
We therefore conclude that only the black hole mass, and not a 
combination of $M_{\rm BH}$ and $L/L_{\rm EDD}$, is responsible for the observed weak X-ray 
luminosity dependence of the clustering. Within the explored $M_{\rm BH}$ range, we estimate 
an average increase of the bias parameter of $\Delta b\sim0.7$ per dex $M_{\rm  BH}$. This corresponds to approximately one dex increase in $M_{\rm DMH}$ per dex $M_{\rm  BH}$.

If our studies had shown that the clustering strength correlates directly 
with $L/L_{\rm EDD}$ but not with $M_{\rm BH}$, then more strongly clustered 
AGNs would accrete more material. In this scenario, because highly accreting AGNs 
would lie in environments with a high density of galaxies, galaxy mergers 
and galaxy--galaxy interactions would likely be the main triggers of AGN activity. 
However, we find that the clustering strength 
depends mainly on $M_{\rm BH}$ and not on $L/L_{\rm EDD}$.  
Thus, high X-ray luminosity AGNs do not necessarily require dense 
environments to accrete more matter. We find that, on average, 
AGNs in dense environments have higher $M_{\rm BH}$ than their counterparts in 
lower density environments. The higher $L_{\rm X}$ that are observed are then a 
direct consequence of having higher $M_{\rm BH}$. 

\begin{figure*}
 \begin{minipage}[b]{0.48\textwidth}
\centering
 \resizebox{\hsize}{!}{
  \includegraphics[bbllx=84,bblly=364,bburx=545,bbury=700]{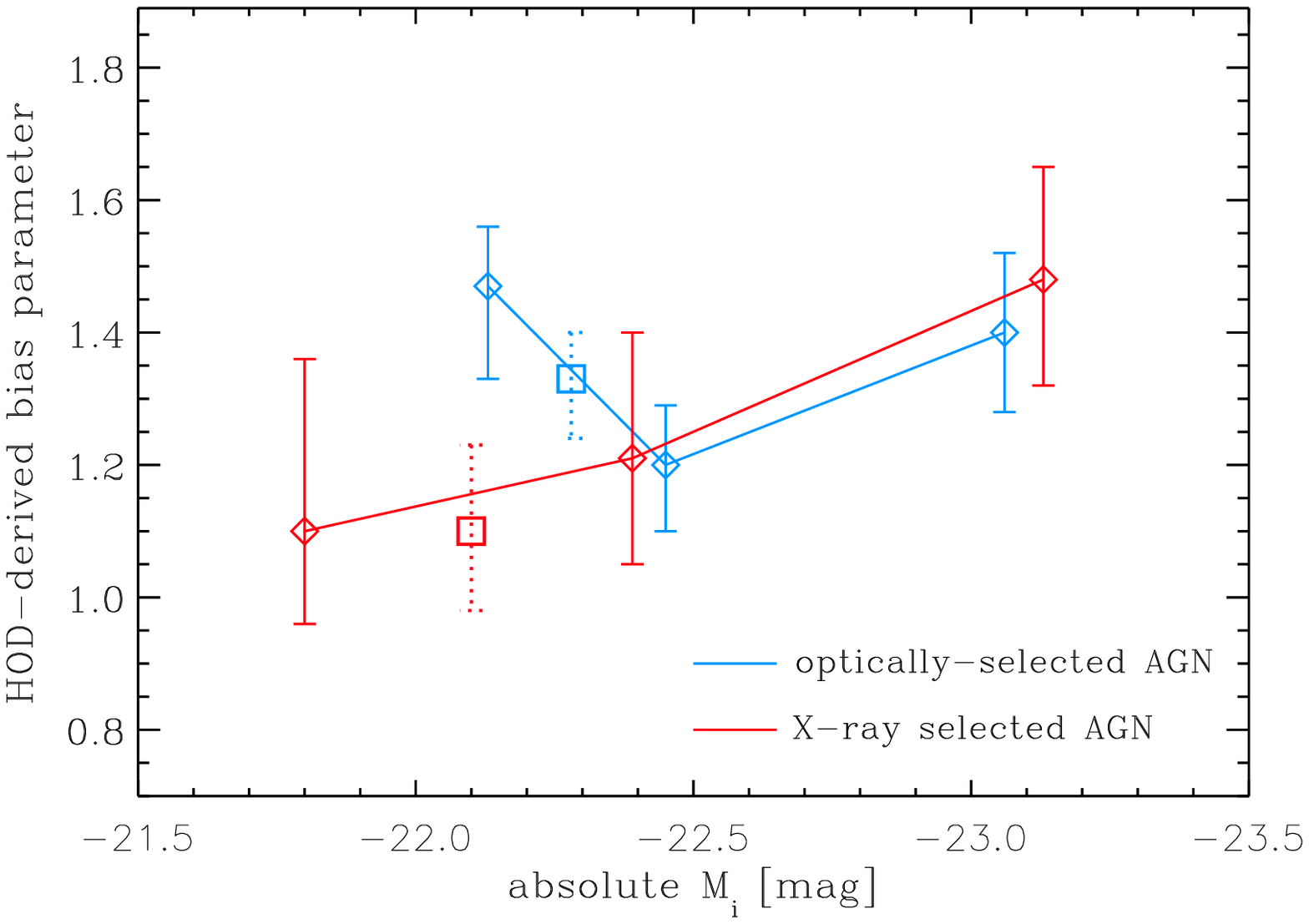}} 
\end{minipage}
\hfill
\begin{minipage}{0.48\textwidth}
\vspace*{-6.05cm}
\centering
 \resizebox{\hsize}{!}{
  \includegraphics[bbllx=84,bblly=364,bburx=545,bbury=700]{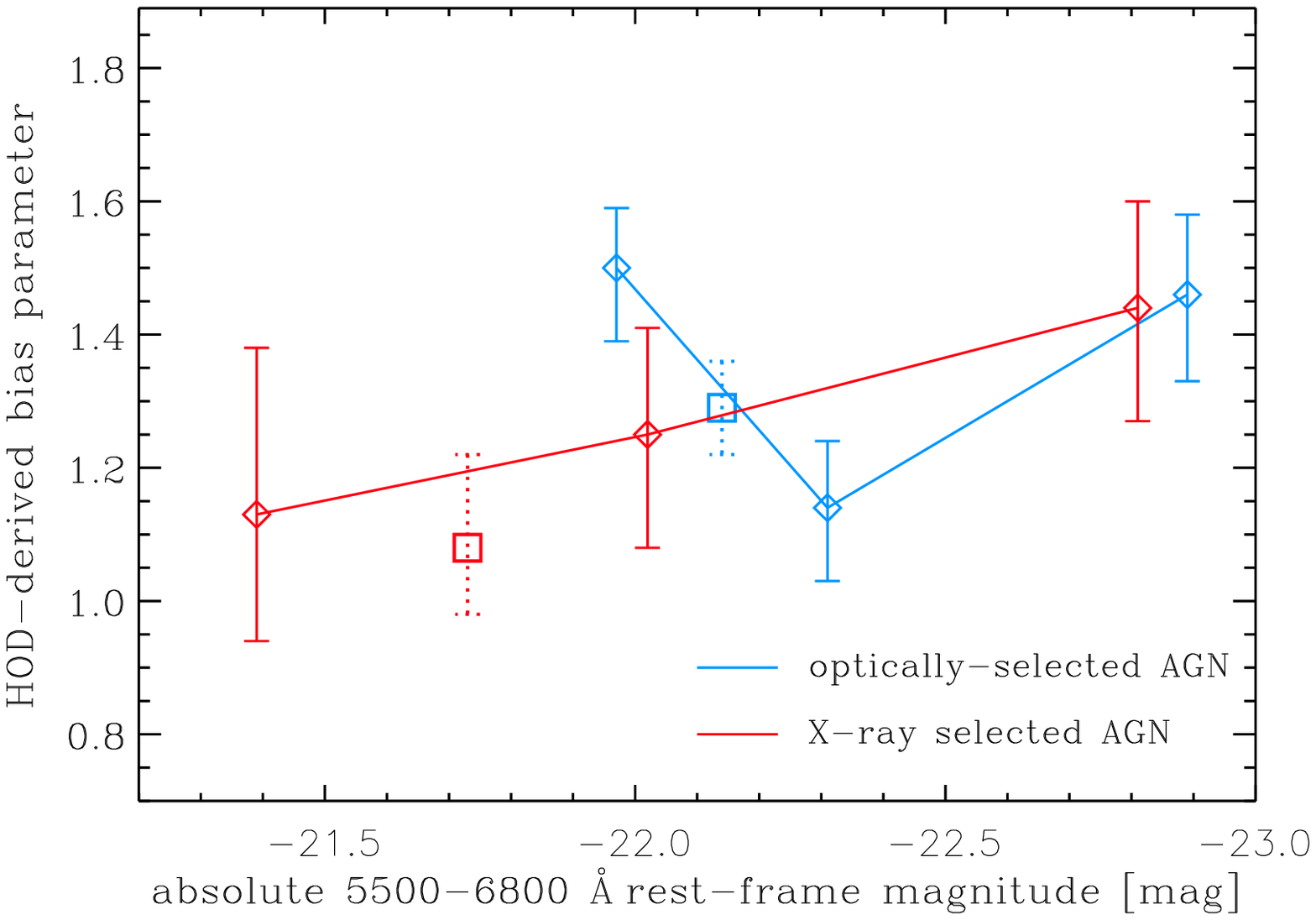}} 
\end{minipage}
\caption{Similar to Fig.~\ref{MBH_LLedd_bias}. \textit{Left:} HOD-derived bias parameter as a function of 
         absolute 
         SDSS $M_i$ magnitude for low (33\%), medium (34\%), and high (33\%) subsamples. 
         \textit{Right:} HOD-derived bias parameter as a function of 
         absolute rest-frame magnitude in the 5500--6800 \AA\ band. The data points shown 
         with a box (with uncertainties shown as dotted lines) correspond 
         to the samples created by combining the low and medium absolute
         5500--6800 \AA\ rest-frame magnitude 
         subsamples.}
         \label{Mi_Mrest_bias}
\end{figure*}

Our findings are also consistent with a scenario in which the AGN luminosity can vary on short 
timescales (e.g., \citealt{hickox_mullaney_2014}), due to a variability 
in the black hole accretion rate ($L/L_{\rm EDD}$). In this view, the 
instantaneous luminosity of an AGN is only a weak indicator of the time-averaged black hole 
accretion rate. 
Since $L/L_{\rm EDD}$ (which reflects the amount of accreted gas) can rapidly fluctuate, 
it will not necessarily show a correlation with the properties of the host galaxy or DMH mass. 
On the other hand, $M_{\rm BH}$ cannot fluctuate on short timescales as it stays nearly constant or grows 
slowly. Therefore, the black hole mass is not subject to stochastic variability and correlates more tightly 
with host galaxy properties and the host DMH mass.

In the hierarchical model of structure formation, more massive galaxies reside in 
more massive DMHs (e.g., \citealt{mostek_coil_2013}; \citealt{skibba_coil_2015}). 
More massive galaxies are also more luminous (e.g., \citealt{tully_fisher_1977}; 
\citealt{mcgaugh_schombert_2000}).
This leads to the well-observed luminosity dependence of the clustering signal for galaxies
(e.g., \citealt{zehavi_zheng_2011}; \citealt{skibba_smith_2014}).
If more massive galaxies also had more massive bulges, the relationship
between the $M_{\rm BH}$ 
and stellar velocity dispersion of the surrounding galactic bulge 
(e.g., \citealt{ferrarese_merritt_2000}; \citealt{gebhardt_bender_2000}; \citealt{kormendy_ho_2013}) 
should then also lead to a correlation of $M_{\rm BH}$ and $M_{\rm DMH}$. 

\begin{figure}
\centering
 \resizebox{\hsize}{!}{
  \includegraphics[bbllx=90,bblly=367,bburx=545,bbury=697]{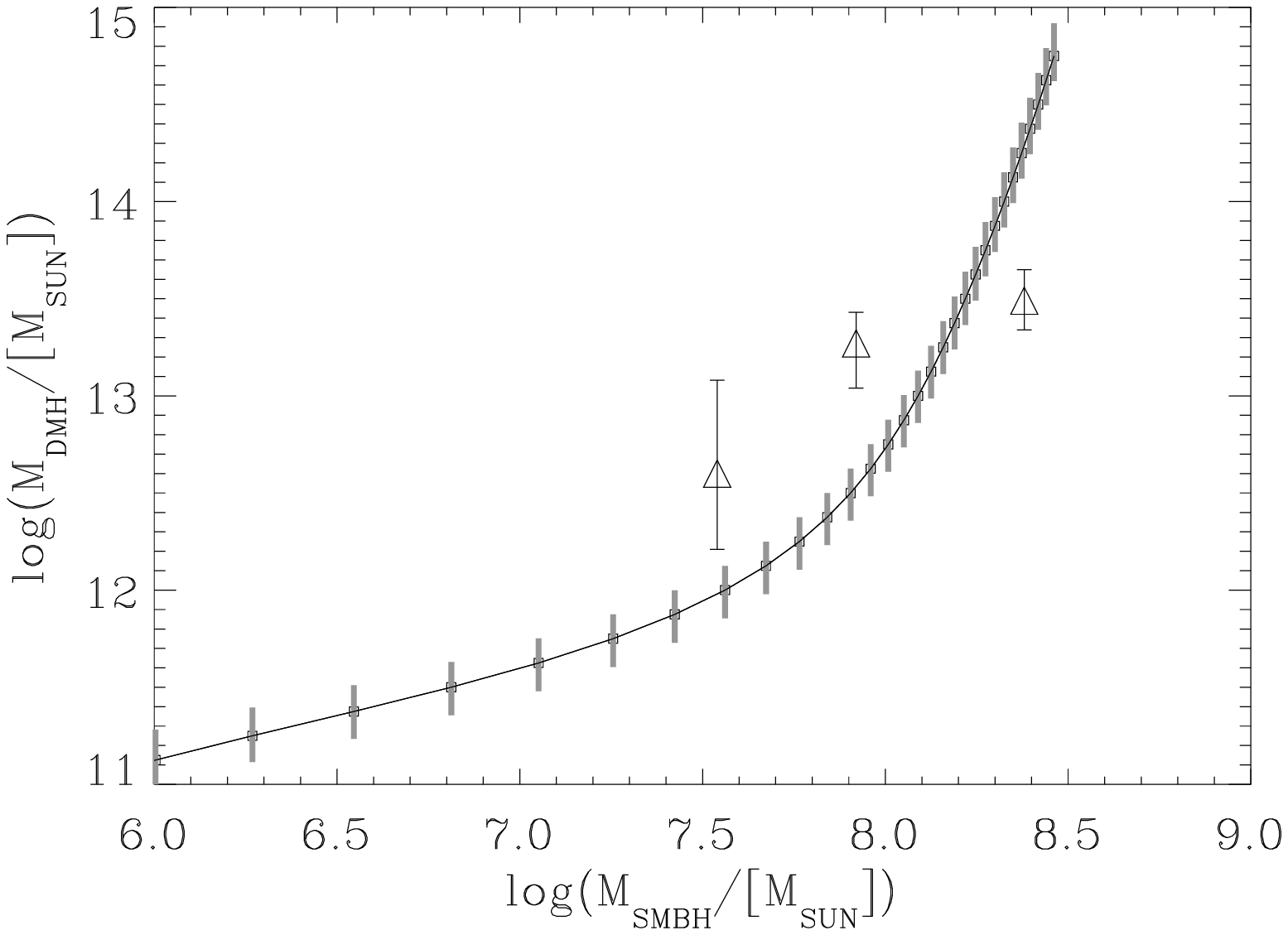}} 
\caption{Simple model prediction of the relation between the dark
 matter halo mass and SMBH mass, using \cite{behroozi_wechsler_2013} and
\cite{conroy_white_2013}. This simple model does not include any potential
effects that an AGN can have on the host galaxy or host halo.
The gray error bars of the
correlation reflect the observational errors in the data used to derive the 
\cite{behroozi_wechsler_2013} relationship between stellar mass and $M_{\rm DMH}$.
The triangles show our
clustering results for the X-ray-selected AGN sample when 
we divide the full sample into low, medium, and high $M_{\rm BH}$ samples.}
         \label{simple_model}
\end{figure}

We first compare our results with a simple empirically motivated model, 
rather than a full semianalytic model. We take the relations between galaxy stellar 
mass and DMH mass at $z \sim 0.3$ from \cite{behroozi_wechsler_2013}.
Next, we assign each galaxy a black hole with a mass that is 
determined by the stellar mass of the host galaxy, following
\cite{conroy_white_2013} (their Eq.~1). No scatter in the stellar mass to
$M_{\rm BH}$ relationship is included in our treatment. 
 
The result of this simple model is shown in Fig.~\ref{simple_model}, along with our
measurements.  There are discrepancies between the model prediction and
measurements: at $M_{\rm BH} < 10^8 M_{\odot}$ the model predicts a
lower $M_{\rm DMH}$ than we find, though the difference is within 3$\sigma$.
At higher $M_{\rm BH}$, the model predicts a stronger relationship between 
$M_{\rm BH}$ and $M_{\rm DMH}$ than we find. While
it is understandable how our measurements could lie about the predicted line,
in that we measure the mass of the parent halo from the large-scale bias 
while the model predicts the mass of the subhalo hosting the AGN, our
measurement at $M_{\rm BH} \sim 10^{8.4} M_{\odot}$ falls below the prediction
(a 3.7$\sigma$ difference).

The shallower slope seen in the data compared to the model here
could be caused by the cumulative scatter in each of the correlations in the 
logical chain ($M_{\rm DMH}$ $\rightarrow$ $M_{\rm stellar}$ $\rightarrow$
$M_{\rm bulge}$ $\rightarrow$ $M_{\rm BH}$). 
We will evaluate the predictions of semianalytic 
models in Sect.~\ref{simulations} and will show that they describe our findings
better than the simple model does.

\subsection{Comparing Our Results to Other Studies}

\cite{zhang_wang_2013} and \cite{komiya_shirasaki_2013} study the projected quasar number density 
around galaxies. Both find evidence that the clustering scale length $r_0$ depends on 
$M_{\rm BH}$. \cite{zhang_wang_2013} detects this trend, however, at a significance level of 
only $\sim$1$\sigma$. They also note that there is no clustering dependence on
optical quasar luminosity. 
Based on SDSS AGN data, \cite{komiya_shirasaki_2013} find 
a trend that $r_0$ increases with $M_{\rm BH} > 10^{8}\,{\rm M}_{\odot}$. Our $r_0$ measurements for the 
$M_{\rm BH}$ subsamples (although at moderately different redshift ranges) agree with their results when 
we consider the 1$\sigma$ uncertainties. At lower black hole masses, \cite{komiya_shirasaki_2013}
do not find an $r_0$ dependence. This might be due to selection biases of the AGN population, as 
we will show below in Sect.~\ref{simulations}. 
Additionally, \cite{komiya_shirasaki_2013} do not find significant $r_0$ dependences on
AGN luminosity ($L_{5100}$). However, the dynamic range in their study is less than a magnitude.   

\cite{shen_strauss_2009} compute the ACF of optically selected 
broad-line SDSS AGNs (DR5) at 
$0.4 \leq z \leq 2.5$. They detect a $\sim$2$\sigma$ difference in the clustering strength 
when they split the AGN sample into the 10\% most massive $M_{\rm BH}$ and the
remaining 90\%, in that AGNs 
with the highest $M_{\rm BH}$ are more strongly clustered. Their results are consistent with our 
findings, although their sample extends up to $z=2.5$ and contains more massive black holes at 
higher redshifts. 
When \cite{shen_strauss_2009} split their sample into luminous and faint quasars, they found that 
both correlation functions agree with each other. Only the 10\% most luminous quasars show a 
larger clustering strength (at the $\sim$2$\sigma$ level) than the remaining 90\% of the sample. 
This result is also consistent with our findings, which we will discuss in the next subsection 
(see also Fig.~\ref{Mi_Mrest_bias}, left).\\


\subsection{Why Do We Not Detect an $M_{\rm i}$ Clustering Dependence in the
  Optically Selected AGN Sample?}
\label{why_optical}

Interestingly, we find very similar clustering dependences for the X-ray and
the optically selected AGN 
samples. However, an important question remains: if there are such similar dependences 
 in the clustering strength with $M_{\rm BH}$, $L/L_{\rm EDD}$, FWHM, and $L_{\rm H\alpha}$, 
why do we not see a clustering dependence with $M_{\rm i}$ when we see a dependence with $L_{\rm X}$? 
Within the X-ray-selected sample, we detect a very weak $M_{\rm i}$ clustering dependence, in that 
the faint and luminous $M_{\rm i}$ RASS AGN samples (see Table~\ref{xagn_acf}) differ by 1.2$\sigma$. 
The direct comparison between the $M_{\rm i}$ clustering dependence of the X-ray and 
optically selected sample in Fig.~\ref{Mi_Mrest_bias} (left) shows that only the faint 
optical $M_{\rm i}$ subsamples deviate from the trend detected in the X-ray-selected sample.
However, the difference is not significant considering the uncertainties of the 
measurements. 

Above $z\sim 0.28$ the H$\alpha$ line is shifted outside the SDSS $i$-band filter. 
The redshift range studied here includes sources above and below this redshift.
Thus, below $z\sim 0.28$ the broad-line H$\alpha$ line contributes 
flux to the $M_{\rm i}$ band, while above $z\sim 0.28$ this 
direct AGN luminosity indicator is not contributing to the $M_{\rm i}$ band flux. 
To explore if this difference in $M_{\rm i}$ contribution is responsible for the observed 
clustering difference in X-ray and optically selected AGN samples, we derive the 
absolute 5500--6800 \AA\ rest-frame magnitude for all X-ray and 
optically selected AGN. The band pass of this rest-frame magnitude is chosen in such 
a way that it includes the H$\alpha$ line for all objects in our samples. We compute the 
clustering dependence as a function of this rest-frame absolute magnitude 
(Fig.~\ref{Mi_Mrest_bias}, right) and find trends very similar to those 
as a function of $M_{\rm i}$ (Fig.~\ref{Mi_Mrest_bias}, left). Thus, 
shifting the H$\alpha$ line outside the $M_{\rm i}$ band does not explain the 
difference in the clustering signals. 

\begin{figure*}
\vspace*{0.4cm}
 \begin{minipage}[b]{0.48\textwidth}
\centering
 \resizebox{\hsize}{!}{
  \includegraphics[bbllx=89,bblly=364,bburx=536,bbury=700]{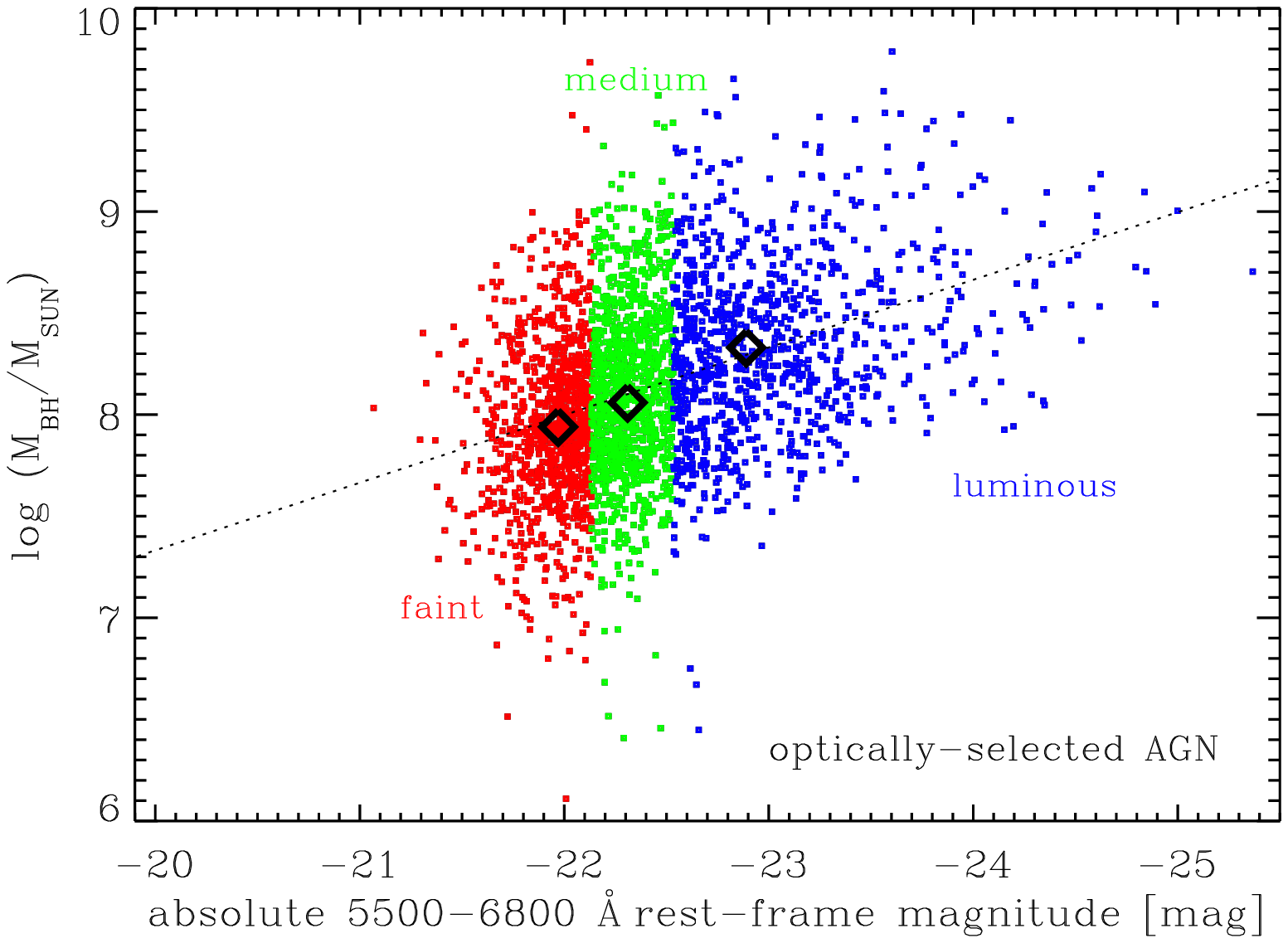}} 
\end{minipage}
\hfill
\begin{minipage}{0.48\textwidth}
\vspace*{-6.07cm}
\centering
\resizebox{\hsize}{!}{
   \includegraphics[bbllx=89,bblly=364,bburx=539,bbury=700]{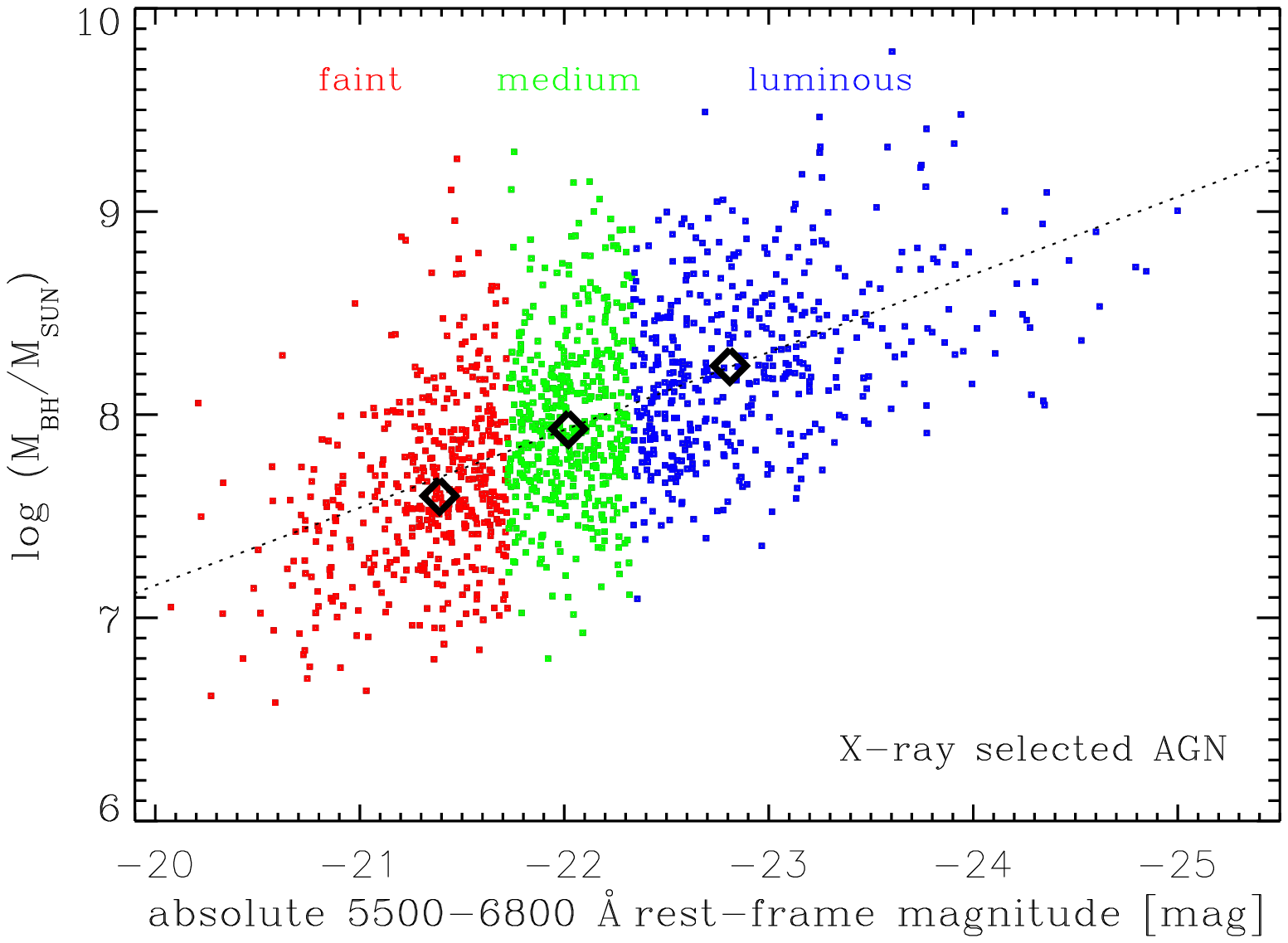}} 
\vspace*{-0.0cm}
\end{minipage}
\caption{
Black hole mass for optically selected SDSS (left) and X-ray-selected RASS/SDSS (right) AGNs
vs. absolute 5500--6800 \AA\ rest-frame magnitude.  
         Different colors denote the division of the full samples  
         into faint (33\%), medium (34\%), and luminous (33\%) objects with respect to the 
         rest-frame absolute magnitude. 
         The black diamonds show the median absolute magnitude and median $M_{\rm BH}$ 
         of the subsamples. We also show linear regression line fits to the full samples as dotted 
         lines.}
         \label{absMagrestframe_MBH}
\end{figure*}

\begin{figure*}
\vspace*{0.4cm}
 \begin{minipage}[b]{0.48\textwidth}
\centering
 \resizebox{\hsize}{!}{
  \includegraphics[bbllx=89,bblly=373,bburx=536,bbury=700]{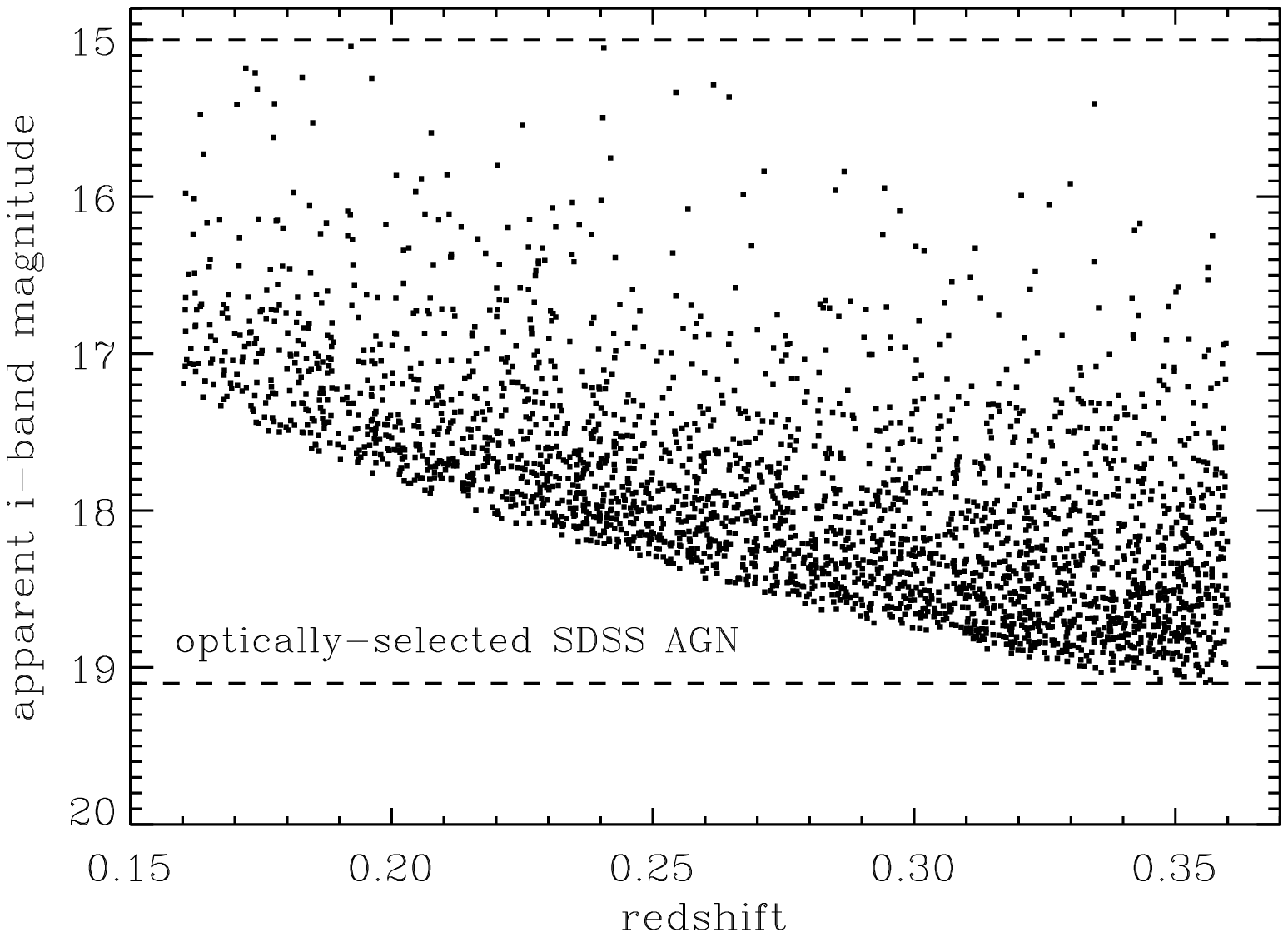}} 
\end{minipage}
\hfill
\begin{minipage}{0.48\textwidth}
\vspace*{-5.87cm}
\centering
\resizebox{\hsize}{!}{
   \includegraphics[bbllx=89,bblly=373,bburx=539,bbury=700]{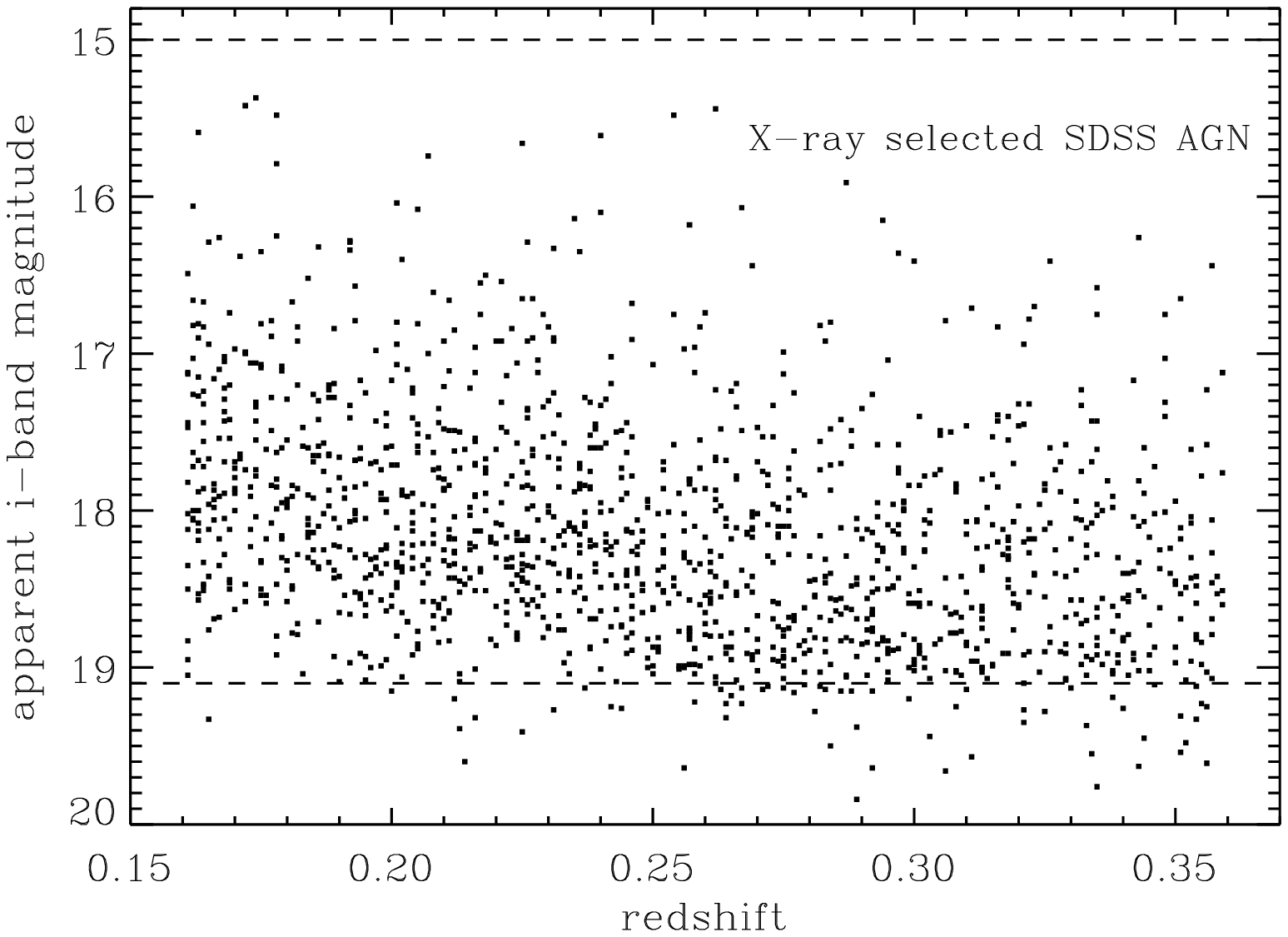}} 
\vspace*{-0.0cm}
\end{minipage}
\caption{Apparent SDSS $i$-band magnitude vs. redshift for optically selected AGNs (left panel) 
         and X-ray-selected AGNs (right panel). The horizontal dashed lines represent the
         lower and upper $i$-band limits for the selection of optical SDSS AGNs. 
         }
         \label{app_imag}
\vspace*{0.4cm}
\end{figure*}

Figure~\ref{absMagrestframe_MBH} shows that the luminous absolute magnitude subsamples for the 
X-ray and optically selected AGN cover a similar parameter space in the 
$M_{\rm BH}$ versus absolute 5500--6800 \AA\ rest-frame magnitude plane.  The median values 
of the samples are almost identical. Thus, it is not very surprising that these samples
have similar clustering strengths given their similar values of $M_{\rm i}$,  H$\alpha$, and 
absolute 5500--6800 \AA\ rest-frame magnitude.
On the other hand, Fig.~\ref{absMagrestframe_MBH} demonstrates that the faint absolute 
magnitude subsamples for the X-ray and optically selected AGN samples are substantially 
different. The X-ray sample extends (as in all other parameters) to much lower magnitudes. 
Thus, the dynamic range 
of the X-ray-selected RASS/SDSS sample is wider than that of the optical AGN sample. As a consequence, 
the medium absolute magnitude subsample of the X-ray-selected AGN sample includes the vast 
majority of the optical AGN in the faint absolute magnitude subsample. If one combines the 
faint and medium absolute magnitude subsamples in the optical to roughly match the medium X-ray subsample, 
the clustering strength between both AGN populations agrees very well 
($b_{\rm opt,faint+med.}=1.29^{+0.07}_{-0.07}$; see Fig.~\ref{Mi_Mrest_bias}). 

Finally, we explore the selection of the X-ray and optical AGNs in more detail. 
The RASS/SDSS sample is an X-ray flux-limited sample. 
\textit{ROSAT}'s soft energy range significantly biases the sample 
toward broad-line AGNs as these sources are unabsorbed (or only mildly absorbed) in the X-rays. 
In addition to the X-ray flux limit, the sample is also subject to an optical flux limit because 
an SDSS spectrum is required for the characterization of a source. However, the 
X-ray and optical luminosities of broad-line AGNs correlate well (e.g.,
\citealt{tananbaum_avni_1979}; \citealt{grupe_komossa_2010}; \citealt{lusso_comastri_2010}). 
Since \textit{ROSAT} has limited 
sensitivity, it is able to detect only the X-ray-brightest AGNs. These objects are typically 
also bright in the optical. Thus, the impact of the optical flux limit on the sample selection 
is moderate.

The optical SDSS AGN selection, on the other hand, is much more complex. The target selection 
relies on several different selection techniques. The main criterion is a color cut and an optical 
flux limit. Broad-line AGNs with a certain amount of extinction will not meet the color-cut criterion.
Additional objects are included in the sample due to their radio emission. Most importantly,
 an absolute magnitude cut is applied. 

Figure~\ref{app_imag} (left) shows 
that the optically selected AGN sample is not a flux-limited sample, although the sample selection 
considers only objects with $15.0<i<19.1$. The absolute $M_{\rm i}$ cut (applied to prevent substantial 
contamination by host galaxy light) causes a serious deficiency of objects above the lower $i$-band limit 
of $i<19.1$. The X-ray-selected AGN sample (Fig.~\ref{app_imag}, right) includes objects down 
to almost the same apparent $i$-band magnitude, independent of redshift. Various SDSS papers 
discuss how the optical SDSS AGN sample is not a uniformly selected sample 
(e.g., \citealt{richards_strauss_2006}). We conclude that the complex optical AGN selection 
might be the origin of the clustering differences between the X-ray and
optically selected 
AGN samples. The strongest differences are detected between the samples of the lowest median 
redshift; this is also the range in which the optical AGN sample deviates most from a clean,  
flux-limited sample. We therefore consider the X-ray-selected RASS/SDSS AGN sample as the 
more complete sample and use only their clustering properties in the next section.


\subsection{Comparison to cosmological simulations}
\label{simulations}
Hydrodynamical (e.g., \citealt{bryan_norman_2014}; \citealt{hirschmann_dolag_2014}; 
\citealt{steinborn_dolag_2015}) and semianalytical 
(e.g., \citealt{malbon_baugh_2007}; \citealt{marulli_bonoli_2008}; 
\citealt{fanidakis_baugh_2011}; \citealt{benson_2012}) state-of-the-art simulations 
have reached a level of complexity such that they can well predict observed quantities such as 
luminosity and stellar mass 
functions (e.g., \citealt{hopkins_hernquist_2006}; \citealt{degraf_dimatteo_2010}; 
\citealt{fanidakis_baugh_2012}). 
The observed clustering properties of galaxies and 
AGNs provide additional important constraints for such models (e.g., \citealt{bonoli_marulli_2009}). 

Theoretical models of galaxy formation that include prescriptions for modeling BH growth 
can provide important constraints on the fueling modes of AGN. For example, models in 
which AGN activity is triggered by major mergers between galaxies or secular processes 
within the host galaxy (e.g., disk instabilities) predict a well-defined correlation function 
for luminous quasars, with an average DMH mass of $M_{\rm DMH} \sim 10^{12}\,h^{-1}\,\rm{M}_\odot$ 
(\citealt{bonoli_marulli_2009,bonoli_2010}; \citealt{fanidakis_maccio_2013}). 
However, the inclusion of an additional mode of growth that is linked to diffuse gas 
accretion in massive halos (\citealt{fanidakis_baugh_2012}) predicts a much higher average 
halo mass for moderate-luminosity AGNs, in contrast to their luminous
counterparts (\citealt{fanidakis_georgakakis_2013}). Comparing 
this specific picture of AGN clustering with the available observations could potentially 
provide important insights into the correct modeling of the growth of black
holes in galaxy-formation models. 

Cosmological simulations and their predicted clustering properties can be also be used 
to test the impact of sample selection biases for observed samples.  
As an example, let us assume that in an unbiased sample there is no clustering 
dependence as a function of a certain AGN
parameter. A particular observational selection bias might cause a 
clustering dependence to be measured. As selection biases can be 
included in the cosmological simulation, one can therefore compare the
predictions for the unbiased and observed (including selection biases) AGN samples. 
Such knowledge is crucial to avoid wrong interpretations 
about the underlying physics when using clustering studies. Thus 
cosmological simulations are not only important to test our theoretical 
understanding of the observed universe but also to explore the impact of 
observational sample biases. 

Here we use the semianalytical galaxy formation model {\tt GALFORM} (\citealt{cole_lacey_2000}) 
to compare with our observed AGN clustering signals.  
\cite{fanidakis_baugh_2011,fanidakis_baugh_2012,fanidakis_georgakakis_2013}
modeled SMBH in these simulations, using an identical model in all three
papers. Only the cosmology has changed from WMAP1 (in the 
first two papers) to WMAP7 (\citealt{fanidakis_georgakakis_2013}). 
In \cite{fanidakis_georgakakis_2013}, the authors incorporate two modes of AGN accretion in the 
{\tt GALFORM} model. The first is the starburst mode (cold accretion), in which 
accretion onto the BH is tightly coupled to the mass of the cold gas available in the galaxy, 
which also contributes to star formation. This mode occurs when the host galaxy encounters a major 
galaxy merger, a minor merger, or a disk instability. These phenomena can take place in
a wide range of halo masses. However, observable AGN luminosities within the range defined in this study
are typically produced in DMHs with $M_{\rm DMH} < 10^{12.5}\,h^{-1}\,\rm{M}_\odot$ 
(\citealt{fanidakis_maccio_2013}). 

The second is the hot-halo mode (hot accretion), which occurs in DMHs with 
higher masses. A fraction of the AGN energy output is used to heat the gas in the DMH and suppress 
cooling of the gas. The SMBH accretes directly from the diffuse gas in the 
DMH. This mode regulates the black hole accretion in very massive halos with 
$M_{\rm DMH} \gtrsim 10^{13}\,h^{-1}\,\rm{M}_\odot$. For more details on both modes and their modeling, 
we refer the reader to \cite{fanidakis_georgakakis_2013} and the references within. 

\begin{figure}
  \centering
 \resizebox{\hsize}{!}{ 
  \includegraphics[bbllx=17,bblly=148,bburx=593,bbury=690]{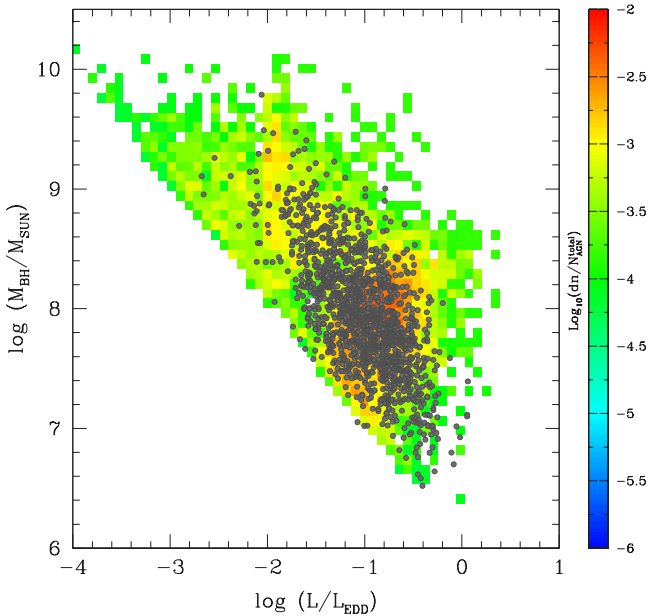}}
      \caption{Eddington ratio vs. mass of the SMBH of simulated (color) and observed (gray) X-ray-selected AGNs. 
The color denotes the number of AGNs above the RASS flux limit divided by 
the total number of AGNs in the simulation, after applying the RASS flux limit. The gray data points 
indicate the location of the observed X-ray-selected RASS/SDSS AGN sample (see Fig.~\ref{high_low_MBH}).}
         \label{sim_MBH_LLedd}
\end{figure}

The model of \cite{fanidakis_maccio_2013} is successful in reproducing several
observational constraints, including the observed luminosity functions of
galaxies and AGNs over a wide range of redshifts (\citealt{fanidakis_baugh_2011,fanidakis_baugh_2012}). 
In addition, they can reproduce reasonably well the observed 
$M_{\rm BH}$--$M_{\rm Gal,bulge}$ relationship and the global and active black hole mass functions.

The primary motivation for running the simulation is to explore how AGN clustering 
depends on the mode of AGN accretion, as described in the two modes mentioned above.  
Here we can thus compare our observational results directly to theoretical predictions. In addition, 
we can explore the effect of the RASS flux limit selection on our observational sample. 

\cite{fanidakis_georgakakis_2013} present the typical DMH masses for 
AGNs observed in a redshift range of $z=0$--$1.3$. Here we repeat this simulation in a redshift range of 
$z=0.16$--$0.36$, identical to the redshift range used for the observed X-ray
and optically selected 
AGN samples. For each galaxy, the simulation allows one to identify when the
AGN is active. An object is included in the AGN sample if its central engine is active in at least one
of the five logarithmically spaced snapshots output between $z=0.16$--$0.36$.
AGN properties such as $M_{\rm BH}$, $L_{\rm bol}$, $L/L_{\rm EDD}$, and redshift are available 
for each of the different snapshots. The parameter $L_{\rm bol}$ is calculated assuming that the accretion 
flow forms a geometrically thin disk for relatively high accretion rates (\citealt{shakura_sunyaev_1973}).
At lower accretion rates, a geometrically thick disk or an advection-dominated 
accretion flow is modeled (ADAF, \citealt{narayan_yi_1994}).

X-ray luminosities are calculated directly from 
$L_{\rm bol}$ by applying the bolometric correction from \cite{marconi_risaliti_2004}.
We consider an object to be an AGN if its intrinsic 2--10 keV rest-frame luminosity 
is $L_{\rm 2-10\,keV} \ge 10^{41.5}$ erg s$^{-1}$. The sample contains
X-ray unabsorbed and absorbed AGNs. Based on the luminosity, we use the empirical formula of \cite{hasinger_2008} to determine the 
probability that the object is absorbed. We then classify statistically if an object is unabsorbed or 
absorbed. Below, we will refer to this AGN sample as the 
``all simulated AGN'' sample.   
More recent estimates of the obscured fraction of AGNs as a function of luminosity have been 
presented by \cite{merloni_bongiorno_2014}, who 
show that the probability that an AGN at high luminosities is absorbed is
higher than that suggested by \cite{hasinger_2008}. We test the impact
of a higher obscured fraction at high luminosities and find insignificant 
differences as far as clustering measurements are concerned. We therefore 
retain the empirical obscuration relation of \cite{hasinger_2008} as presented in 
\cite{fanidakis_georgakakis_2013}.

\begin{figure*}
\vspace*{0.5cm}
 \begin{minipage}[b]{0.31\textwidth}
\centering
 \resizebox{\hsize}{!}{
  \includegraphics[bbllx=17,bblly=148,bburx=574,bbury=690]{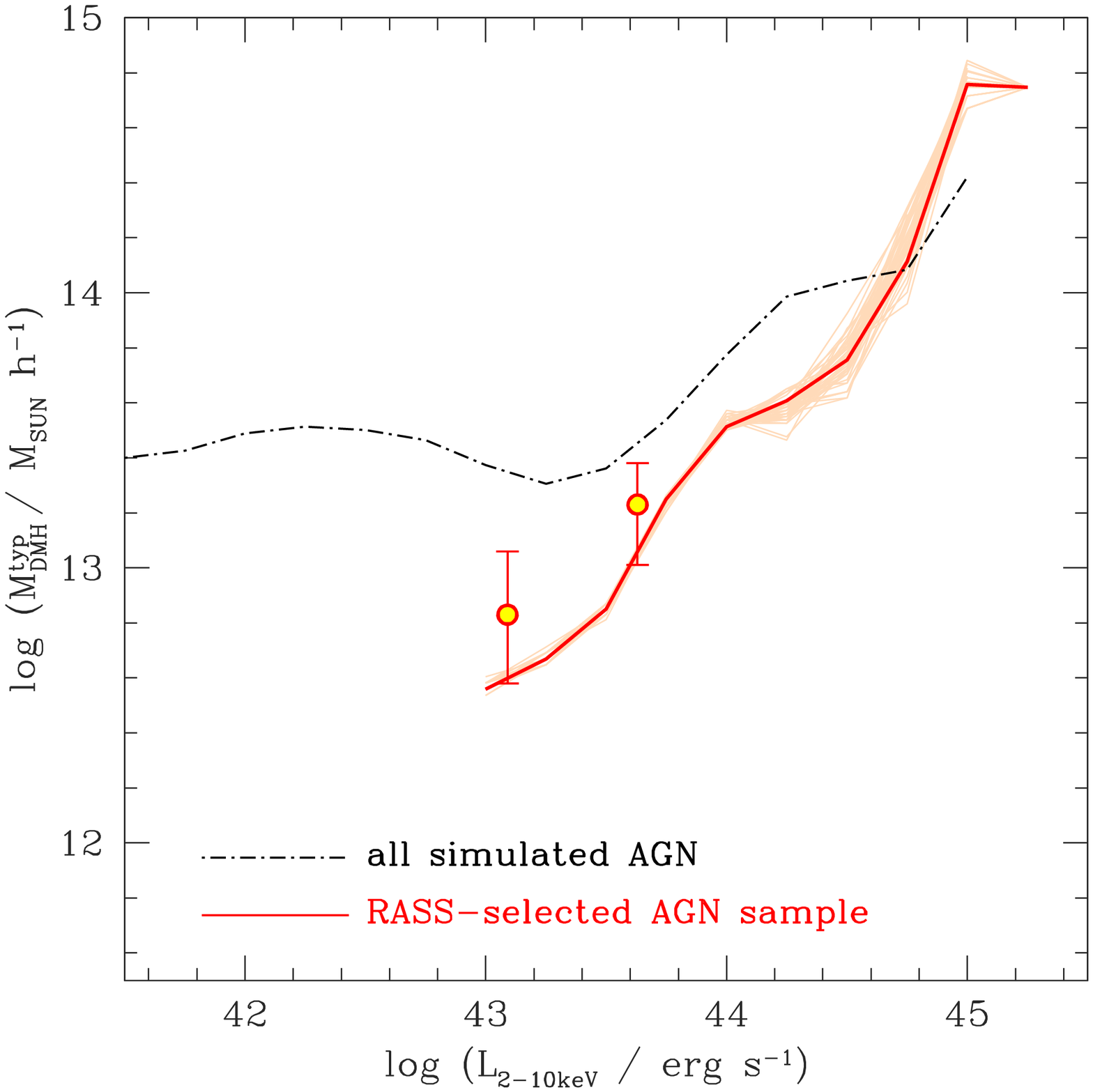}}
\end{minipage}
\hfill
\begin{minipage}{0.31\textwidth}
\vspace*{-5.3cm}
\centering
\resizebox{\hsize}{!}{
  \includegraphics[bbllx=17,bblly=148,bburx=574,bbury=690]{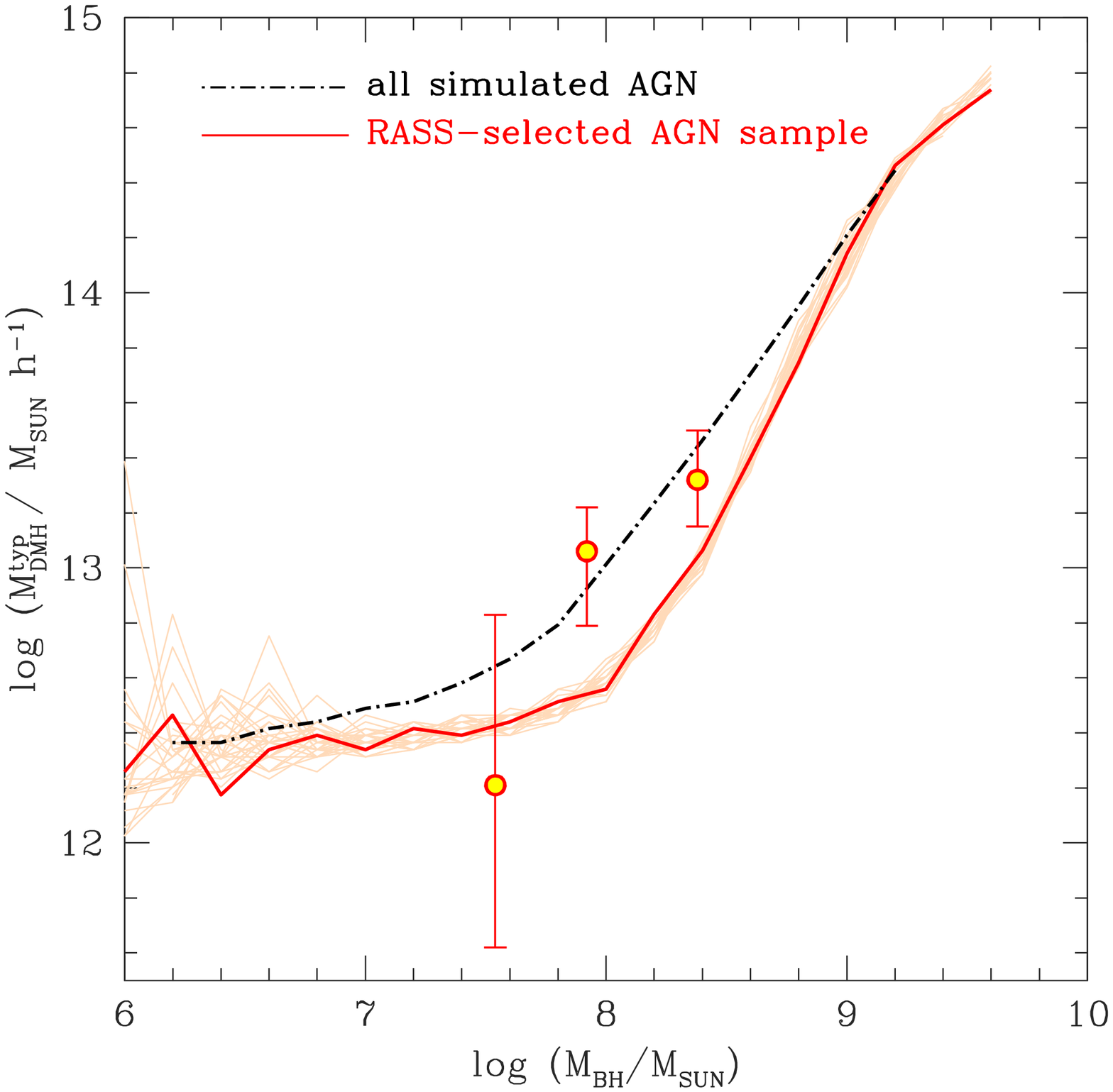}}
\end{minipage}
\hfill
\begin{minipage}{0.31\textwidth}
\vspace*{-5.3cm}
\centering
\resizebox{\hsize}{!}{
  \includegraphics[bbllx=17,bblly=148,bburx=574,bbury=690]{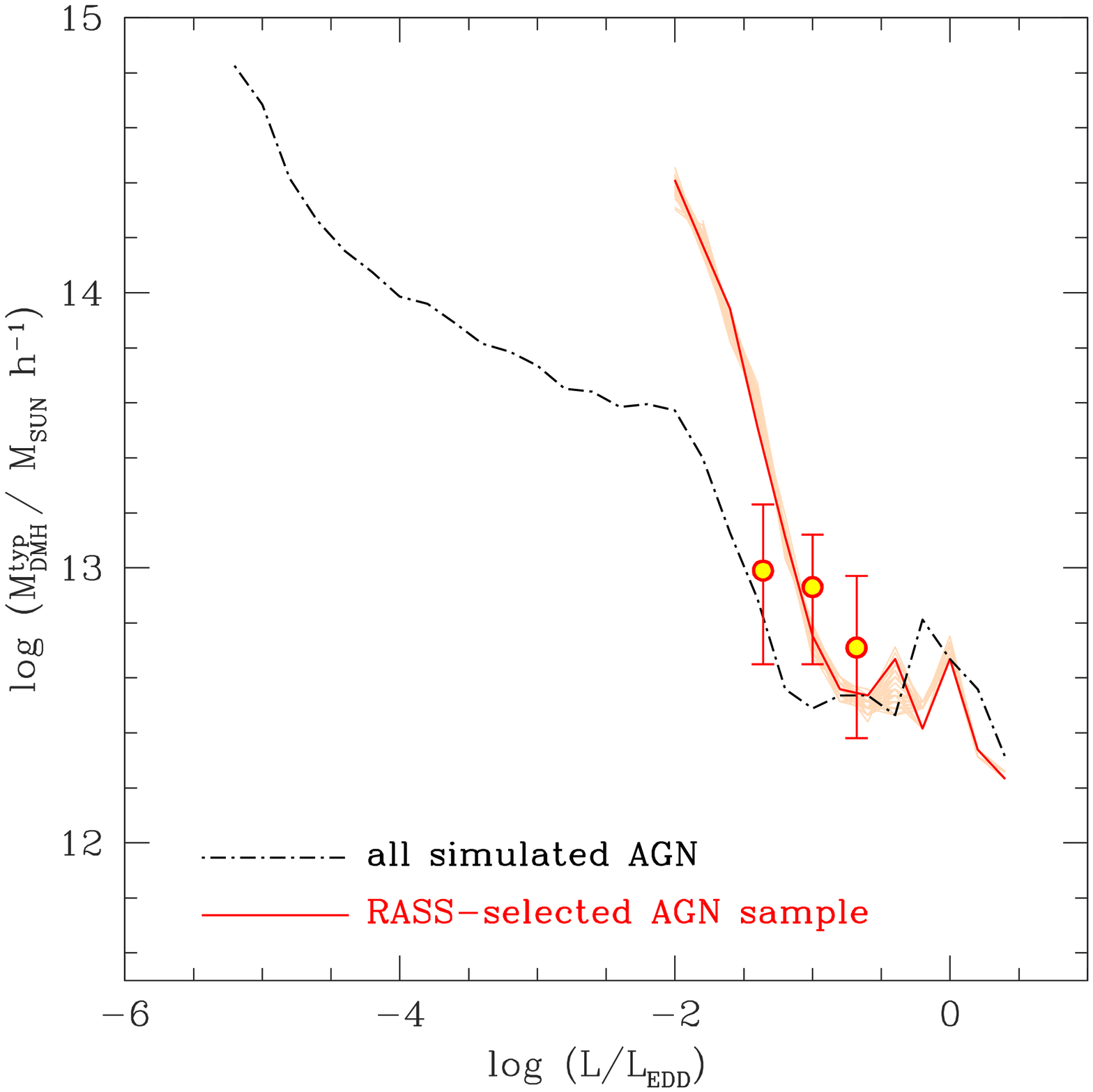}}
\end{minipage}
\caption{
\textit{Left:} comparison between the predicted (lines) and observed (data points) RASS AGN 
typical host dark matter halo mass as a function of hard (2--10 keV) X-ray luminosity.
The red line shows the prediction of a simulated RASS flux-limited sample
(and log $(L/L_{\rm EDD}) > -2$), 
and the black dash-dotted line shows the expected dark matter halo mass 
for the total (unbiased) AGN sample with $L_{\rm 2-10\,keV} \ge 10^{41.5}$ erg
s$^{-1}$. 
We recalculate the relations after adding a random error from a Gaussian distribution with
$\sigma = 0.3$ dex to the simulated SMBH masses. We run 50 such
realizations; a random subset of these is shown here as orange lines.
The data points are plotted at the corresponding median $L_{\rm 2-10\,keV}$ values 
of the observed low and high 
AGN subsamples (see Table~\ref{xagn_acf}, using the conversion presented in Sect.~\ref{RASS_sample}).
\textit{Middle:} similar to the left panel, here showing the predicted and observed halo mass 
as a function of $M_{\rm BH}$. 
\textit{Right:} similar to the left panel, here showing the predicted and observed halo mass 
as a function of Eddington ratio ($L/L_{\rm EDD}$).}
\label{fig:simulation}
\vspace{0.7cm}
\end{figure*}

Since RASS is essentially a flux-limited survey (see also Fig.~\ref{app_imag}, right), 
only AGNs above a certain X-ray flux 
are contained in the RASS AGN catalog. However, due to the characteristics of the 
RASS, different regions of the sky have different exposure times. 
To mimic the RASS flux selection, we also apply an average flux cut to the 
simulated AGN sample by using the X-ray luminosities and redshifts 
of objects in the simulation. We refer to this sample as the ``RASS-selected AGN sample.'' 
Its location in the 
$M_{\rm BH}$ versus $L/L_{\rm EDD}$ plane is shown in Fig.~\ref{sim_MBH_LLedd}
(colored grid points). 
This sample also includes the effect that RASS is sensitive only to unabsorbed X-ray AGNs. 
However, since the luminosities of the RASS AGNs are high, the simulation classifies the vast 
majority of the objects as X-ray unabsorbed. 

In the same plane, we also show the observed X-ray-selected 
broad-line RASS/SDSS AGN sample. In general, the simulation agrees well with the 
observations, in that sources lie in a similar region of this space. 
However, the simulation slightly overestimates the number of AGNs with $M_{\rm BH} > 10^{8.5}\,\rm{M}_\odot$ 
and log $(L/L_{\rm EDD}) < -2$. Such objects should be detectable given the RASS 
flux limit, but they do not exist in large numbers (at the observed 
redshift range). However, these are the objects in the simulation for which the accretion flow
is modeled with an ADAF. 
These AGNs are observationally associated with radio-bright,
mechanical feedback-dominated SMBHs (e.g., \citealt{churazov_sazonov_2005};
\citealt{hickox_jones_2009}; \citealt{smolcic_2009}).
The optical spectra of these objects do not contain 
any high excitation lines typical of AGNs (e.g., \citealt{best_heckman_2012}). Thus, one might argue 
that these objects would not be classified as AGNs in an SDSS spectrum. We
therefore limit the simulated RASS-selected AGN sample (and the predictions in
the following paragraphs) to those sources with log $(L/L_{\rm EDD}) > -2$.

Despite these minor differences, 
we calculate the bias parameter for the simulated AGNs and the observed RASS-selected AGN sample. 
The bias is calculated in bins of 0.2 in log $(M_{\rm BH}/\rm{M}_\odot)$ and log $(L/L_{\rm EDD})$
using Eq.~4 given in \cite{fanidakis_georgakakis_2013} (see also Eq.~8 in this paper). 
We follow the same procedure as for the observed AGN samples. Thus, 
the bias values are converted into typical DMH masses 
(log $M_{\rm DMH}^{\rm typ}$). Since the simulation uses the conversion given by 
\cite{sheth_mo_2001} without applying the improved fitting formula by 
\cite{tinker_weinberg_2005}, we recompute for the observed $L_{\rm X}$, $M_{\rm BH}$, and 
$L/L_{\rm EDD}$ AGN subsamples in Table~\ref{xagn_acf} $M_{\rm DMH}^{\rm typ}$ based only 
on the ellipsoidal collapse model of \cite{sheth_mo_2001}. This allows us 
to directly compare the predictions with the observations, as shown in 
Fig.~\ref{fig:simulation}.

In the case of the simulated RASS-selected AGN sample, 
we also assign to every BH mass a random error from a Gaussian distribution with
$\sigma = 0.3$ dex. We recalculate all of the relations (red lines) shown in
Fig.~\ref{fig:simulation} for 50 realizations and show a subset of these as orange
lines. This exercise does not only allows for a more realistic comparison to the data, but
also allows us to explore the impact of the intrinsic scatter of 
approximately 0.3 dex
in the $M_{\rm BH}$ estimate of the observed sample. Fig.~\ref{fig:simulation}
shows that the 0.3 dex uncertainty in the BH mass estimates has no significant
impact on the prediction of the correlations shown. Only at very low $M_{\rm BH}$
and high $L/L_{\rm EDD}$ does the scatter between different realizations
increase. However, this is caused by the low number of objects with such
properties in the simulations.

For the all-simulated AGN sample (black line), the simulation predicts a positive clustering 
dependence on X-ray luminosity only above log $(L_{\rm 2-10}/[\rm{erg}\,\rm{s}^{-1}]) \sim 44$. 
This might explain why studies of moderate X-ray luminosity do not see 
a clustering dependence on luminosity and conflicting results are presented in the literature. 
The RASS selection results in a luminosity dependence that extends down to 
log $(L_{\rm 2-10}/[\rm{erg}\,\rm{s}^{-1}]) \sim 43$ (red line). The data and the predictions for
the RASS sample agree remarkably well. 

For the $M_{\rm BH}$ clustering dependence (Fig.~\ref{fig:simulation}, middle),
the simulation predicts a strong clustering dependence with $M_{\rm BH}$ for the all
simulated AGN sample; a 
steady increase of typical DMH mass with $M_{\rm BH}$ is found (see also Fig.~7 in 
\citealt{fanidakis_baugh_2012}). Above 
$M_{\rm BH} \sim 10^{8}\,\rm{M}_\odot$ the correlation is even stronger. 
The RASS flux limit selection of the all-simulated AGN sample has a 
moderate impact on the $M_{\rm BH}$ clustering dependence. It amplifies 
the correlation at $M_{\rm BH} \gtrsim 10^{8}\,\rm{M}_\odot$ and weakens it below 
$M_{\rm BH} \sim 10^{8}\,\rm{M}_\odot$ compared to an unbiased AGN 
sample. The model shows that, in flux-limited samples below 
$M_{\rm BH} \sim 10^{8}\,\rm{M}_\odot$, no clustering correlation with 
$M_{\rm BH}$ should be detectable, although an unbiased AGN sample will 
still show a very weak $M_{\rm BH}$ clustering dependence in this $M_{\rm BH}$ range. 
This is consistent with the result found by \cite{komiya_shirasaki_2013}. 

The observed $M_{\rm BH}$ clustering dependence agrees reasonably well with the model 
prediction. The data hint that the observed clustering dependence might not 
be as strong as predicted by the simulated RASS-selected AGN sample. 

The predicted $L/L_{\rm EDD}$ clustering dependence also matches the observations 
well (Fig.~\ref{fig:simulation}, right).  
Below $L/L_{\rm EDD} \sim 10^{-2}$ the model predicts higher DMH masses
than are currently detected for AGNs, but our data do not probe Eddington ratios that low. 
The model predicts that, at low $L/L_{\rm EDD}$ values, the RASS flux limit selection should cause a very 
strong negative $L/L_{\rm EDD}$ dependence to the clustering amplitude. 
The data point with the lowest $L/L_{\rm EDD}$ in the 
RASS/SDSS AGN sample does not allow us to verify or reject this prediction. 
Although our samples contain objects down to $L/L_{\rm EDD} \sim 10^{-2}$ 
(see Fig.~\ref{RASS_AGN_MBH_LLEDD}), AGN samples with even lower $L/L_{\rm EDD}$ 
are needed to critically test the prediction of high $M_{\rm DMH}$ at low $L/L_{\rm EDD}$.

Such tests between observations and state-of-the-art simulations offer a 
unique opportunity to constrain the physical mechanisms included in galaxy and AGN formation 
models. For example, a mechanism for the hot-halo mode in which black holes with 
low $L/L_{\rm EDD}$ and high $M_{\rm BH}$ reproduce the real RASS-selected sample might 
agree better with the observed AGN clustering dependences. Thus, a channel 
has to be found to remove the simulated objects with 
$M_{\rm BH}\gtrsim10^{8.5}\,{\rm M}_{\odot}$ and log $(L/L_{\rm EDD}) \lesssim -2$.
In the current model, the hot-halo AGNs arise from the necessity for AGN feedback in 
massive DMHs. The accretion rate in this mode is calculated indirectly 
from the cooling properties of the host halo and corresponds to an accretion flow 
that is able to reproduce powerful AGN outflows that suppress gas cooling 
and star formation. The tuning of the free parameters in this calculation is done by 
fitting the model to the observed galaxy luminosity function at $z=0$. Obviously, 
more observational data are needed to contrain the model. The comparison between 
the simulated and real RASS-selected AGN samples in Figs.~\ref{sim_MBH_LLedd} 
and \ref{fig:simulation} could be 
used as additional constraints for constraining the free parameters in the model.

When cosmological simulations such as the one presented here are adjusted to
match the observed 
AGN clustering properties, to be considered successful they still have to match other observational 
constraints, such as the luminosity function of galaxies and AGNs at different redshifts. 
Changes in the physical treatment of AGN accretion might also solve other challenges 
such as the observed deficiency of low-luminosity AGN at high redshifts when compared 
to predictions from simulations (e.g., \citealt{miyaji_2015}). Including various 
constraints from galaxy and AGN clustering measurements has thus the power not only 
to improve the simulations but also to enhance our general understanding of AGN physics 
and AGN and galaxy coevolution. \\


\section{Conclusions}

Motivated by the detection of a weak X-ray luminosity dependence of the clustering strength of AGNs 
in the first paper of this series, here we explore the physical origin of this dependence. Using the 
optical spectra of our soft X-ray-selected (RASS/SDSS) luminous, broad-line AGN sample at $0.16<0.36$, 
we estimate black hole masses and Eddington ratios and calculate the clustering dependence on each.

Since $M_{\rm BH}$ and $L/L_{\rm EDD}$ are correlated, we create subsamples in $M_{\rm BH}$ and 
$L/L_{\rm EDD}$ that have matched distributions in the other parameter of interest. We compute 
the clustering strength for the subsamples and find a weak clustering dependence with $M_{\rm BH}$
and no significant correlation with $L/L_{\rm EDD}$. Various adjustments in how the subsamples are created 
do not change the results. We also study the clustering of the observed parameters (luminosity and 
FWHM of the broad H$\alpha$ line) that are used to derive $M_{\rm BH}$. 
We find a weak FWHM$_{\rm H\alpha}$ clustering correlation in that AGNs with low 
FWHM$_{\rm H\alpha}$ are less clustered than their high FWHM$_{\rm H\alpha}$ counterparts.

We also study the clustering properties of an optically selected SDSS AGN sample. 
This sample has 29\% of its objects in common with the X-ray-selected RASS/SDSS AGN sample.  
We detect the same trends as found for the X-ray AGN sample with respect to 
$M_{\rm BH}$, $L/L_{\rm EDD}$, $L_{\rm H\alpha}$, and FWHM$_{\rm
  H\alpha}$. The X-ray and optically selected 
AGN samples show divergent clustering signals for the lowest $M_i$ (absolute $i$-band magnitude) samples. 
We argue that this is caused by various complex selection effects in the optical sample and 
that the X-ray-selected RASS/SDSS AGN sample is more uniformly selected. 

From our correlation function measurements, we conclude that $M_{\rm BH}$ is the origin of the 
observed weak X-ray luminosity clustering dependence. The confidence contours
of our AGN HOD modeling parameters 
further support this finding, as low and high $M_{\rm BH}$ and $L_{\rm X}$ samples show extremely 
similar contours, while the low and high $L/L_{\rm EDD}$ sample contours differ significantly 
from the $L_{\rm X}$ contours. 

The $M_{\rm BH}$ clustering dependence is detected at a significance level of 2.7$\sigma$.
In both the X-ray and optical AGN samples, the highest clustering strength is found for 
AGNs with the 30\% highest $M_{\rm BH}$. Thus, at a redshift range of $0.16<z<0.36$ 
luminous broad-line AGNs with more massive 
$M_{\rm BH}$ reside, on average, in more massive DMHs. In this context, the 
DMH mass refers to the single largest (parent) halo mass. 

Since the observed clustering strength does not depend on $L/L_{\rm EDD}$ in 
a statistically significant way, AGNs with high accretion rates do not require large-scale 
dense environments with high galaxy density. This provides evidence that major or minor mergers 
play only a limited role in the AGN accretion processes in the low-redshift Universe. 
Internal processes such as 
disk instabilities could be the dominant AGN triggering mechanism at late cosmic times.  

Empirically motivated models that include simple monotonic relationships
between $M_{\rm DMH}$, $M_{\rm stellar}$, and $M_{\rm BH}$ and without AGN
feedback or treatments of AGN triggering do not match well the observed
relationships between $M_{\rm BH}$ and $M_{\rm DMH}$ derived from clustering.
Thus, we use a semianalytical model to study the predicted clustering dependences as a function 
of $L_{\rm X}$, $M_{\rm BH}$, and $L/L_{\rm EDD}$. This model included AGN
feedback and two modes of AGN accretion and fits the observed $M_{\rm BH}$
vs. $M_{\rm DMH}$ relation better than the simple model. Comparing the simulated full AGN sample with a simulated 
RASS flux-limited AGN sample, we show that observational selection effects moderately change the 
expected clustering dependences. The simulation predicts that higher $M_{\rm BH}$ are found in 
more massive halos, and the predicted correlations with $L_{\rm X}$, $M_{\rm BH}$, and $L/L_{\rm EDD}$ 
agree reasonably well with our observations presented here. 

Clustering measurements with small uncertainties offer a unique opportunity, 
after considering the effects of selection biases, to identify missing physics 
in our current understanding of the galaxy and AGN evolution by comparing 
theoretical predictions with observations. Thus, 
future clustering measurements as a function of various galaxy and AGN parameters 
will deliver needed additional constraints that will improve upon and distinguish between
current theoretical models. 


\acknowledgments
\section{acknowledgments}
We would like to thank Andreas Schulze, Scott Croom, and 
Antonis Georgakakis for helpful discussions. We also like to thank Peter
Behroozi for providing the relation between galaxy stellar mass and dark
matter halo mass at $z\sim 0.3$. 

This work has been supported by DFG grant KR 3338/3-1, NASA grant NNX07AT02G,  CONACyT Grant Cient\'ifica B\'asica 
\#179662, UNAM-DGAPA Grants PAPIIT IN104113 and IN109710.
The research leading to these results has received funding from the European
Community's Seventh Framework Programme (/FP7/2007-2013/) under grant agreement
No 229517. ALC acknowledges support from NSF CAREER award AST-1055081.

The \textit{ROSAT} Project was supported by the Bundesministerium f{\"u}r Bildung 
und Forschung (BMBF/DLR) and the Max-Planck-Gesellschaft (MPG).
Funding for the Sloan Digital Sky Survey (SDSS) has been 
provided by the Alfred P. Sloan Foundation, the participating 
institutions, the National Aeronautics and Space Administration, 
the National Science Foundation, the U.S. Department of Energy, 
the Japanese Monbukagakusho, and the Max Planck Society. 
The SDSS Web site is http://www.sdss.org/.

The SDSS is managed by the Astrophysical Research Consortium (ARC) 
for the Participating Institutions. The participating institutions 
are The University of Chicago, Fermilab, the Institute for Advanced 
Study, the Japan Participation Group, The Johns Hopkins University, 
Los Alamos National Laboratory, the Max-Planck-Institute for 
Astronomy (MPIA), the Max-Planck-Institute for Astrophysics (MPA), 
New Mexico State University, University of Pittsburgh, Princeton 
University, the United States Naval Observatory, and the 
University of Washington. 

This research also made use of a computing facility 
available from Departmento de Superc\'omputo, DGSCA, UNAM. \\



\end{document}